\documentclass[submission, LectureNotes]{SciPost}

\usepackage{braket}
\usepackage{cancel}
\usepackage{enumerate}
\usepackage{bbm}
\usepackage{bm}
\usepackage{nicematrix}
\usepackage{tikz}
\usepackage{mathtools}
\usepackage{import}

\usepackage{rotating}
\usepackage{dutchcal}
\usepackage{hyperref}
\usepackage{cleveref}
\usepackage{amsfonts}
\usepackage[most]{tcolorbox}
\usepackage{orcidlink}

\hypersetup{
    colorlinks=true,
    linkcolor=blue,
    citecolor=blue,
    urlcolor=blue
}

\usetikzlibrary{
    3d,
    angles,
    arrows.meta,
    automata,
    backgrounds,
    calc,
    decorations.pathreplacing,
    fadings,
    fit,
    intersections,
    petri,
    quotes,
    snakes,
    shadings,
    shadows,
    shapes,
    shapes.geometric,
    topaths
    }

\newcommand{\totenergy}{E}
\newcommand{\intenergy}{\varepsilon}

\newcommand{\mymat}[1]{\mathbf{#1}}%
\newcommand{\myvec}[1]{\vec{\bm{#1}}}%
\newcommand{\norm}[1]{\lVert #1 \rVert}

\newcommand{\multiindex}[1]{\mathbf{#1}}
\newcommand{\kasymp}[2]{\ket{#2}\otimes\ket{\multiindex{#1}}}
\newcommand{\basymp}[2]{\bra{#2}\otimes\bra{\multiindex{#1}}}
\newcommand{\obasymp}[2]{\bra{#2}\otimes\langle\multiindex{#1}}
\newcommand{\kint}[1]{\ket{\Psi_{\multiindex{#1}}^{k_{\ell}}}}%
\newcommand{\bint}[1]{\bra{\Psi_{\multiindex{#1}}^{k_{\ell}}}}
\newcommand{\obint}[1]{\langle\Psi_{\multiindex{#1}}^{k_{\ell}}}
\newcommand{\op}[1]{\mathbf{\hat{#1}}}
\newcommand{\myvecop}[1]{\myvec{\hat{#1}}}
\newcommand{\infin}[1]{\mathrm{d}#1}

\def\tot{_{\mathrm{tot}}}

\def\matnumvar{\mymat{y}}
\def\matnumlinsc{\mymat{s}}

\def\matnumratiosc{\mymat{q}}
\def\matnumdiffeq{\mymat{w}}

\def\numlinsc{s}

\def\numratiosc{q}

\def\opell{\pmb{\hat{\ell}}}

\def\jcsshift{-5.5cm}

\def\qmat{\matnumratiosc}

\def\channelwavenumber{k_{\alpha\beta}}
\def\matchannelwavenumber{\mymat{k}}
\def\channelwavevec{\myvec{k}_{\alpha\beta}}

\def\ampminus{a_{\multiindex{n'}\leftarrow \multiindex{n}}^{(-)}} %
\def\ampplus{a_{\multiindex{n'}\leftarrow \multiindex{n}}^{(+)}}
\def\amppm{a_{\multiindex{n'}\leftarrow \multiindex{n}}^{(\pm)}}

\def\Acoords{\left\{\mathcal{R}_{A}\right\}}
\def\Bcoords{\left\{\mathcal{R}_{B}\right\}}

\DeclareMathOperator{\sinc}{sinc}

\newcounter{keyresult}[section]
\renewcommand{\thekeyresult}{\thesection.\arabic{keyresult}}
\crefname{keyresult}{key result}{key results}
\Crefname{keyresult}{Key result}{Key results}
\newcommand{\kr}[1]{%
  \refstepcounter{keyresult}%
  \par\smallskip\noindent
  \textbf{Key result~\thekeyresult.}\enspace #1\par}
\newtcolorbox{keyresults}{%
  enhanced, breakable,
  colback=blue!3, colframe=blue!45!black,
  boxrule=0.6pt, arc=2pt,
  left=8pt, right=8pt, top=6pt, bottom=8pt,
  fonttitle=\bfseries\sffamily, coltitle=white,
  title={Key results}}

\definecolor{revnewcolor}{HTML}{C2185B}

\tikzstyle{widthrect} = [
rectangle,
text width=4cm,
line width=1.0mm
]

\tikzstyle{insidefit} = [
widthrect,text centered]

\tikzstyle{startstop} = [
widthrect, rounded corners, 
text centered, 
draw=black, 
]

\tikzstyle{process} = [
widthrect,
text centered, 
draw=ForestGreen,
rounded corners]

\tikzstyle{decision} = [regular polygon, regular polygon sides=8, yscale=1.0,
line width=1mm,
rounded corners,
minimum width=2cm, 
minimum height=2cm, 
text width=3.5cm,
text centered, 
draw=red]

\tikzstyle{morph} = [widthrect, 
minimum width=4 cm,
minimum height=3 cm,
text centered,
rounded corners,
draw=blue]

\tikzstyle{exp} = [widthrect, 
minimum width=4 cm,
minimum height=3 cm,
text centered,
draw=black]

\tikzstyle{done} =[widthrect, rounded corners, 
minimum width=3cm, 
minimum height=1cm,
text centered,
draw=green, 
fill=green!20]

\tikzstyle{arrow} = [->,>=stealth,line width=1.0mm]

\tikzset{
pot_curve/.style={
    xshift=-150,
    yshift=-80,
    xscale=5.6, 			
    yscale=1.0, 
    domain=0.97:2.4, 
    smooth,           	              
    variable=\x, 
    samples=100, 
    anchor=center,                 	      
    opacity=1.0, 
    line cap=round, 
    shorten >=-2pt,                  	      
    line width=1.5pt},
penning_lower_curve/.style={
    pot_curve,
    xscale=0.18,
    xshift=-35,
    yshift=90,
    domain=4.8:11},
penning_upper_curve/.style={
    pot_curve,
    xscale=1.,
    xshift=10,
    yshift=145,
    domain=0.88:1.4},
gaussian/.style={
    xscale=0.3,
    yscale=0.4,
    domain=-1.:1.,
    smooth,           	              
    variable=\x, 
    samples=100, 
    anchor=center,                 	      
    opacity=1.0, 
    line cap=round, 
    shorten >=-2pt,                  	      
    line width=1.1pt},
->/.style={-stealth,line width=0.8pt},
<->/.style={stealth-stealth,line width=0.8pt}
}

\usepackage{import}

\begin{document}

\begin{center}{\Large \textbf{
A tutorial on quantum scattering theory and coupled-channel calculations of bimolecular collisions
}}\end{center}

\begin{center}
    K. P. Horn\textsuperscript{1,2},
    C. P. Koch\textsuperscript{1*}\,\orcidlink{0000-0001-6285-5766}
\end{center}

\begin{center}
{\bf 1} Freie Universität Berlin, Dahlem Center for
Complex Quantum Systems and Fachbereich Physik, 14195
Berlin, Germany
\\
{\bf 2}  Department of Physics, Technische
Universität Dortmund, 44227 Dortmund, Germany
\\
* christiane.koch@fu-berlin.de
\end{center}

\begin{center}
\today
\end{center}

\markboth{}{}
\section*{Abstract}
{\bf
Collisions between atoms and molecules govern how energy is exchanged and redistributed in the gas phase, underpinning phenomena from spectral line broadening and interstellar molecular excitation to the cooling and trapping of molecules.
A quantitative account of these collisions requires connecting the theoretical description of the interaction between the collision partners to measurable quantities such as cross sections.
However, scattering theory as presented in quantum mechanics textbooks does not readily translate %
into numerical methods. %
Here, we formulate the scattering problem  
in Dirac notation, as customary in present-day quantum mechanics courses, 
and %
show how to build a practical numerical framework from it. 
We derive the `coupled channel' representation of the scattering problem with 
correct asymptotic behaviour and introduce Numerov's method to obtain 
time-independent wavefunctions, from which we calculate cross sections, the observables measured in scattering experiments.
Starting from single channel scattering, we build up to the realistic 
collision of a ground-state atom with a diatomic molecule.

}

\vspace{10pt}
\noindent\rule{\textwidth}{1pt}
\tableofcontents\thispagestyle{fancy}
\noindent\rule{\textwidth}{1pt}
\vspace{10pt}

\section*{About this tutorial}
\addcontentsline{toc}{section}{About this tutorial}
\markboth{About this tutorial}{About this tutorial}

\paragraph{Motivation}
The aim of this tutorial is to bridge the gap between the abstract treatment of quantum scattering theory found in many textbooks and the hands-on numerical methods required for taking on real world collision simulation problems.
Readers interested in textbooks on scattering theory have a rich choice,
ranging from formal treatments~\cite{taylor2006scattering} to texts dedicated to molecular collisions~\cite{child1996molecular,krems2018molecules}.
Whilst the review \cite{hutson1994coupled} sets an emphasis on contrasting
different available coupled channel methods for the evaluation of bound states,
the documentation of publicly available scattering programmes 
such as MOLSCAT \cite{hutson2019molscat} and Hibridon \cite{alexander2023hibridon}
can also prove to be a rich resource 
to the engaged learner.
Our hope is that our tutorial is set apart from these other approaches by 
offering a single, consistent derivation which leads all the way from the 
abstract formulation of the quantum mechanical scattering problem in Dirac
notation to the formulas that actually get implemented.
We aim to state all conventions, such as normalisation, phases and index ordering, 
as explicitly as possible, allowing each step to be followed and verified.
To this end, the content of this tutorial is divided into 4 sections, two of which introduce readers with a background in quantum mechanics (but not specifically scattering theory) to the abstract formulation and solution of a two-particle collision problem.  The next two sections then give concrete prescriptions for turning the formulated mathematical problem (either single channel elastic or atom-diatom collision) into a numerical one that actually can be solved in practice.

\paragraph{Prerequisites}
To get the most out of this tutorial, the reader should be fluent in quantum mechanical topics including the Schrödinger equation (SE), Dirac notation, angular momentum coupling and spherical harmonics, which are covered in most textbooks on quantum mechanics~\cite{sakurai2011modern, cohen1986quantumv2}. The mathematical prerequisites consist of a firm grasp on calculus, specifically the treatment of ordinary differential equations (ODEs) as well as the solution of boundary value problems. In addition the reader should be familiar with techniques from linear algebra, for the treatment of coupled differential equations~\cite{arfken2013methods, strang2016introduction}. Finally in order to actually implement the derived equations in code, readers should be sufficiently comfortable in numerics, such as discretising an ODE on a grid~\cite{press2007numerical}.

\paragraph{Outcome}
By the end of this tutorial, the reader should have a conceptual understanding of all the bits and pieces of machinery required to write a programme that simulates a collision. They will know which equations to implement in code, how they connect and what external data are required to simulate a particular collision system. More importantly, they will have an understanding of the underlying mathematical principles that allows an interpretation of the simulated wavefunctions and cross sections.

\paragraph{Scope}
Tackling even the simplest scattering problems can prove to be a demanding challenge, requiring a toolbox of mathematical, physical, programming and even chemistry skills. As such, this tutorial does not aim to cover more specialised situations one might encounter in advanced scattering problems, such as reactive collisions or accounting for external electromagnetic fields. Additionally, this tutorial will not cover the topic of obtaining interaction potentials of the two colliding particles as this is a huge subject in its own right as well as being the subject of very active research.
We limit our description of integral and differential cross sections to the space-fixed frame, whilst noting
that, amongst others, the body-fixed (helicity) frame is a commonly used alternative, discussed elsewhere \cite{pack1974space}.

\paragraph{Tutorial structure}
In \cref{sect:formalism} we start with the time-independent SE for two interacting particles and introduce all the machinery needed to derive a set of coupled channels equations appropriate for the numerical description of elastic and inelastic scattering processes. In particular, the formulation of the problem entails an expansion of the product wavefunction general enough to treat particles with intricate internal structures.

Next, \cref{sect:solution} takes the set of coupled channel differential equations derived in the former section and
mathematically abstractly treats the associated boundary value problem. This is achieved by dividing space into four
distinct regions and ensuring that collisions between particles obey the correct asymptotic behaviour.
We examine the properties of scattering states and introduce the scattering amplitude and cross section, expressed
in terms of the $\mymat{S}$-matrix, which corresponds to the experimentally
detectable observable describing the outcome of collision events and allows the comparison of real world scattering and theory.

Having dealt with the preliminaries, the second half of this tutorial supplies the reader with the ingredients required
for the implementation of a computer simulation of the scattering problem. \Cref{sect:single_channel_scattering}
specialises the abstract formalism to the more simple case of single channel, elastic scattering
(which could, for example be used to treat the scattering between two atoms, whose internal structure do not
influence the outcome of the collision). After discussing quantities characteristic of reactance, scattering
and transition ($K$, $S$ and $T$ respectively), we introduce Numerov's method as a way of solving the second-order
differential equation starting from small inter-particle separations and working our way outwards to where the
asymptotic boundary conditions can be matched.

In the final part, \cref{sect:atom_diatom_collisions}, we look at a realistic elastic and inelastic collision scenario between an
atom and a diatomic molecule.
This is a true multi-dimensional problem that can be solved using the techniques introduced in this tutorial.
We break the multidimensional problem down to a set of one-dimensional coupled equations using the Jacobi-coordinate
system and an expansion of the potential energy surface in both the angular and diatomic bases.
To this end an additional numerical algorithm, so-called discrete variable representation (DVR) is introduced
to obtain the bound states of the diatomic molecule. Consequently Numerov's method is generalised to the multi-channel case.
Obtaining the reactance matrix $\mymat{K}$ from propagation and converting it to $\mymat{S}$, we can explicitly
evaluate the integral cross section in Arthurs--Dalgarno form.
We conclude by detailing a mathematical trick enabling the propagation of the overlap with an input wavepacket
(as opposed to scattering states sharp in energy) without the necessity of explicitly reconstructing the wavefunction.

Readers interested in mathematical details can find a derivation of Bessel functions in \cref{apdx:bessel} where power series
solutions to second order ordinary differential equations are discussed, along with the difficulties of treating singularities.
Using a technique due to Frobenius, we arrive at a series representation for the spherical Bessel functions of first kind.
Using the powerful technique of generating functions we arrive at a recursion relation that allows for the evaluation
of Bessel functions.

\section{Formal derivation of the bimolecular collision problem}

\label{sect:formalism}

There is a multitude of different approaches to quantum scattering theory
and a good starting point for the description of the quantum collisions
we will be dealing with here, is to begin by considering two separate quantum objects, as
Krems does in his textbook \cite{krems2018molecules}.
Consequently, we will begin by first concentrating on the wavefunctions of
separated two-particle systems and then explore the effect of subsequently adding
a mutual interaction between the particles.
Individually, each particle satisfies its own Schr{\"o}dinger equation
\begin{align}
    \op{H}_{A,\mathrm{internal}} \ket{\alpha} &= \intenergy_{\alpha} \ket{\alpha}\,\,, \\
    \op{H}_{B,\mathrm{internal}} \ket{\beta}  &= \intenergy_{\beta} \ket{\beta}\,\,.
\end{align}
Above, we consider both $A$ and $B$ as particles (e.g. atoms or molecules) with their own respective (internal) Hamiltonians, and corresponding sets of eigenstates and eigenvalues $\left\{\ket{\alpha}\right\}$ and $\left\{\intenergy_{\alpha}\right\}$ for particle $A$ and
$\left\{\ket{\beta}\right\}$ and $\left\{\intenergy_{\beta}\right\}$ for particle $B$  respectively.
If they have some internal structure relevant to the collision process, then $\alpha, \beta$
can correspond to either a single or a whole set of several quantum numbers.
Otherwise, if the internal structure is unaffected by the collision,
the individual terms corresponding to either particle does not need to
appear in the total Hamiltonian explicitly.
Formally, we shall assume that for the internal structure of the particles
that must be regarded for the collision process, we possess the respective eigenvalues and
eigenstates but will have to confront the fact that, for numerical purposes,
in practice, a basis of infinite size has to be truncated to a finite basis
according to some convergence criterion.
This can be accomplished, for example, by starting with the smallest
possible basis, corresponding to the energetically lowest lying
quantum states and continually increasing the basis size
until the calculated result no longer changes.

In anticipation of considering only the relative motion of the two particles and not their
centre-of-mass motion, the kinetic parts of the Hamiltonians above have already been split off, i.e., the full Hamiltonian for each particle is given as
a sum of the kinetic and internal parts
\begin{align}
    \op{H}_{A} &= \op{E}_{A,\mathrm{kin}} + \op{H}_{A,\mathrm{internal}} \\
    \op{H}_{B} &= \op{E}_{B,\mathrm{kin}} + \op{H}_{B,\mathrm{internal}}\,\,.
\end{align}
Above the kinetic energy operators are given in terms of the respective
particle's momentum operators $\op{p}_{A}$, $\op{p}_{B}$ and masses
$m_{A}$, $m_{B}$
\begin{align}
  \op{E}_{A,\mathrm{kin}}= \frac{\myvecop{p}_{A}^2}{2m_{A}}\,\,,\\
  \op{E}_{B,\mathrm{kin}}= \frac{\myvecop{p}_{B}^2}{2m_{B}}\,\,.
\end{align}
Treating $A$ and $B$ as individual objects is justified for infinite separation, but as the
distance between the two is decreased, their mutual interaction is accounted for by a potential
$\op{V}$ within the total Hamiltonian.
We should note that in the presence of $\op{V}$,
the two quantum objects are no longer strictly separated physical systems
but form a combined system.
Therefore the combination of $\left\{\ket{\alpha}\right\}$ and $\left\{\ket{\beta}\right\}$
is no longer necessarily a good basis for the interacting system.
$\op{V}$ usually depends on the exact configuration of
both particles (and therefore both of their internal coordinates
and relative orientations in space, i.e., we should not regard them
as simple point particles).
To be concrete, in the case of atoms and molecules, evaluation of $\op{V}$ requires
sophisticated electronic structure calculations, which we will assume we have the luxury of taking for granted.
To make this point explicit, we could aim
to hide all the complexity caused by the electrons
in the $\op{V}$ operator, under the assumption that we
might then treat the nuclear motion separately.
The next section will briefly make this idea more explicit.

\subsection{Born-Oppenheimer approximation}

From a quantum physics standpoint, dealing with molecules is difficult because our description above can account for all the internal configurations of the particle, potentially consisting of several to hundreds of nuclei. Atoms, however, can provide a similarly complex challenge. Whilst we typically treat the nucleus of an atom as a single particle in this context (even though it can consist of many protons and neutrons) as we traverse the periodic table of elements the number of electrons quickly grows. The approach of treating each constituent of the system as a particle in its own right therefore quickly becomes intractable, even for the smallest of atoms and molecules. Luckily for us, an intrinsic property of electrons, when compared to the nuclei, can help us out of this bind and Born and Oppenheimer~\cite{born1927quantentheorie} were the first to introduce the machinery to make this idea work.

The electron mass is dwarfed by the masses of protons and neutrons, with %
$m_p/m_e \approx 1836$.
On account of their much lower mass, electrons can very quickly be accelerated and decelerated, i.e. pushed around by the movement of the nuclei. From the perspective of the nuclei, the electrons adapt very quickly (to good approximation `instantaneously') to any new configuration. From the perspective of the electrons the nuclei are `clamped' in space at any given moment of time, the only information they require to adapt their configuration is the spatial distribution of the nuclear charges.

We can express this idea mathematically in terms of the Hamiltonian of a given system
\begin{align}
	\op{H} = \op{E}_{\mathrm{nuc,kin}} + \op{E}_{\mathrm{el,kin}} + \op{V}_{\mathrm{Coulomb}}\,\,.
\end{align}
Above, the total Hamiltonian of a system consisting of multiple nuclei and electrons is partitioned into kinetic parts for the nuclei and electrons respectively, as well as the potential due to electron-electron, electron-nucleus and nucleus-nucleus Coulomb interactions, which can be written down exactly if all of the constituent particles were known. We now collect the electronic kinetic energy operator and the entire set of Coulomb interaction potentials (including nuclear-nuclear) into a Hamiltonian depending on the set of electronic coordinates $\{\myvec{r}_{\mathrm{el}}\}$ and also parametrically on the set of nuclear coordinates $\{\myvec{R}_{\mathrm{nuc}}\}$
\begin{align}
	\op{H} = \op{E}_{\mathrm{nuc,kin}} + \op{H}_{\mathrm{el}}(\{\myvec{r}_{\mathrm{el}}\};\{\myvec{R}_{\mathrm{nuc}}\})\,\,.
\end{align}
Now, assuming the two quantities above now operate on two different timescales, with the electronic part much much faster, we adopt the factorisation \textit{ansatz} for the wavefunction
\begin{align}
	\Psi(\{\myvec{r}_{\mathrm{el}}\},\{\myvec{R}_{\mathrm{nuc}}\}) \approx \psi_{\mathrm{nuc}}(\{\myvec{R}_{\mathrm{nuc}}\}) \phi_{\mathrm{el}}(\{\myvec{r}_{\mathrm{el}}\};\{\myvec{R}_{\mathrm{nuc}}\})
\end{align}
keeping in mind the parametric dependence of the electronic wavefunctions $\phi_{\mathrm{el}}$ on the nuclear coordinates $\{\myvec{R}_{\mathrm{nuc}}\}$.
Under these conditions, given any particular nuclear geometry (represented by $\{\myvec{R}_{\mathrm{nuc}}\}$) we can solve the SE for the electronic part of the wavefunction separately, i.e. obtain $\intenergy_{\mathrm{el}}(\{\myvec{R}_{\mathrm{nuc}}\})$ from

\begin{align}
	\op{H}_{\mathrm{el}} \phi_{\mathrm{el}}(\{\myvec{r}_{\mathrm{el}}\};\{\myvec{R}_{\mathrm{nuc}}\}) = \intenergy_{\mathrm{el}}(\{\myvec{R}_{\mathrm{nuc}}\}) \phi_{\mathrm{el}}(\{\myvec{r}_{\mathrm{el}}\};\{\myvec{R}_{\mathrm{nuc}}\})\,\,.
    \label{eq:electronic_eigenequation}
\end{align}
This is the essence of the Born-Oppenheimer approximation. If we sample many geometries and the eigenenergy $\intenergy_{\mathrm{el}}(\{\myvec{R}_{\mathrm{nuc}}\})$ varies smoothly as a function of the nuclear coordinates, we can stitch the points together to create a surface, which we refer to as the potential energy surface ($\op{V}$ above) and obtain this interaction potential for any particular geometry we need by interpolating between points. A lot of research has gone into obtaining these potential energy points at various different levels of theory (Hartree-Fock~\cite{hartree1928wavei}, coupled cluster~\cite{bartlett2007coupled}, configuration interaction~\cite{szalay2012multiconfiguration}) and the creation of fitted surfaces (e.g.\ using RKHS~\cite{ho1996general,unke2017toolkit,meuwly_rkhs_software}) and it continues to be an area of active and fruitful research.

Having obtained the potential energy surface, what remains to be solved is the nuclear problem, which can now be expressed as follows
\begin{align}
	\op{H}\Psi(\{\myvec{r}_{\mathrm{el}}\},\{\myvec{R}_{\mathrm{nuc}}\}) \approx & \left(\op{E}_{\mathrm{nuc,kin}} + \op{H}_{\mathrm{el}}(\{\myvec{r}_{\mathrm{el}}\};\{\myvec{R}_{\mathrm{nuc}}\}) \right) \psi_{\mathrm{nuc}}(\{\myvec{R}_{\mathrm{nuc}}\}) \phi_{\mathrm{el}}(\{\myvec{r}_{\mathrm{el}}\};\{\myvec{R}_{\mathrm{nuc}}\}) \nonumber \\
	\approx& \phi_{\mathrm{el}}(\{\myvec{r}_{\mathrm{el}}\};\{\myvec{R}_{\mathrm{nuc}}\})\underbrace{\left(\op{E}_{\mathrm{nuc,kin}} +\op{V} \right)\psi_{\mathrm{nuc}}(\{\myvec{R}_{\mathrm{nuc}}\})}_{=\totenergy_{\mathrm{tot}}\psi_{\mathrm{nuc}}(\{\myvec{R}_{\mathrm{nuc}}\})} \,\,.
\end{align}
It should be noted above, that the second equality is also only approximate since acting on $\phi_{\mathrm{el}}(\{\myvec{r}_{\mathrm{el}}\};\{\myvec{R}_{\mathrm{nuc}}\})$ with $\op{E}_{\mathrm{nuc,kin}}$ technically requires taking both adiabatic and non-adiabatic coupling terms into account.
Specifically since $\phi_{\mathrm{el}}$ parametrically depends on $\{\myvec{R}_{\mathrm{nuc}}\}$, the electronic wavefunction does not commute with the derivative operator in $\op{E}_{\mathrm{nuc,kin}}$. We neglect these terms however, since these are presumed to be small on account of the large mass ratio between nuclei and electrons.

The validity of a single electronic state, Born-Oppenheimer potential energy
surface can depend on different factors. 
Even for a one-electron molecule (e.g. the hydrogen molecular ion H$_2^+$)
the mass ratio between nuclei and electron is 
large enough to legitimately exploit the time-scale separation, 
but near-degeneracy in different electronic states
(as they occur for avoided crossings or conical intersections) can cause the approximation to break down requiring so-called `beyond-Born-Oppenheimer' corrections. In the context of this tutorial we will consider well isolated electronic surfaces and collision energies sufficiently low to stay in the non-reactive inelastic regime.

Having developed this useful tool to treat interacting nuclei, we can return to the problem at hand of treating two interacting particles $A$ and $B$ having conveniently wrapped up the thorny issue of handling electrons as a separate problem.

\subsection{Centre-of-mass frame}
Returning to our abstract interacting particles, we can note that the Schr{\"o}dinger equation for the total Hamiltonian is given by
\begin{align}
    \op{H}\tot \ket{\Psi\tot} &= \left( \op{H}_{A} + \op{H}_{B} + \op{V}\right) \ket{\Psi\tot}  \\
    &= \left(
    \op{E}_{A,\mathrm{kin}} + \op{E}_{B,\mathrm{kin}}+ \op{H}_{A,\mathrm{internal}} + \op{H}_{B,\mathrm{internal}} + \op{V}\right)\ket{\Psi\tot} = \totenergy\tot \ket{\Psi\tot}\,\,. \nonumber
\end{align}
Above, $\ket{\Psi\tot}$ is a (non-asymptotic) eigenstate corresponding
to a certain eigenenergy $\totenergy_{\mathrm{tot}}$ to the total Hamiltonian.
In this work, we will exclusively be considering collisions between
particles in the absence of external fields, which affords us certain
symmetries that we can take advantage of.
The first simplification we can perform is separating centre-of-mass
and relative motion.

We shall proceed by following the treatment in \cite{messiah1974quantum},
though similar derivations may
also be found elsewhere, such as in \cite{nolting2012grundkurs}.
Assuming that $\myvec{r}_{A}$ and $\myvec{r}_{B}$ point from the origin of our chosen
coordinate system to the respective centres of mass for each particle
A and B, we introduce the relative and centre-of-mass position vectors %
\begin{align}
    \myvec{R}_{\mathrm{rel}}&=\myvec{r}_{A}-\myvec{r}_{B}, \\
    \myvec{r}_{\mathrm{com}}&=\frac{m_{A}\myvec{r}_{A}+m_{B}\myvec{r}_{B}}{m_{A}+m_{B}}\,\,.
\end{align}
Above the subscripts $\mathrm{rel}$ and $\mathrm{com}$ respectively
indicate relative and centre of mass motion.
Similarly, since the two momentum operators in position space are given by
$\myvecop{p}_{A}=-i\hbar \myvecop{\nabla}_{A}$
and $\myvecop{p}_{B}=-i\hbar \myvecop{\nabla}_{B}$,
we shall define
\begin{align}
    \myvecop{P}_{\mathrm{rel}} &= \frac{m_{B}\myvecop{p}_{A}-m_{A}\myvecop{p}_{B}}{m_{A}+m_{B}}\,\,, \\
    \myvecop{p}_{\mathrm{com}} &= \myvecop{p}_{A}+\myvecop{p}_{B}
\end{align}
along with the reduced and total masses,
$\mu=\frac{m_{A}m_{B}}{m_{A}+m_{B}}$ and $m_{AB}=m_{A}+m_{B}$.
In Cartesian coordinates, $\myvecop{r}_{A}$, $\myvecop{r}_{B}$ and
$\myvecop{p}_{A}$, $\myvecop{p}_{B}$ obey the commutation relations
\begin{align}
    [\myvecop{r}\cdot\myvecop{e}_{i}, \myvecop{r}\cdot\myvecop{e}_{j}] &= 0\,\,, \\
    [\myvecop{r}\cdot\myvecop{e}_{i}, \myvecop{p}\cdot\myvecop{e}_{j}] &= i\hbar \delta_{ij}\,\,, \\
    [\myvecop{p}\cdot\myvecop{e}_{i}, \myvecop{p}\cdot\myvecop{e}_{j}] &= 0
\end{align}
for
$i,j \in \left\{x,y,z\right\}$.
Therefore, by making use of the various identities of commutators,
we may allow ourselves to perform the canonical substitutions
\begin{align}
    m_{A}r_{A}^2 + m_{B}r_{B}^2 &\rightarrow \mu R_{\mathrm{rel}}^2 + m_{AB} r_{\mathrm{com}}^2
\end{align}
and
\begin{align}
    \frac{p_{A}^2}{2m_{A}} + \frac{p_{B}^2}{2m_{B}} &\rightarrow \frac{P_{\mathrm{rel}}^2}{2\mu} + \frac{p_{\mathrm{com}}^2}{2m_{AB}}\,\,.
\end{align}
Making the (reasonable) assumption that the interaction between
the two particles can only depend on $\myvec{R}_{\mathrm{rel}}$
(of course this might not always be the case, for instance when
external fields are involved but there are ways around this
\cite{krems2018molecules}), performing this substitution
allows us to obtain a separable Schr{\"o}dinger equation.
\begin{align}
    \bra{\myvec{R}_{\mathrm{rel}},\myvec{r}_{\mathrm{com}}} \op{H}\tot \ket{\Psi\tot}
    =& \Big(\underbrace{\frac{\myvecop{P}_{\mathrm{rel}}^2}{2\mu}
    +\op{H}_{A,\mathrm{internal}}+\op{H}_{B,\mathrm{internal}}+\op{V}}_{=\op{H}_{\mathrm{rel}}}
    + \underbrace{\frac{\myvecop{p}_{\mathrm{com}}^2}{2m_{AB}}}_{=\op{H}_{\mathrm{com}}} \Big) \Psi_{\mathrm{tot}}(\myvec{R}_{\mathrm{rel}},\myvec{r}_{\mathrm{com}}) \nonumber \\
    =& \underbrace{(\totenergy_{\mathrm{rel}} + \totenergy_{\mathrm{com}})}_{=\totenergy_{\mathrm{tot}}}
    \Psi_{\mathrm{tot}}(\myvec{R}_{\mathrm{rel}},\myvec{r}_{\mathrm{com}})
    \,\,.
\end{align}
As such, we can adopt the separability \textit{ansatz} for the total wavefunction
\begin{align}
    \Psi_{\mathrm{tot}}(\myvec{R}_{\mathrm{rel}},\myvec{r}_{\mathrm{com}}) =
    \Psi_{\mathrm{rel}}(\myvec{R}_{\mathrm{rel}}) \Psi_{\mathrm{com}}(\myvec{r}_{\mathrm{com}})\,\,,
\end{align}
with each factor
individually satisfying its own eigenvalue equation,
\begin{align}
    \op{H}_{\mathrm{rel}} \Psi_{\mathrm{rel}}(\myvec{R}_{\mathrm{rel}}) &= \totenergy_{\mathrm{rel}} \Psi_{\mathrm{rel}}(\myvec{R}_{\mathrm{rel}})\,\,, \\
    \op{H}_{\mathrm{com}} \Psi_{\mathrm{com}}(\myvec{r}_{\mathrm{com}}) &= \totenergy_{\mathrm{com}} \Psi_{\mathrm{com}}(\myvec{r}_{\mathrm{com}})\,\,.
\end{align}
As we will exclusively be concerned with the relative motion, from this point onward we will drop
the subscript `$\mathrm{rel}$' and implicitly mean the relative motion degrees of freedom whenever
a Hamiltonian, energy, wavefunction etc. appears
\begin{align}
    \myvec{R}_{\mathrm{rel}}&\rightarrow \myvec{R}\,\,, \\
    \op{H}_{\mathrm{rel}} &\rightarrow \op{H}\,\,, \\
    \totenergy_{\mathrm{rel}} &\rightarrow \totenergy\,\,, \\
    \ket{\Psi_{\mathrm{rel}}} &\rightarrow \ket{\Psi}\,\,.
\end{align}
Please note that the capitalisation of $\myvec{R}$ has
been chosen deliberately in anticipation of our treatment of collisions between atoms and molecules in
\cref{sect:atom_diatom_collisions}.

\subsection{Expansion of non-asymptotic eigenstates}

\label{sect:channel_basis}

Whilst obtaining the asymptotic eigenstates of the separate
particles is by no means a simple task for all but the most
simple of cases and will in general require resorting to numerical
solutions, as for example discussed in \cref{subsec:dvr},
actually solving the Schr{\"o}dinger equation for the composite system can present
a challenge necessitating the use of extensive numerical resources.
We assume that in the asymptotic limit
$\norm{\myvec{R}_{\mathrm{rel}}}=\norm{\myvec{R}}= R\rightarrow\infty$, $\op{V}\rightarrow 0$
and consequently
\begin{align}
  \op{H} \xrightarrow[R\rightarrow\infty]{} \op{H}_0=\op{E}_{\mathrm{kin}}+\op{H}_{A,\mathrm{internal}}+\op{H}_{B,\mathrm{internal}}\,\,.
\end{align}
In accordance with the convention established above,
$\op{H}$, $\op{H}_{0}$ and $\op{E}_{\mathrm{kin}}=\myvecop{P}^2/(2\mu)$ all refer only
to the relative motion under omission of the subscript
`$\mathrm{rel}$'.

In the absence of $\op{V}$, by expressing the Laplace
operator in (spatial) spherical coordinates
\begin{align}
  \nabla^2_R =
  \frac{1}{R^2} \partial_R \left(R^2 \partial_{R}\right)
  + \frac{1}{R^2\sin\theta} \partial_{\theta} \left(\sin\theta \partial_{\theta}\right)
  + \frac{1}{R^2\sin^2\theta} \partial_{\varphi}^2
  = \frac{1}{R^2} \partial_R \left(R^2 \partial_{R}\right) - \frac{\op{\ell}^2}{\hbar^2 R^2}\,\,,
  \label{eq:spherical_laplace}
\end{align}
we have a simultaneously commuting
set of operators $\op{H}_0$, $\op{H}_{A,\mathrm{internal}}$, $\op{H}_{B,\mathrm{internal}}$, $\opell^2$ and $\opell_z$.
As usual, the operators $\opell^2$ and $\opell_z$ are the square
and $z$-axis projection of the orbital angular momentum operator
corresponding to eigenstates $\ket{\ell,m_{\ell}}$ and
resulting in eigenvalues $\hbar^2 \ell(\ell+1)$ and $\hbar m_{\ell}$.
With the polar and azimuthal spherical coordinates
$\theta$ and $\varphi$, the eigenstates
are given by the spherical harmonics
\begin{align}
  Y_{\ell m_{\ell}}(\theta,\varphi) = \braket{\theta,\varphi|\ell,m_{\ell}}\,\,.
\end{align}
If the particles $A$ and $B$ are each outfitted with their own
sets of internal coordinates $\Acoords$
and $\Bcoords$ (in position space
$\op{H}_{A,\mathrm{internal}} = \op{H}_{A,\mathrm{internal}}(\Acoords)$ and
$\op{H}_{B,\mathrm{internal}} = \op{H}_{B,\mathrm{internal}}(\Bcoords)$
),
the differential equation
\begin{align}
  \left(\op{E}_{\mathrm{kin}}+\op{H}_{A,\mathrm{internal}}+\op{H}_{B,\mathrm{internal}}\right) \psi(R,\Acoords,\Bcoords,\theta,\varphi) = \totenergy\psi(R,\Acoords,\Bcoords,\theta,\varphi)
\end{align}
becomes separable with the product wavefunction
\begin{align}
  \psi(R,\Acoords,\Bcoords,\theta,\varphi)
  &=
  \braket{R,\Acoords,\Bcoords,\theta,\varphi|k_{\ell},\alpha,\beta,\ell,m_{\ell}} \nonumber \\
  &=\underbrace{\braket{R|k_{\ell}}}_{=\psi_{k}^{\ell}(R)}
    \underbrace{\braket{\Acoords|\alpha}}_{=\mathcal{A}_{\alpha}(\Acoords)}
    \underbrace{\braket{\Bcoords|\beta}}_{=\mathcal{B}_{\beta}(\Bcoords)}
    \underbrace{\braket{\theta, \varphi|\ell m_{\ell}}}_{=Y_{\ell m_{\ell}}(\theta,\varphi)}\,\,.
\end{align}
Above, $\ket{k_{\ell},\alpha,\beta,\ell,m_{\ell}}$ is the tensor
product of the relative wave vector eigenstate
$\ket{k_{\ell}}$ associated with
the $\ell^{\mathrm{th}}$ partial wave,
with the internal single-particle eigenstates
$\ket{\alpha}$, $\ket{\beta}$ and
the angular momentum eigenstate $\ket{\ell, m_{\ell}}$.
We have introduced $\mathcal{A}_{\alpha}(\Acoords)$
and $\mathcal{B}_{\beta}(\Bcoords)$
as the position space probability amplitudes associated with
$A$ and $B$ in their respective internal states $\alpha$ and $\beta$.
As we shall see later, the radial functions $\psi_{\ell}^{k}(R)$
consist of spherical Bessel functions
which are derived in detail in \cref{apdx:bessel}
and which will show up in numerous of places throughout this work.
Introducing the (multi-)index $\multiindex{n}$ to label channels
comprised of quantum numbers $\alpha,\beta,\ell,m_{\ell}$,
we can
express the eigenvalue equation for $\op{H}_{0}$ as
\begin{align}
  \op{H}_{0} \kasymp{n}{k_{\ell}} = \totenergy_{\multiindex{n}}^{k_{\ell}} \kasymp{n}{k_{\ell}}\,\,.
\end{align}
By using the completeness relation in $R$, whose states are orthonormalised according to $\braket{R|R'} = \delta(R-R')/R^2$, we see that
the abstract eigenstates can be expressed as
\begin{align}
  \kasymp{n}{k_{\ell}} = \int \infin{R'} R'^2 \ket{R'}\braket{R'|k_{\ell}} \ket{\multiindex{n}} = \int \infin{R'} R'^2 \psi_{\ell}^{k}(R') \kasymp{n}{R'}\,\,,
  \label{eq:asymp_exp}
\end{align}
giving us either the $\left\{\kasymp{n}{k_{\ell}}\right\}$ or
$\left\{\kasymp{n}{R}\right\}$ bases to work with.
Now, accounting for the presence of
a non-vanishing anisotropic interaction potential
$\op{V}$ (i.e., $V=V(R,\Acoords,\Bcoords,\theta,\varphi)$)
the corresponding Schr{\"o}dinger equation for
the entire Hamiltonian $\op{H}$ is
\begin{align}
  \left(\op{E}_{\mathrm{kin}}+\op{H}_{A,\mathrm{internal}}+\op{H}_{B,\mathrm{internal}}+\op{V}\right) & \Psi(R,\Acoords,\Bcoords,\theta,\varphi) \nonumber \\    &=\totenergy\Psi(R,\Acoords,\Bcoords,\theta,\varphi)\,\,.
    \label{eq:non_asym_SE_function}
\end{align}
In general, the presence of $\op{V}$ will lead to
a differential equation which is not separable.
Instead, we may seek solutions as linear combinations
of states in the asymptotic basis.
We may first express an arbitrary state $\ket{\Psi}$ in terms of the original asymptotic basis
\begin{align}
  \ket{\Psi}
  = \left(\int \infin{k_{\ell'}'} \sum_{\multiindex{n'}}
  \kasymp{n'}{k_{\ell'}'} \basymp{n'}{k_{\ell'}'}
  \right) \ket{\Psi}\,\,.
\end{align}
Then, using the Hermitian transpose of \cref{eq:asymp_exp}
\begin{align}
  \ket{\Psi}
  &= \int \infin{k_{\ell'}'} \sum_{\multiindex{n'}} \kasymp{n'}{k_{\ell'}'} \int \infin{R'} R'^2 \underbrace{{\psi_{\ell'}^{k'}}^*(R')}_{=\braket{k_{\ell'}'|R'}} \left(\basymp{n'}{R'}\right) \ket{\Psi} \nonumber \\
  &= \int \infin{R'} R'^2 \sum_{\multiindex{n'}} \left( \int \infin{k_{\ell'}'} \ket{k_{\ell'}'}\bra{k_{\ell'}'} \right) \kasymp{n'}{R'}  \left(\basymp{n'}{R'}\right) \ket{\Psi} \nonumber \\
  &= \int \infin{R'} R'^2 \sum_{\multiindex{n'}} \kasymp{n'}{R'} \left(\basymp{n'}{R'}\right) \ket{\Psi}
  \,\,.
  \label{eq:arbitrary_state_expansion}
\end{align}
gives rise to the $R$-dependent coefficients
\begin{align}
  \left(\basymp{n''}{R}\right)\ket{\Psi} = \Psi_{\multiindex{n''}}(R)\,\,.
\end{align}
Instead of regarding arbitrary states, however, we are interested
in specifically seeking eigenstates of $\op{H}$
\begin{align}
    \op{H} \kint{n} = \totenergy
    \kint{n}\,\,.
    \label{eq:non_asym_SE_braket}
\end{align}
Even in the knowledge that we may not assume that the corresponding
quantum numbers are meaningful/descriptive in the interaction region,
we still label our basis states with the channel index $\multiindex{n}$.
This can be justified by considering that in the limit of large $R$,
as the effects of $\op{V}$ become less and less pronounced, there has
to be a correspondence between each basis state $\kint{\multiindex{n}}$ and an
asymptotic basis state $\kasymp{n}{k_{\ell}}$. %
Finally, using the position representation in the $R$ degree of freedom
obtained in \cref{eq:arbitrary_state_expansion}, we are able
to perform an ($R$-dependent) expansion of non-asymptotic basis states
using the asymptotic state basis with the relation
\begin{align}
    \kint{n} = \int \infin{R'} R'^2 \sum_{\multiindex{n'}} \kasymp{n'}{R'}
    \underbrace{\obasymp{n'}{R'} \kint{\multiindex{n}}}_{=U_{\multiindex{n'}\multiindex{n}}^{\totenergy}(R')}
    \,\,.
    \label{eq:non_asymptotic_expansion}
\end{align}

\subsubsection*{A note on the expansion coefficients $U_{\multiindex{n'}\multiindex{n}}^{\totenergy}(R)$}

The $R$-dependent coefficients $U_{\multiindex{n'}\multiindex{n}}^{\totenergy}$ possess
a double `channel-index' subscript, effectively 
`connecting' the $\multiindex{n'}^{\mathrm{th}}$ asymptotic solution with the $\multiindex{n}^{\mathrm{th}}$ non-asymptotic
solution for a given energy $\totenergy$ (the first index labels the channel, the second the solution).
We will see later what possible interpretations this
expansion has for scattering problems, where population
in a given input channel may be distributed amongst 
several exit channels, but for now we will simply
regard it as a mathematical fact.

\subsection{Derivation of coupled channel equations}

To obtain the coupled channels equations,
we can now substitute the expansion of the non-asymptotic eigenstates
\cref{eq:non_asymptotic_expansion} into
\cref{eq:non_asym_SE_braket} and multiply by $\basymp{n''}{R}$ from the left
\begin{align}
    \basymp{n''}{R} \op{H}\kint{n} = \basymp{n''}{R} \totenergy\kint{n}
\end{align}
to obtain a coupled set of differential equations
\begin{align}
  0&=\basymp{n''}{R} \left(-\frac{\hbar^2}{2\mu} R^{-2}\partial_R(R^2 \partial_R ) + \frac{\opell^{2}}{2\mu R^{2}} + \op{H}_{A,\mathrm{internal}} + \op{H}_{B,\mathrm{internal}} +\op{V} - \totenergy \right) \nonumber \\
        &\quad\quad \times \int \infin{R'} R'^2 \sum_{n'} U_{\multiindex{n'}\multiindex{n}}^{\totenergy}(R') \kasymp{n'}{R'} \nonumber \\
    &=\int \infin{R'} R'^2 \delta(R-R')/R'^2 \sum_{\multiindex{n'}} \bra{\multiindex{n''}} \Big(-\frac{\hbar^2}{2\mu}R^{-2}\partial_R(R^2 \partial_R ) + \frac{\opell^{2}}{2\mu R^{2}} \nonumber \\
        &\quad\quad + \op{H}_{A,\mathrm{internal}} + \op{H}_{B,\mathrm{internal}} +\op{V} - \totenergy \Big) U_{\multiindex{n'}\multiindex{n}}^{\totenergy}(R') \ket{\multiindex{n'}} \nonumber \\
    &=\sum_{\multiindex{n'}} \bra{\multiindex{n''}} \Big(-\frac{\hbar^2}{2\mu}R^{-2}\partial_R(R^2 \partial_R ) + \frac{\opell^{2}}{2\mu R^{2}} \nonumber \\
        &\quad\quad + \op{H}_{A,\mathrm{internal}} +\op{H}_{B,\mathrm{internal}} +\op{V} - \totenergy \Big) U_{\multiindex{n'}\multiindex{n}}^{\totenergy}(R) \ket{\multiindex{n'}} \nonumber \\
    &=\left(-\frac{\hbar^2}{2\mu}\left[ R^{-2}\partial_R(R^2 \partial_R ) - \ell''(\ell''+1)R^{-2} \right] + \intenergy_{\multiindex{n''}} - \totenergy \right) U_{\multiindex{n''}\multiindex{n}}^{\totenergy}(R) \nonumber \\
         &\quad\quad+ \sum_{\multiindex{n'}} \bra{\multiindex{n''}} \op{V} \ket{\multiindex{n'}} U_{\multiindex{n'}\multiindex{n}}^{\totenergy}(R)
\end{align}
We can simplify the radial part of the Laplace operator by
defining $U_{\multiindex{n''}\multiindex{n}}^{\totenergy}(R) = R^{-1} u_{\multiindex{n''}\multiindex{n}}^{\totenergy}(R)$,
applying the chain rule
above and multiplying by $R$ from the left
\begin{align}
    0&=R (-\frac{\hbar^2}{2\mu})\left[ R^{-1} \partial_{R}^2 (R R^{-1} u_{\multiindex{n''}\multiindex{n}}^{\totenergy}(R)) - \ell''(\ell''+1)R^{-2} R^{-1} u_{\multiindex{n''}\multiindex{n}}^{\totenergy}(R) \right] \nonumber \\
    &\quad\quad+ R(\intenergy_{\multiindex{n''}} - \totenergy)R^{-1} u_{\multiindex{n''}\multiindex{n}}^{\totenergy}(R)
    + R \sum_{\multiindex{n'}} \bra{\multiindex{n''}} \op{V} \ket{\multiindex{n'}} R^{-1} u_{\multiindex{n'}\multiindex{n}}^{\totenergy}(R) \nonumber \\
    &=\left((-\frac{\hbar^2}{2\mu})\left[ \partial_{R}^2  - \ell''(\ell''+1)R^{-2}  \right] +(\intenergy_{\multiindex{n''}} - \totenergy) \right)u_{\multiindex{n''}\multiindex{n}}^{\totenergy}(R) \nonumber \\
    &\quad\quad+ \sum_{\multiindex{n'}} \bra{\multiindex{n''}} \op{V} \ket{\multiindex{n'}} u_{\multiindex{n'}\multiindex{n}}^{\totenergy}(R)
    \label{eq:cc}
\end{align}
This is the same coupled channels equation as obtained
in \cite{pack1974space,lester1976dynamics}.

\subsection{Discussion}

In this section, we have framed the quantum scattering problem in a way that the wavefunction of two interacting particles (such as atoms or molecules) can be determined mathematically.
This treatment, beginning from first principles with the SE of the system, culminated in a set of coupled channel equations obeyed by the radial wavefunction of the collision complex, taking all the relevant quantum states, encoded into so-called channels, into account.

Arriving at this result required a closer look at the origin of the mutual interaction between the particles.
The choice of applying the Born-Oppenheimer approximation limits us to describing low-energy problems where the corresponding electronic surface is well isolated.
New research, developing with `beyond-Born-Oppenheimer' corrections, can help to address these issues, whilst maintaining the advantages gained by reducing the problem's complexity \cite{tiemann2020beyond}.
In order to further reduce complexity, we had to assume that the collision takes place in the absence of external electromagnetic fields.
This assumption allowed us to disregard the centre-of-mass motion of the complex, reducing the dimensionality of the problem to solve.
The reader interested in exploring the important influence external fields have on collisions, which can lead to a whole host of physical effects, is referred to the thorough textbook on the subject by Krems \cite{krems2018molecules}.
Deriving the coupled channels equation further required the expansion of the total wavefunction in terms of the system's various eigenstates, which we can label as so-called channels.
Whilst these labels are descriptive when the colliding particles are far apart, since individual quantum numbers are well defined, we should bear in mind that the mutual interaction mixes the quantum states at close distances, leading channel labels to lose their meaning.

With this set of differential equations in hand, we can proceed in the next section, where we will actually cast the collision as a boundary condition problem, taking the correct physical behaviour into account for different regions of space.
In the following, readers more familiar with a non-numerical perspective on quantum scattering will recognise several concepts such as the plane wave expansion, cross sections and so on and will learn how the coupled channels treatment of collisions connects with these.

\subsection{Section summary}

\begin{keyresults}

    \kr{\label{kr:ch1-pes}%

    Applying the Born-Oppenheimer approximation reduces our description of the system to a single electronic surface
    for which the electronic interaction can be solved separately, yielding an interaction potential or potential energy surface (PES)
    $\op{V}$ by evaluating many nuclear geometries
    \begin{equation*}
        \op{H}_{\mathrm{el}} \phi_{\mathrm{el}}(\{\myvec{r}_{\mathrm{el}}\};\{\myvec{R}_{\mathrm{nuc}}\}) = \intenergy_{\mathrm{el}}(\{\myvec{R}_{\mathrm{nuc}}\}) \phi_{\mathrm{el}}(\{\myvec{r}_{\mathrm{el}}\};\{\myvec{R}_{\mathrm{nuc}}\}) .
    \end{equation*}
    \emph{Cf.}~\cref{eq:electronic_eigenequation}.}

    \kr{\label{kr:ch1-cc}%
    The coupled-channel differential equations obeyed by the components of the radial wavefunction $u(R)$, dependent on $\op{V}$ and
    with channels labelled by the (multi-)index $\multiindex{n}$:
    \begin{equation*}
        \left(\left(-\frac{\hbar^2}{2\mu}\right)
        \left[ \partial_{R}^2 - \ell''(\ell''+1)R^{-2} \right]
        + \left(\intenergy_{\multiindex{n''}} - \totenergy\right) \right)
        u_{\multiindex{n''}\multiindex{n}}^{\totenergy}(R)
        + \sum_{\multiindex{n'}} \bra{\multiindex{n''}} \op{V} \ket{\multiindex{n'}}
        u_{\multiindex{n'}\multiindex{n}}^{\totenergy}(R) = 0 .
    \end{equation*}
    \emph{Cf.}~\cref{eq:cc}.}

\end{keyresults}

\section{Formal solution of the bimolecular collision problem}
\label{sect:solution}

\subsection{Asymptotic behaviour}

\label{subsec:boundary_conditions}

\def\lineupper{0}
\def\linelower{-6}
\def\iib{-0.5}
\def\ib{0.25}
\def\mb{4.5}
\def\ob{6.5}
\def\oob{9}
\def\margin{0.2}
\def\csm{0.5}
\begin{figure}
	\begin{center}
	\begin{tikzpicture}[line width=0.5pt]
		\node [] (barrier) at (0,0) {};
		\node [] (min) at (2,-5) {};
		\node [] (asymptote1) at (3.8,-3) {};
		\node [] (asymptote2) at (7,-2.5) {};
		\node (textpos) at (3,0.5) {};

        \draw [->,line width=0.75pt] (\iib-\csm,\linelower-\csm) -- (\oob+\csm,\linelower-\csm) node [midway,below] {$R$};
        \draw [->,line width=0.75pt] (\iib-\csm,\linelower-\csm) -- (\iib-\csm,\lineupper+\csm) node [midway,left] {$V$};
        
        \node (iibll) at (\iib,\linelower) {};
        \node (iblu) at (\ib,\lineupper) {};
        \node (ibmlu) at (\ib+\margin,\lineupper) {};
        \node (mbll) at (\mb,\linelower) {};
        \node (mbmll) at (\mb+\margin,\linelower) {};
        \node (oblu) at (\ob,\lineupper) {};
        \node (obmlu) at (\ob+\margin,\lineupper) {};
        \node (oobll) at (\oob,\linelower) {};
        \node [fill=red,opacity=0.1,fit=(iibll)(iblu),inner sep=-5*\margin] {};
        \node [fill=blue,opacity=0.1,fit=(ibmlu)(mbll),inner sep=-5*\margin] {};
        \node [fill=green,opacity=0.1,fit=(mbmll)(oblu),inner sep=-5*\margin] {};
        \node [fill=orange,opacity=0.1,fit=(obmlu)(oobll),inner sep=-5*\margin] {};
        \draw [red] (\ib,\lineupper) -- (\ib,\linelower);
        \draw [red] (\iib,\linelower) -- (\iib,\lineupper);
        \draw [red] (\iib,\linelower) -- (\ib,\linelower);
        \draw [red] (\iib,\lineupper) -- (\ib,\lineupper);

        \draw [blue] (\ib+\margin,\lineupper) -- (\ib+\margin,\linelower);
        \draw [blue] (\mb,\linelower) -- (\mb,\lineupper);
        \draw [blue] (\mb,\linelower) -- (\ib+\margin,\linelower);
        \draw [blue] (\mb,\lineupper) -- (\ib+\margin,\lineupper);
        
        \foreach \i in {-6,...,-1}
            \draw [red] (\iib,\i) -- (\ib,\i+1);
        \draw [orange] (\ob+\margin,\lineupper) -- (\ob+\margin,\linelower);
        \draw [orange] (\oob,\lineupper) -- (\oob,\linelower);
        \draw [orange] (\ob+\margin,\lineupper) -- (\oob,\lineupper);
        \draw [orange] (\ob+\margin,\linelower) -- (\oob,\linelower);
        \draw [green] (\mb+\margin,\lineupper) -- (\ob,\lineupper);
        \draw [green] (\mb+\margin,\linelower) -- (\ob,\linelower);
        \draw [green] (\mb+\margin,\linelower) -- (\mb+\margin,\lineupper);
        \draw [green] (\ob,\linelower) -- (\ob,\lineupper);

        \draw[pot_curve] plot ({\x},{10*\x^-12 - 10*\x^-6});
        \node [fill=white,draw=red] at (-0.125,1*\linelower) {\huge I};
        \node [fill=white,draw=blue] at (2.5,1*\linelower) {\huge II};
        \node [fill=white,draw=green] at (5.5,1*\linelower) {\huge III};
        \node [fill=white,draw=orange] at (7.75,1*\linelower) {\huge IV};
	\end{tikzpicture}
	\end{center}
    \caption{Overview of the four regions of space (as a one-dimensional
    slice in the $R$ degree of freedom) for which the coupled channels
    problem is solved.
    From left to right, the strongly repulsive region
    (I) - where the wavefunction is assumed to vanish,
    the interaction region (II) - where the potential energy surface
    $V$ dominates the interaction, the centrifugal region (III) -
    where the potential energy surface is negligible but the
    centrifugal term still contributes to the interaction and 
    the free particle region (IV) - where no interaction between the
    particles is felt, are shown.    
    }
    \label{fig:boundary_conditions}
\end{figure}
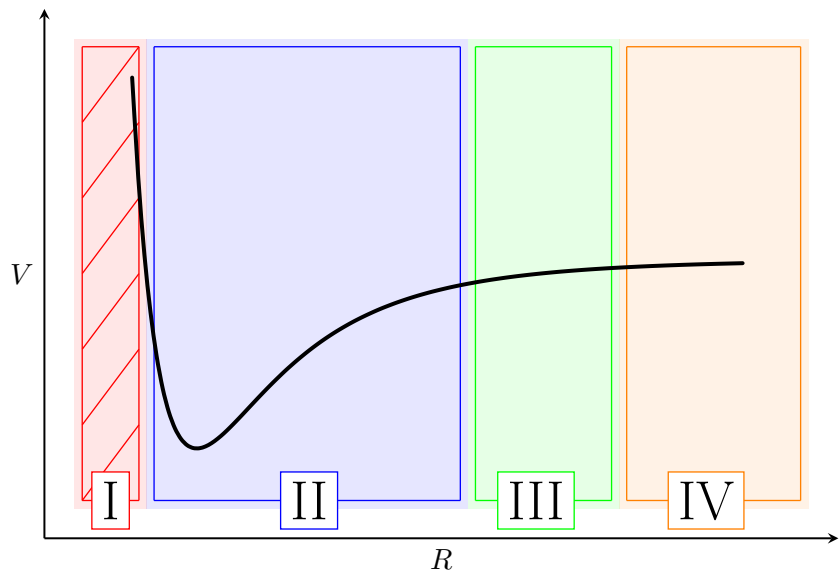

Given a potential $\op{V}$ that increases exponentially as the inter-particle separation $R$ tends
towards zero and falls off faster with $R$, than the rotational barrier falloff
$\opell^2/R^{2}$, space
can be divided up into four regions
(as rings along the axis of inter-particle separation $R$, as sketched
in \cref{fig:boundary_conditions}):
\begin{enumerate}[(I)]
    \item $R\rightarrow 0$ - hard repulsion, wavefunction vanishes
    \item $R\sim R_{0}$ - interaction region ($R_{0}$ being the range of $\op{V}$), in general require numerics to evaluate wavefunction
    \item $R \gg R_{0}$ - wavefunctions satisfy the spherical Bessel differential equation
    \item $R \rightarrow \infty$ - wavefunction satisfies the free particle Schr{\"o}dinger equation \\
    $\left(\partial_{R}^{2}+k_{\alpha'\beta'}^{2}\right)u_{\multiindex{n'}\multiindex{n}}^{\totenergy}(R)=0$, solved by free plane/spherical waves
\end{enumerate}
In the truly asymptotic region (IV), the differential equation
\begin{equation}
    \left(\partial_{R}^{2}+k_{\alpha'\beta'}^{2}\right)u_{\multiindex{n'}\multiindex{n}}^{\totenergy}(R)=0
\end{equation}
is solved by
\begin{equation}
    u_{\multiindex{n'}\multiindex{n}}^{\totenergy}(R\rightarrow\infty) =
    \ampminus e^{-i(k_{\alpha'\beta'} R - \phi_{-})} + \ampplus e^{+i(k_{\alpha'\beta'}R-\phi_{+})}
    \label{eq:fully_asymptotic_solution}
\end{equation}
with as-of-yet undetermined amplitudes and phases
$\amppm$ and $\phi_{\pm}$.
In region (IV), we recognise that the wavenumber cannot
depend on the partial wave quantum number $\ell$.
Rather, the kinetic energy corresponding to each channel
$\multiindex{n}$ can in general depend on the internal
quantum numbers $\alpha$ and $\beta$, so we shall outfit
the wavenumber $\channelwavenumber=\sqrt{\frac{2\mu}{\hbar^2} \left( \totenergy-\intenergy_{\alpha} -\intenergy_{\beta}\right)}$
with the corresponding subscripts.
We refer to channels with $\totenergy > \intenergy_{\alpha} + \intenergy_{\beta}$ as open
(i.e. they can be populated asymptotically), whilst those with
$\totenergy < \intenergy_{\alpha} + \intenergy_{\beta}$ are referred to as closed.

Conversely, in the region (III), i.e. when $R\gg R_{0}$
is sufficiently large to neglect the interaction potential
but before the centrifugal term $\opell^2/R^2$ becomes
negligibly small, $u_{\multiindex{n'}\multiindex{n}}^{\totenergy}(R)$ satisfy the
differential equation
\begin{align}
    \left(\partial_{R}^{2} - \ell'(\ell'+1)/R^{2} + k_{\alpha'\beta'}^{2}\right) u_{\multiindex{n'}\multiindex{n}}^{\totenergy}(R) = 0\,\,.
    \label{eq:riccati_bessel}
\end{align}
At this point, the mathematically inclined
reader is referred to \cref{apdx:bessel} for
a derivation of the solution to the above
equation utilising Frobenius' method.
For now, however, we will be content to note
that the above equation has two linearly
independent solutions $k_{\alpha\beta}R\cdot j_{\ell}(k_{\alpha\beta}R)$
and $k_{\alpha\beta}R\cdot y_{\ell}(k_{\alpha\beta}R)$, which we refer to as the
Riccati--Bessel functions. These differ from the ordinary spherical Bessel
functions due to the prefactor $k_{\alpha\beta}R$.

\subsection{Spherical Hankel functions}
\label{subsec:hankel}

Mathematically speaking, the spherical Bessel functions $j_{\ell}(x)$ and $y_{\ell}(x)$ are perfectly valid solutions to the spherical Bessel differential
equation. In terms of physics, however, it is advantageous to define a related set of functions with which solutions
to the coupled-channel differential equations in region (III) can be built.
Using these new functions, we can separate out the incoming and outgoing parts of the wavefunctions.
The spherical Hankel functions of first and second kind are given by
\begin{align}
    h_{\ell}^{(1)}(x) &= j_{\ell}(x) + iy_{\ell}(x)
    \label{eq:spherical_Hankel_first} \\
    h_{\ell}^{(2)}(x) &= j_{\ell}(x) - iy_{\ell}(x)\,\,.
    \label{eq:spherical_Hankel_second}
\end{align}
In analogy to the spherical Bessel functions above, we introduce the Riccati versions
of the spherical Hankel functions as $k_{\alpha\beta}R\cdot h_{\ell}^{(i)}(k_{\alpha\beta}R)$ for $i \in \{1,2\}$, which we shall refer to as Riccati--Hankel functions accordingly.
For $x \in \mathbb{R}$ the spherical Hankel function
of first kind (and its complex conjugate) can be
expressed in closed-form as
\begin{align}
    h_{\ell}^{(1)}(x) =& (-i)^{\ell+1} \frac{e^{ix}}{x} \sum_{m=0}^{\ell} \frac{i^{m}}{m!(2x)^{m}} \frac{(\ell+m)!}{(\ell-m)!} \\
    h_{\ell}^{(2)}(x) =& i^{\ell+1} \frac{e^{-ix}}{x} \sum_{m=0}^{\ell} \frac{(-i)^{m}}{m!(2x)^{m}} \frac{(\ell+m)!}{(\ell-m)!}\,\,.
\end{align}
This means that asymptotically (i.e.,
in region (IV), %
the above sum is dominated by $m=0$ term with
$\frac{i^0}{0!(2x)^{0}}\frac{\ell!}{\ell!}=1$)
they behave as
\begin{align}
    h_{\ell}^{(1)}(x\gg 1) &\sim (-i)^{\ell+1} \frac{e^{ix}}{x}
    = e^{-i(\ell+1)\frac{\pi}{2}} \frac{e^{ix}}{x} \\
    h_{\ell}^{(2)}(x\gg 1) &\sim i^{\ell+1} \frac{e^{-ix}}{x} = e^{i(\ell+1)\frac{\pi}{2}} \frac{e^{-ix}}{x}\,\,.
\end{align}
Returning to the interacting particle boundary value
problem, the arbitrary phases $\phi_{\pm}$
in \cref{eq:fully_asymptotic_solution} can
now be fixed in terms of the orbital angular momentum quantum number $\ell$
by comparing the wavefunctions to the long-range behaviour
of the Riccati--Hankel functions to obtain
\begin{align}
    \phi_{+} &=(\ell+1)\pi/2 \\
    \phi_{-} &=(\ell+1)\pi/2\,\,.
\end{align}
Since we have the luxury of arbitrarily choosing
the overall phase of the wavefunction,
and as some other authors do too \cite{krems2018molecules,johnson1977new},
we will introduce an overall phase shift of $-\pi/2$
such that
\begin{align}
    \phi_{+} &=\ell\pi/2 +\pi \\
    \phi_{-} &=\ell\pi/2 \,\,.
\end{align}
leading to
\begin{equation}
    u_{\multiindex{n'}\multiindex{n}}^{\totenergy}(R\rightarrow\infty) =
    \ampminus e^{-i(k_{\alpha'\beta'} R - \ell'\pi/2)} - \ampplus e^{+i(k_{\alpha'\beta'}R-\ell'\pi/2)}\,\,,
    \label{eq:asymptotic_solution_fixed_phases}
\end{equation}
Above, the relative phase of $\pi$ appears as
the minus sign between the two terms.

Apart from their very tangible significance in the context of quantum scattering,
which we will discuss in detail later, the importance of these asymptotic
boundary conditions lies in uniquely determining the entire wavefunction,
as is customary when solving a boundary value problem comprised of
a differential equation and a set of boundary conditions.

\subsection{Amplitudes}

\label{subsec:amplitudes}

The advantage of expressing the asymptotic solution in terms of Riccati--Hankel
functions is that by some careful bookkeeping, we are able to
unravel the dynamical collision process, just by knowing the stationary
states.
We shall first note that the two terms in \cref{eq:fully_asymptotic_solution}
are spherical waves propagating in opposite directions (remembering that
$U_{\multiindex{n'}\multiindex{n}}^{\totenergy}(R)=u_{\multiindex{n'}\multiindex{n}}^{\totenergy}(R)/R$), one corresponding
to inwards propagation (negative sign in the exponent), the other
corresponding to outwards propagation (positive sign in the exponent).
From what we have seen so far,
asymptotically the $\multiindex{n'}$ components
of the $R$-dependent wavefunction behave as
\begin{align}
    \ampminus e^{-i(k_{\alpha'\beta'}R-\ell'\pi/2)} - \ampplus e^{i(k_{\alpha'\beta'}R-\ell'\pi/2)}\,\,.
\end{align}
The ultimate objective of performing scattering calculations
is to determine $\ampplus$ amplitudes in each outgoing channel
depending on the initial occupancy of the incoming channels.
This can be expressed by the transition probability
amplitudes $P_{\multiindex{n'}\multiindex{n}}$ between an input channel $\multiindex{n}$ and
output channel $\multiindex{n'}$
\begin{align}
    \ampplus = P_{\multiindex{n'}\multiindex{n}} \,\,.
    \label{eq:amplitude_outgoing}
\end{align}
The incoming wave is a superposition of different channels
$\multiindex{n}$ and from solving the coupled channels differential
equations we obtain as many (linearly independent) solutions as we have channels ($N$).
From these, we can form a linear combination of solutions in order to obtain the output state of the
system corresponding to a given input channel (or superposition thereof).
Since the incoming amplitudes are fixed by how we prepare the system
(no information from $\op{V}$ can enter), we can set
\begin{align}
    \ampminus = \delta_{\multiindex{n'}\multiindex{n}}\,\,.
    \label{eq:amplitude_incoming}
\end{align}
Next, we shall turn to the famous `plane wave expansion',
which will be instrumental for the upcoming considerations.

\subsection{The plane wave expansion}

Though it also appears elsewhere \cite{cohen1986quantumv2},
we shall follow the nice treatment of the plane wave expansion
given in \cite{haber2018physics}.
The plane wave expansion arises from the
observation that the
three dimensional
free particle wave equation
\begin{align}
    \left(\nabla^{2} + \channelwavenumber^2\right) \psi_{\channelwavevec}(\myvec{R})=0
\end{align}
can both be solved by plane waves
in Cartesian coordinates
\begin{align}
    \psi_{\channelwavevec}(\myvec{R}) = \frac{1}{(2\pi)^{3/2}} e^{i\channelwavevec\cdot \myvec{R}}
\end{align}
or in spherical coordinates and momentum
eigenstates in terms of the spherical
Bessel functions we have
already become acquainted with
\begin{align}
    \psi_{\channelwavevec}(R,\theta,\varphi)
    =\braket{R,\theta,\varphi|\totenergy_{\alpha\beta} \ell m}
    =i^{\ell} \left(\frac{2\mu \channelwavenumber}{\pi \hbar^2}\right)^{1/2} j_{\ell}(\channelwavenumber R)Y_{\ell m}(\theta,\varphi)\,\,.
\end{align}
Above, $Y_{\ell m}(\theta,\varphi)=Y_{\ell m}(\myvec{R}/\norm{\myvec{R}})=\braket{\theta,\varphi|\ell m}$ is the
spherical harmonic of degree $\ell$ and order $m$ evaluated for a certain orientation of $\myvec{R}$ expressed
by the angles $\theta$ and $\varphi$.
The respective prefactors are due to the
normalisation conditions
\begin{align}
  \int \infin{{}^3 R}~\psi_{\channelwavevec}^{*}(\myvec{R}) \psi_{\channelwavevec'}(\myvec{R}) = \delta^{3}(\channelwavevec-\channelwavevec')
\end{align}
and
\begin{align}
  \braket{\totenergy_{\alpha \beta}' \ell' m'| \totenergy_{\alpha \beta} \ell m} = \delta_{\ell \ell'} \delta_{m m'}\delta(\totenergy_{\alpha \beta} -\totenergy_{\alpha \beta}')
\end{align}
in each of the two cases.

Since both variants are valid solutions
to the same differential equation
(and form orthonormal bases), we
are able to express one solution
(fixed energy $\totenergy_{\alpha\beta}=\frac{\hbar^2 \channelwavenumber^2}{2\mu}$
eigenstate) as a
linear combination of the other
\begin{align}
    e^{i\channelwavevec\cdot\myvec{R}} = \sum_{\ell=0}^{\infty} \sum_{m=-\ell}^{\ell} c_{\ell m}(\channelwavevec) j_{\ell}(\channelwavenumber R)Y_{\ell m}\left(\frac{\myvec{R}}{\norm{\myvec{R}}}\right)\,\,.
    \label{eq:linear_combination_plane_waves}
\end{align}
This  expression can be simplified
significantly by choosing the coordinate
system such that $\channelwavevec$ points
along the $z$-axis.
In that case $e^{i\channelwavevec\cdot\myvec{R}}=e^{i\channelwavenumber R\cos\theta}$
is independent of $\varphi$, therefore the right hand side
of \cref{eq:linear_combination_plane_waves} must also
be independent of $\varphi$.
In this case, the expansion coefficients
must vanish whenever $m\neq 0$ and we
are left with a single sum over Legendre
polynomials
\begin{align}
    e^{i\channelwavenumber R\cos\theta} = \sum_{\ell=0}^{\infty} \left(\frac{2\ell+1}{4\pi}\right)^{1/2} A_{\ell}(\channelwavenumber) j_{\ell}(\channelwavenumber R)P_{\ell}(\cos\theta)\,\,.
\end{align}
Multiplying both sides by $P_{\ell'}(\cos\theta)$ and integrating over
$\cos\theta$ allows us to exploit the orthogonality of Legendre
polynomials to solve for the coefficients $A_{\ell}(\channelwavenumber)$
\begin{align}
    \int_{-1}^{1} \infin{\cos} &\theta~e^{i\channelwavenumber R\cos\theta} P_{\ell'}(\cos\theta) \nonumber \\
    =& \int_{-1}^{1} \infin{\cos\theta}~\sum_{\ell=0}^{\infty} \left(\frac{2\ell+1}{4\pi}\right)^{1/2} A_{\ell}(\channelwavenumber) j_{\ell}(\channelwavenumber R)P_{\ell}(\cos\theta) P_{\ell'}(\cos\theta) \nonumber \\
    =& \sum_{\ell=0}^{\infty} \left(\frac{2\ell+1}{4\pi}\right)^{1/2} A_{\ell}(\channelwavenumber) j_{\ell}(\channelwavenumber R) \frac{2}{2\ell+1} \delta_{\ell\ell'}
    = \frac{2A_{\ell'}(\channelwavenumber)j_{\ell'}(\channelwavenumber R)}{\sqrt{4\pi(2\ell'+1)}}
\end{align}
We see that the integral on the left hand side
has to be proportional to the $\ell'^{\mathrm{th}}$
spherical Bessel function.
Without proof, it will be noted that
\begin{align}
  \int_{-1}^{1} \infin{\cos}~\theta~e^{i\channelwavenumber R\cos\theta} P_{\ell}(\cos\theta) = 2 i^{\ell}j_{\ell}(\channelwavenumber R)\,\,,
\end{align}
which can be shown
by integrating $j_{0}(\channelwavenumber R)$ by parts $\ell$ times
and recognising Rodrigues' formula for the
Legendre polynomials
\begin{align}
  P_{\ell}(x) = \frac{1}{2^{\ell}\ell!}\partial_x^{\ell} \left(x^2-1\right)^{\ell}\,\,.
\end{align}
It follows that
\begin{align}
    A_{\ell}(\channelwavenumber) = i^{\ell} \sqrt{4\pi(2\ell+1)}\,\,.
\end{align}
With the spherical harmonic addition theorem
\begin{align}
    P_{\ell}(\cos\theta) = \frac{4\pi}{2\ell+1} \sum_{m=-\ell}^{\ell}
    Y_{\ell m}\left(\frac{\myvec{R}}{\norm{\myvec{R}}}\right)
    Y_{\ell m}^{*}\left(\frac{\channelwavevec}{\norm{\channelwavevec}}\right)
\end{align}
we finally obtain
\begin{align}
    e^{i\channelwavevec \cdot\myvec{R}} = 4\pi \sum_{\ell=0}^{\infty} \sum_{m=-\ell}^{\ell} i^{\ell} j_{\ell}(\channelwavenumber R)
    Y_{\ell m}\left(\frac{\myvec{R}}{\norm{\myvec{R}}}\right)
    Y_{\ell m}^{*}\left(\frac{\channelwavevec}{\norm{\channelwavevec}}\right)\,\,.
\end{align}
Significantly, for asymptotically large $R$
we identify the spherical Bessel as a superposition
of incoming and outgoing spherical waves
\begin{align}
    j_{\ell}(\channelwavenumber R) \rightarrow \frac{1}{\channelwavenumber R} \sin{(\channelwavenumber R-\frac{\ell\pi}{2})}
    = \frac{1}{\channelwavenumber R} \frac{e^{i(\channelwavenumber R-\frac{\ell\pi}{2})} - e^{-i(\channelwavenumber R-\frac{\ell\pi}{2})}}{i2}
\end{align}
leading to
\begin{align}
    e^{i\channelwavevec \cdot\myvec{R}} \rightarrow~&
    4\pi \sum_{\ell=0}^{\infty} \sum_{m=-\ell}^{\ell}
    \frac{i^{\ell}}{\channelwavenumber R} \frac{e^{i(\channelwavenumber R-\frac{\ell\pi}{2})}}{i2}
    Y_{\ell m}\left(\frac{\myvec{R}}{\norm{\myvec{R}}}\right)
    Y_{\ell m}^{*}\left(\frac{\channelwavevec}{\norm{\channelwavevec}}\right) \nonumber \\
    &-
    4\pi \sum_{\ell=0}^{\infty} \sum_{m=-\ell}^{\ell}
    \frac{i^{\ell}}{\channelwavenumber R} \frac{e^{-i(\channelwavenumber R-\frac{\ell\pi}{2})}}{i2}
    Y_{\ell m}\left(\frac{\myvec{R}}{\norm{\myvec{R}}}\right)
    Y_{\ell m}^{*}\left(\frac{\channelwavevec}{\norm{\channelwavevec}}\right)
    \,\,.
    \label{eq:plane_wave_asymptotic}
\end{align}
We note the special cases for $\cos\theta=\pm 1$
\begin{align}
    e^{\pm i\channelwavenumber R} &= \sum_{\ell=0}^{\infty} (2\ell+1) i^{\ell} j_{\ell}(\channelwavenumber R) (\pm 1)^{\ell} \\
    &\rightarrow \sum_{\ell=0}^{\infty} (2\ell+1) i^{\ell} \frac{1}{\channelwavenumber R} \frac{e^{i(\channelwavenumber R-\frac{\ell\pi}{2})} - e^{-i(\channelwavenumber R-\frac{\ell\pi}{2})}}{i2} (\pm 1)^{\ell}\,\,.
\end{align}

\subsection{Scattering state}

\label{subsec:scattering_state}

So far we know that asymptotically we should observe an incoming part
of the wavefunction (terms proportional to
$e^{-ikR}$) and an outgoing part
(terms proportional to $e^{ikR}$).
The outgoing part can further be decomposed into
a plane wave part (which itself is comprised of superposition of incoming and outgoing waves)
and a scattering part (which is the incoming part
converted to outgoing spherical waves and therefore
can finally be interpreted as the part of the
wavefunction attributed to the collision).
Roughly speaking, we can express this idea as
\begin{align}
    \mathrm{scattered}(\mathrm{WF}) =
    \mathrm{outgoing}(\mathrm{WF}) -
    \mathrm{outgoing}(\mathrm{PW})\,\,.
\end{align}
Where above the `$\mathrm{outgoing}$' operation
corresponds to dropping the $e^{-i\channelwavenumber R}$ terms
with a negative exponent.
Using the plane wave expansion, we are now able to isolate
the scattering part of the wavefunction by explicitly
evaluating
\begin{align}
    \mathrm{outgoing}&\left(e^{i\channelwavevec \cdot\myvec{R}}\Big|_{R\rightarrow\infty}\right) =
    -\frac{i2\pi}{\channelwavenumber R} \sum_{\ell=0}^{\infty} \sum_{m=-\ell}^{\ell} i^{\ell}e^{i(\channelwavenumber R-\frac{\ell\pi}{2})}
    Y_{\ell m}\left(\frac{\myvec{R}}{\norm{\myvec{R}}}\right)
    Y_{\ell m}^{*}\left(\frac{\channelwavevec}{\norm{\channelwavevec}}\right)\,\,.
    \label{eq:outgoing_plane_wave}
\end{align}
Without loss of generality we can choose $\channelwavevec$ to point along $z$, in which case
$Y_{\ell m}^{*}(\theta=0) = \delta_{m0}\sqrt{(2\ell+1)/(4\pi)}$, collapsing the inner sum to just $m=0$.
Meanwhile, from \cref{eq:asymptotic_solution_fixed_phases} we see that
\begin{align}
    \mathrm{outgoing}\left( u_{\multiindex{n'}\multiindex{n}}^{\totenergy}(R\rightarrow\infty)\right)
    = -\ampplus e^{i(k_{\alpha'\beta'} R-\ell'\frac{\pi}{2})}\,\,.
    \label{eq:outgoing_radial}
\end{align}
In analogy to our analysis of \cref{subsec:amplitudes}, in three dimensions we can
express the component of the asymptotic total wavefunction going from one set of internal
states ($\alpha$, $\beta$) to another ($\alpha'$, $\beta'$) as
\begin{align}
    \Psi_{\alpha'\beta'\leftarrow\alpha\beta}(R\rightarrow\infty,\theta,\varphi) =
    \delta_{\alpha\alpha'}\delta_{\beta\beta'}e^{ik_{\alpha\beta}z} + f_{\alpha'\beta'\leftarrow\alpha\beta}(\theta,\varphi)\frac{e^{i k_{\alpha'\beta'} R}}{R}\,\,,
    \label{eq:scattering_amplitude_def}
\end{align}
for some amplitude $f_{\alpha'\beta'\leftarrow\alpha\beta}(\theta,\varphi)$, which we shall refer to
as `scattering amplitude' and again assuming $z=R\cos\theta$ to be the axis of incidence.
For general $R$, in \cref{subsec:amplitudes} we have seen that solutions can be built by combining different
radial channel solutions with $\multiindex{n}=(\alpha,\beta,\ell,m=0)$ and $\multiindex{n'}=(\alpha',\beta',\ell',m')$
multiplied by their respective spherical harmonic
\begin{align}
    \Psi_{\alpha'\beta'\leftarrow\alpha\beta}(R,\theta,\varphi) =
    \sum_{\ell=0}^{\infty} c_{\ell} \sum_{\ell'=0}^{\infty}\sum_{m'=-\ell'}^{\ell'} R^{-1} u_{\multiindex{n'}\multiindex{n}}^{\totenergy}(R)Y_{\ell'm'}(\theta,\varphi)\,\,.
\end{align}
Since the scattered term in \cref{eq:scattering_amplitude_def} is purely outgoing, the incoming part of the
wavefunction ansatz has to match the incoming part of the plane wave expansion \cref{eq:plane_wave_asymptotic}.
Then, since \cref{eq:amplitude_incoming} is only nonzero for $\multiindex{n}=\multiindex{n'}$, we can match $c_{\ell}$ coefficients
\begin{align}
    c_{\ell} = - \frac{i^{\ell}\sqrt{4\pi(2\ell+1)}}{i2 k_{\alpha\beta}}\,\,.
\end{align}
The scattering part is obtained from subtracting the outgoing part of the plane wave (\cref{eq:outgoing_plane_wave}, with $\channelwavevec$ along $z$)
from the outgoing part of the ansatz (\cref{eq:outgoing_radial}), weighted with $c_{\ell} R^{-1} Y_{\ell'm'}(\theta,\varphi)$
\begin{align}
\mathrm{scattered}(\mathrm{WF}) =
\sum_{\ell=0}^{\infty} c_{\ell} \sum_{\ell'=0}^{\infty}\sum_{m'=-\ell'}^{\ell'}
\left(\delta_{\multiindex{n'}\multiindex{n}} - P_{\multiindex{n'}\multiindex{n}}\right)
\frac{e^{i(k_{\alpha'\beta'}R-\ell'\pi/2)}}{R}\, Y_{\ell'm'}(\theta,\varphi)\,\,.
\end{align}
Finally, by comparing to \cref{eq:scattering_amplitude_def} (and inserting $c_{\ell}$) we obtain an expression for
the scattering amplitude
\begin{align}
    f_{\alpha'\beta'\leftarrow\alpha\beta}(\theta,\varphi) = \sum_{\ell=0}^{\infty}\sum_{\ell'=0}^{\infty}\sum_{m'=-\ell'}^{\ell'}
    \frac{\sqrt{4\pi(2\ell+1)}}{i2k_{\alpha\beta}}\, i^{\ell-\ell'}
    \left(P_{\multiindex{n'}\multiindex{n}} - \delta_{\multiindex{n'}\multiindex{n}}\right) Y_{\ell'm'}(\theta,\varphi)\,\,.
    \label{eq:scattering_amplitude_P}
\end{align}
It should be noted here that the scattered part converges,
since $P_{\multiindex{n'}\multiindex{n}}-\delta_{\multiindex{n'}\multiindex{n}}\rightarrow 0$ as $\ell$ is increased,
whereas the incoming and outgoing sums of the plane wave do not converge individually.
This makes it necessary to perform the matching on a per-partial-wave basis, i.e. for each $Y_{\ell m}(\theta,\varphi)$ separately,
as opposed to the full sums as functions of $\theta$ and $\varphi$.

\subsection{The collision cross section}

\label{subsec:cross_sect}

A quantum collision cross section is somewhat more abstract than
a cross section in the classical sense,
which is an inherently geometric
quantity (imagine bouncing particles off
of a hidden object and then tracing back
the trajectory of each deflected particle
in order to gradually reconstruct the shape
of the unknown target).
To make this more concrete, a detector is
arranged far away from a target (the object of
study), ideally surrounding it in all directions.
Repeatedly firing particles at the target can
either result in a hit, in which case
particle's trajectory is changed and the
particle impacts the detector at some
polar angle $\theta$ and some
azimuthal angle $\varphi$
or a miss otherwise, in which case the
trajectory remains unchanged.
Knowing the characteristics of an object
and determining a prediction for the outgoing trajectory
of an incoming particle, is sometimes referred
to as a direct scattering problem, whilst
its complement, i.e. deriving characteristics
of the scatterer from the recorded trajectories
is referred to as the inverse scattering problem.

Fundamentally, the quantum mechanical
cross section is still closely tied to
experimental observations
and so far, our treatment of collisions has been from
the theory-oriented perspective of individual
pairs of interacting particles.
In the more realistic setting, such as we will be studying
in the upcoming sections, a scattering
experiment might involve
a beam of particles
directed at a stationary target
or two intersecting beams of particles.
Then, instead of detecting each collision in isolation,
an experiment might only be able to gather statistics on
the outcome of a vast number of collision events.
In this setting it is appropriate to consider a flux of
particles with a probability current vector given by
\begin{align}
  \myvec{j} = \frac{\hbar}{i2\mu}\left(\Psi^{*}\myvecop{\nabla} \Psi - \Psi \myvecop{\nabla} \Psi^{*}\right)\,\,.
\end{align}
If we now consider the ratio of outgoing particle flux,
scattered in a particular direction (imagine a spherical screen
surrounding the intersection point of the beams at
some large distance) to the incoming particle flux,
we arrive at a quantity corresponding to what an experiment
would measure, which we define as the differential cross section.
We already know that the incident wavefunction in a certain
internal state $\alpha \beta$ behaves as a plane wave
\begin{align}
    \mathrm{incident}(\mathrm{WF}) = e^{ik_{\alpha\beta}z}
\end{align}
leading to
\begin{align}
    \myvec{j}_{\mathrm{incident}} = \frac{\hbar}{\mu} \myvec{k}_{\alpha\beta}\,\,.
\end{align}
On the other hand, as established in \cref{subsec:scattering_state},
the scattered part of the wavefunction in an internal state $\alpha'\beta'$
takes the form of a spherical wave
\begin{align}
    \mathrm{scattered}(\mathrm{WF}) = f_{\alpha'\beta'\leftarrow\alpha\beta}(\theta,\varphi) \frac{e^{ik_{\alpha'\beta'}R}}{R}
\end{align}
with a radial probability current given by
\begin{align}
    \myvec{j}_{\mathrm{scattered}}\cdot\frac{\myvec{R}}{\norm{\myvec{R}}} = \lvert f_{\alpha'\beta'\leftarrow\alpha\beta}(\theta,\varphi)\rvert^2 \frac{\hbar k_{\alpha'\beta'}}{\mu R^2}\,\,.
\end{align}
The number of outgoing particles passing through an infinitesimal
patch of the surface of a sphere of radius $R$, given by
$R^2\sin\theta \infin{\theta} \infin{\varphi} = R^2 \infin{\Omega}$ per unit time
is therefore
\begin{align}
    \lvert f_{\alpha'\beta'\leftarrow\alpha\beta}(\theta,\varphi)\rvert^2 \frac{\hbar k_{\alpha'\beta'}}{\mu} \infin{\Omega}\,\,.
\end{align}
The differential, or per unit angle, cross section can be expressed by \cite{krems2018molecules}
\begin{align}
    \frac{\infin{\sigma}}{\infin{\Omega}}
    = R^2\frac{|\myvec{j}_{\mathrm{scattered}}|}{|\myvec{j}_{\mathrm{incident}}|}
    = \frac{k_{\alpha'\beta'}}{k_{\alpha\beta}}\lvert f_{\alpha'\beta'\leftarrow\alpha\beta}(\theta,\varphi)\rvert^{2}
    \label{eq:differential_cross_section}
\end{align}
Above, we have implicitly adopted the quantum mechanical
convention that the asymptotic radial wavefunctions are normalised
by unit density.
In the context of a flux of particles, it may be more appropriate
to normalise them by probability current instead.
In this case we require
\begin{align}
    \frac{\hbar}{\mu}\,\mathrm{Im}\left(u^{*}\partial_R u\right) = \pm 1\,\,,
\end{align}
which applies separately to the incoming and outgoing waves of each channel.
For a given channel, a wave $c e^{\pm i k_{\alpha\beta} R}$
carries the probability current $\pm v_{\alpha\beta}|c|^2$,
where $v_{\alpha\beta}=\hbar k_{\alpha\beta}/\mu$ is the magnitude of the relative velocity
in channels with $\alpha$ and $\beta$.
To carry unit flux, the coefficients must therefore be given by $v_{\alpha\beta}^{-1/2}$.
Consequently, the asymptotic radial wavefunction takes the form
\begin{align}
    u_{\multiindex{n'}\multiindex{n}}(R\rightarrow\infty)
    = v_{\alpha'\beta'}^{-\frac{1}{2}} \left( \delta_{\multiindex{n'}\multiindex{n}}e^{-i(k_{\alpha'\beta'} R - \ell'\pi/2)}
    -  e^{i(k_{\alpha'\beta'} R - \ell'\pi/2)} S_{\multiindex{n'}\multiindex{n}} \right)\,\,.
    \label{eq:s_matrix_def}
\end{align}
Above, the $S_{\multiindex{n'}\multiindex{n}}$ form the $\mymat{S}$-matrix,
which has the advantageous properties of being unitary and symmetric \cite{krems2018molecules},
unlike the $P_{\multiindex{n'}\multiindex{n}}$ encountered in \cref{subsec:amplitudes}.
By rescaling the solutions to unit incoming flux, the relation
\begin{align}
    S_{\multiindex{n'}\multiindex{n}} = \sqrt{\frac{v_{\alpha'\beta'}}{v_{\alpha\beta}}}\, P_{\multiindex{n'}\multiindex{n}}\,\,
\label{eq:S_from_P}
\end{align}
holds.
Also, on account of the asymptotic degeneracy in $\ell$
we refer to $k_{\alpha\beta}=k_{\multiindex{n}}$ whilst $\ell=\ell_{\multiindex{n}}$
is the partial wave of the channel $\multiindex{n}$.
Consequently, the explicit matrix form of \cref{eq:s_matrix_def} is given by
\begin{align}
    \mymat{u}(R\rightarrow\infty)  = \mymat{v}^{-\frac{1}{2}}\left(
        e^{-i(\matchannelwavenumber R - \mymat{L}\pi/2)}
    - e^{i(\matchannelwavenumber R - \mymat{L}\pi/2) }\mymat{S} \right)\,\,,
\end{align}
with $\mymat{v}=\mathrm{diag}(v_{\multiindex{n}})$,
$\mymat{L}=\mathrm{diag}(\ell_{\multiindex{n}})$
and $\matchannelwavenumber=\mathrm{diag}(k_{\multiindex{n}})$.
The rows of the matrices refer to the channels whilst the columns contain different linearly independent solutions.
Specifically the rows of the $\mymat{S}$-matrix correspond to different exit channels.
By integrating \cref{eq:differential_cross_section} over the solid angle
$\Omega$ (exploiting the orthonormality of $Y_{\ell m}(\theta,\varphi)$)
and substituting $S_{\multiindex{n'}\multiindex{n}}$ for
$P_{\multiindex{n'}\multiindex{n}}$ we finally obtain the integral cross section
\begin{align}
    \sigma_{\alpha'\beta'\leftarrow\alpha\beta} = \int \infin{\Omega}\,\frac{\infin{\sigma}}{\infin{\Omega}}
    = \frac{\pi}{k_{\alpha\beta}^2}\sum_{\ell'=0}^{\infty}\sum_{m'=-\ell'}^{\ell'}
    \left\lvert\sum_{\ell=0}^{\infty} i^{\ell}\sqrt{2\ell+1}\left(\delta_{\multiindex{n'}\multiindex{n}} - S_{\multiindex{n'}\multiindex{n}}\right)\right\rvert^2\,\,.
\label{eq:integral_cross_section}
\end{align}
As a short aside, we shall next compare different ways
that the asymptotic wavefunction can be expressed in.

\subsection{Expressing the asymptotic solutions with $\mymat{S}$ and $\mymat{K}$ matrices}
\label{subsec:K_and_S_matrices}

There are two different ways of expressing asymptotic solutions.
Either using the Riccati--Bessel functions introduced above
or using their related counterparts, the Riccati--Hankel functions.
When dealing with multiple channels, it is convenient to
express the asymptotic solution in its matrix form
\begin{align}
    \mymat{u}(R \gg R_{0}) = -i \matchannelwavenumber R \mymat{v}^{-\frac{1}{2}} \left( \mymat{h}^{(2)}(\matchannelwavenumber R) + \mymat{h}^{(1)}(\matchannelwavenumber R)\mymat{S}\right)\,\,.
\end{align}
Above, $\mymat{h}^{(i)}(\matchannelwavenumber R)$ are diagonal matrices of
Riccati--Hankel functions in all open channels, i.e., the channels whose
asymptotic eigenenergies are below the collision energy, leaving an excess
of kinetic energy for the two particles in the corresponding
channel scattering state to be able to separate after the interaction.
We have dropped the subscripts of $\channelwavenumber$ in the argument
of the matrix form, since each element of these matrices has its own
channel-dependent wavenumber associated with it.
Closed channels, accordingly, are those whose scattering states are
inaccessible for the given collision energy, therefore their
bound states can only be occupied at close inter-particle separations,
for instance through inter-channel couplings.
As we will see below, $\mymat{S}$ is a symmetric but complex-valued
matrix encoding the `conversion' of the ingoing part of wavefunction into
its respective outgoing part.
Alternatively, we can introduce a corresponding matrix $\mymat{K}$
to connect the two (linearly independent) types of Riccati--Bessel functions in an analogous
way.
\begin{align}
    &(\mymat{F}(\matchannelwavenumber R)-i\mymat{G}(\matchannelwavenumber R)) + (\mymat{F}(\matchannelwavenumber R)+i\mymat{G}(\matchannelwavenumber R))\mymat{S} \nonumber \\
    &= \mymat{F}(\matchannelwavenumber R)(\mathbbm{1}+\mymat{S}) - \mymat{G}(\matchannelwavenumber R)i(\mathbbm{1}-\mymat{S}) \nonumber \\
    &= \Big(\mymat{F}(\matchannelwavenumber R) - \mymat{G}(\matchannelwavenumber R)\underbrace{i(\mathbbm{1}-\mymat{S})(\mathbbm{1}+\mymat{S})^{-1}}_{=\mymat{K}}\Big)(\mathbbm{1}+\mymat{S})\,\,.
    \label{eq:asymptotic_boundary_matrix}
\end{align}
Explicitly we can now express the asymptotic solution in terms of the Riccati--Bessel functions as
\begin{align}
    \mymat{u}^{(\mymat{K})}(R\gg R_{0}) = \mymat{F}(\matchannelwavenumber R) - \mymat{G}(\matchannelwavenumber R) \mymat{K}\,\,.
    \label{eq:asymptotic_bessel_wavefunction}
\end{align}
Above, we have introduced $\mymat{F}$ and $\mymat{G}$ as the
matrices consisting of the Riccati--Bessel functions of first and second kind
related to the Riccati--Hankel functions by \cref{eq:spherical_Hankel_second,eq:spherical_Hankel_first}.
Specifically, we can express the coefficients of regular (well-behaved at the
origin, i.e., as $\channelwavenumber R \rightarrow 0$) and irregular
(tend to $-\infty$ as $\channelwavenumber R \rightarrow 0$)
wave open channel boundary conditions by
\begin{align}
    \label{eq:F_matrix}
	F_{\multiindex{n}\multiindex{n'}}^{(\mathrm{open})}(\channelwavenumber R) &= \delta_{\multiindex{n}\multiindex{n'}} \sqrt{\frac{\mu k_{\alpha\beta}}{\hbar}} R j_{\ell}(\channelwavenumber R)
 \end{align}
 and
 \begin{align}
    \label{eq:G_matrix}
	G_{\multiindex{n}\multiindex{n'}}^{(\mathrm{open})}(\channelwavenumber R) &= \delta_{\multiindex{n}\multiindex{n'}} \sqrt{\frac{\mu k_{\alpha\beta}}{\hbar}} R y_{\ell}(\channelwavenumber R)\,\,,
\end{align}
respectively.
For closed channels with an internal energy of $\intenergy_{\mathbf{n}}$ we have
$\channelwavenumber^{2} = \frac{2\mu}{\hbar^2} \left(\totenergy - \intenergy_{\mathbf{n}}\right) < 0$, leading to a purely imaginary $\channelwavenumber$,
so effectively we're solving the same differential equation
\begin{align}
	x^2 y''(x) + (x^2-\ell(\ell+1))y(x) = 0
\end{align}
twice - once for
$x \in \mathbb{R}$ and once for $x=i\kappa R$, with $\kappa R > 0$ and $\kappa=|k_{\alpha\beta}|$,
which are solved by the Riccati--Bessel and the Riccati form of the modified Bessel functions, respectively.
In the imaginary case, the modified (cylindrical) Bessel functions of first and second kind
are referred to as $I_{\ell+1/2}(\kappa R)$ and $K_{\ell+1/2}(\kappa R)$ respectively.
Since they are cylindrical instead of spherical, their Riccati versions are obtained through
multiplication with $\sqrt{(\pi \kappa R)/2}$, in analogy to \cref{eq:spherical_to_cylindrical}.
Consequently
the boundary conditions can be expressed by  \cite{karman2014renormalized}
\begin{align}
  F_{\multiindex{n}\multiindex{n'}}^{(\mathrm{closed})}(\kappa R) &=
    -\delta_{\multiindex{n}\multiindex{n'}}\sqrt{\frac{2\mu R}{\pi\hbar}} I_{\ell+1/2}(\kappa R)\\
  G_{\multiindex{n}\multiindex{n'}}^{(\mathrm{closed})}(\kappa R) &=
    -\delta_{\multiindex{n}\multiindex{n'}}\sqrt{\frac{2\mu R}{\pi\hbar}} K_{\ell+1/2}(\kappa R)\,\,.
\end{align}
The $\mymat{K}$-matrix defined above is sometimes
referred to as the (augmented) reactance matrix, where `augmented'
specifies that both the open and closed channels are included.
In the following, we will show that the open-open block $\mymat{K}_{\mathrm{oo}}$
of the reactance matrix (corresponding to open incoming and outgoing channels only)
is both real and symmetric.
First, we can express the open-open block $\mymat{S}_{\mathrm{oo}}$ in terms of $\mymat{K}_{\mathrm{oo}}$
\begin{align}
    \mymat{S}_{\mathrm{oo}} = (\mathbbm{1}+i\mymat{K}_{\mathrm{oo}})(\mathbbm{1}-i\mymat{K}_{\mathrm{oo}})^{-1}\,\,.
\end{align}
We can show that $\mymat{K}_{\mathrm{oo}}=i(\mathbbm{1}-\mymat{S}_{\mathrm{oo}})(\mathbbm{1}+\mymat{S}_{\mathrm{oo}})^{-1}$ is hermitian, using $\mymat{S}_{\mathrm{oo}}^{\dagger}=\mymat{S}_{\mathrm{oo}}^{-1}$ and the fact that $\mathbbm{1}\pm\mymat{S}_{\mathrm{oo}}$ commute
\begin{align}
\mymat{K}_{\mathrm{oo}}^{\dagger} &= -i\left(\mathbbm{1}+\mymat{S}_{\mathrm{oo}}^{-1}\right)^{-1}\left(\mathbbm{1}-\mymat{S}_{\mathrm{oo}}^{-1}\right) \nonumber \\
&= -i\left(\mathbbm{1}+\mymat{S}_{\mathrm{oo}}\right)^{-1}\mymat{S}_{\mathrm{oo}}\mymat{S}_{\mathrm{oo}}^{-1}\left(\mymat{S}_{\mathrm{oo}}-\mathbbm{1}\right) \nonumber \\
&= i\left(\mathbbm{1}+\mymat{S}_{\mathrm{oo}}\right)^{-1}\left(\mathbbm{1}-\mymat{S}_{\mathrm{oo}}\right) = \mymat{K}_{\mathrm{oo}}\,\,.
\end{align}
In subsequent sections we will also encounter the transition operator $\op{T}$ with its
corresponding matrix representation $\mymat{T}$ (and its open-open block $\mymat{T}_{\mathrm{oo}}$) which we introduce here as
\begin{align}
    \mymat{S}_{\mathrm{oo}} = \mathbbm{1} + i2\mymat{T}_{\mathrm{oo}}\,\,.
\end{align}
Time-reversal symmetry is a complicated topic but for our
purposes we will restrict ourselves to the following observations.
If the potential $V(R)$ is real and the time dependent Schr{\"o}dinger equation
\begin{align}
    i \hbar \partial_t \Psi(R,t) &= \left(-\frac{\hbar^2}{2\mu}\nabla^2 + V(R)\right) \Psi(R,t)
\end{align}
is solved by $\Psi(R,t)$,
then it is also solved by $\Psi^*(R,-t)$, which follows from complex conjugation
\cite{littlejohn2021time}
\begin{align}
    (-i) \hbar \partial_t \Psi^*(R,\tau(t)) &= \left(-\frac{\hbar^2}{2\mu}\nabla^2 + V(R)\right) \Psi^*(R,\tau(t))
\end{align}
and noting when $\tau(t)=-t$, the chain rule prefactor
$\partial_t \tau(t)= -1$ is required to balance out
the negation of the prefactor $i$ on the right hand side.
Having navigated that short aside, and knowing that (roughly speaking)
$\Psi_{\mathrm{out}}=\mymat{S}\Psi_{\mathrm{in}}$ and therefore
$\Psi^*_{\mathrm{out}}=\mymat{S}^*\Psi^*_{\mathrm{in}}$,
along with the exchange of $\Psi_{\mathrm{in}}$ and
$\Psi_{\mathrm{out}}$ in the time reversed case, i.e.
$\Psi^*_{\mathrm{in}}=\mymat{S}\Psi^*_{\mathrm{out}}$
we can conclude $\Psi^*_{\mathrm{out}}=\mymat{S}^*\mymat{S}\Psi^*_{\mathrm{out}}$.
From $\mymat{S}_{\mathrm{oo}}^{\dagger}\mymat{S}_{\mathrm{oo}}={\mymat{S}_{\mathrm{oo}}^{T}}^{*}\mymat{S}_{\mathrm{oo}}=\mathbbm{1}=\mymat{S}_{\mathrm{oo}}^*\mymat{S}_{\mathrm{oo}}$
it follows that $\mymat{S}_{\mathrm{oo}}^{T}=\mymat{S}_{\mathrm{oo}}$, which is to say that
the $\mymat{S}$-matrix is symmetric.
The reader should be aware that this symmetry condition is tied to suitably phased angular functions
and that time reversal can give different results when complex $Y_{\ell m}$ are used, e.g. in an uncoupled angular momentum basis.
Since $\mathbbm{1}\pm\mymat{S}_{\mathrm{oo}}$ commute, the symmetry of $\mymat{S}_{\mathrm{oo}}$ directly implies that
\begin{align*}
    \mymat{K}_{\mathrm{oo}}&=i(\mathbbm{1}-\mymat{S}_{\mathrm{oo}})(\mathbbm{1}+\mymat{S}_{\mathrm{oo}})^{-1}\,\,, \\
    \mymat{T}_{\mathrm{oo}}&=(\mymat{S}_{\mathrm{oo}}-\mathbbm{1})/(i2)
\end{align*}
are also symmetric.
Finally, combining the symmetry and hermiticity properties
we are able to conclude that $\mymat{K}_{\mathrm{oo}}=\mymat{K}_{\mathrm{oo}}^{\dagger}={\mymat{K}_{\mathrm{oo}}^{*}}^T=\mymat{K}_{\mathrm{oo}}^{T}$
making the $\mymat{K}_{\mathrm{oo}}$-matrix real.
We note here that the (augmented) $\mymat{K}$ is real on account of $\mymat{F}$ and $\mymat{G}$ being real (please see \cref{subsec:Q_to_K}),
whereas its symmetry depends on the normalisation of the closed-channel parts of $\mymat{F}$ and $\mymat{G}$.

\subsection{Collisions with input wavepackets}

\label{subsec:wavepacket_expansion}

In theory, accounting for the fact that the initial state
appears in the form of a wavepacket in $R$, is relatively
straightforward.
Having obtained the non-asymptotic eigenstates (or specifically
the coefficients $U_{\multiindex{n'}\multiindex{n}}^{\totenergy}(R)$), we can expand the initial
wavepacket in the non-asymptotic eigenbasis.
Let's assume that the system is initially in a state that is
sharp in a certain input channel $\multiindex{m}$, which we denote by
$\ket{\Phi_{\mathrm{init},\multiindex{m}}}$, then the overlap for a
given energy $\totenergy$ is given by
\begin{align}
  c_{\multiindex{n}}^{k(\totenergy)} &= \obint{n} \ket{\Phi_{\mathrm{init},\multiindex{m}}} \nonumber \\
    &= \bint{n}
    \underbrace{\int \infin{R} R^2 \sum_{\multiindex{n'}} \kasymp{n'}{R} \obasymp{\multiindex{n'}}{R}}_{=\mathbbm{1}\otimes\mathbbm{1}}
    \ket{\Phi_{\mathrm{init},\multiindex{m}}} \nonumber \\
    &= \sum_{\multiindex{n'}} \int \infin{R} R^2 \underbrace{\obint{n}\kasymp{n'}{R}}_{={U_{\multiindex{n'}\multiindex{n}}^{\totenergy}}^*(R)}
    \underbrace{\obasymp{n'}{R}\ket{\Phi_{\mathrm{init},\multiindex{m}}}}_{=\delta_{\multiindex{n'}\multiindex{m}} \Phi_{\mathrm{init}}(R)} \nonumber \\
    &= \int \infin{R}~{u_{\multiindex{m}\multiindex{n}}^{\totenergy}}^*(R) \phi_{\mathrm{init}}(R)\,\,,
\end{align}
where, for convenience, we have defined the initial
wavefunction as $\Phi_{\mathrm{init}}(R)=\bra{R} \otimes \braket{\multiindex{m} | \Phi_{\mathrm{init},\multiindex{m}}}$
as well as its counterpart $\phi_{\mathrm{init}}(R)=R\cdot \Phi_{\mathrm{init}}(R)$.
We assume that the scattering states are energy normalised, i.e.\ $\braket{\Psi_{\multiindex{n}}^{\totenergy}|\Psi_{\multiindex{n'}}^{\totenergy'}}=\delta_{\multiindex{n}\multiindex{n'}}\delta(\totenergy-\totenergy')$.
  The full expansion is given by
\begin{align}
  \ket{\Phi_{\mathrm{init},\multiindex{m}}} &= \int \infin{\totenergy} \sum_{\multiindex{n}} c_{\multiindex{n}}^{k(\totenergy)} \kint{n} + \sum_{\multiindex{b}} c_{\multiindex{b}} \ket{\Psi_{\multiindex{b}}}\,\,,
\end{align}
where the first sum over $\multiindex{n}$ runs over all channels open at energy
$\totenergy$ whilst the second sum runs over the bound states $\ket{\Psi_{\multiindex{b}}}$ 
of the collision complex.
The terms with coefficients $c_{\multiindex{b}}=\braket{\Psi_{\multiindex{b}}|\Phi_{\mathrm{init},\multiindex{m}}}$, 
complete the set of eigenstates of $\op{H}$. 
In the following we shall neglect the contribution of the bound states under the assumption
that our initial wavepacket is prepared at large $R$, where both the interparticle interaction and 
bound state overlaps become small. This leads to
\begin{align}
    \phi_{\mathrm{init}}(R) &= R\obasymp{m}{R}\ket{\Phi_{\mathrm{init},\multiindex{m}}} \nonumber \\
    &= \int \infin{\totenergy} \sum_{\multiindex{n}} R \basymp{m}{R} c_{\multiindex{n}}^{k(\totenergy)} \kint{n}
    = \int \infin{\totenergy} \underbrace{\sum_{\multiindex{n}} c_{\multiindex{n}}^{k(\totenergy)} u_{\multiindex{m}\multiindex{n}}^{\totenergy}(R)}_{=\phi_{\mathrm{prop},\multiindex{m}}^{\totenergy}(R)}\,\,.
    \label{eq:wavepacket_expansion}
\end{align}
Above, in analogy to $\phi_{\mathrm{init}}(R)$, we have introduced $\phi_{\mathrm{prop},\multiindex{m}}^{\totenergy}(R) = R\cdot \Phi_{\mathrm{prop},\multiindex{m}}^{\totenergy}(R)$
as the $R$-dependent initial wavepacket component for a specific
energy in the non-asymptotic basis.
Since the solutions of \cref{subsec:K_and_S_matrices} are $\mymat{K}$-normalised instead
of energy normalised, after restricting contributions to the open channels the energy-normalised 
solutions are given by
\begin{align}
    \mymat{u}^{(\totenergy)} = \sqrt{\frac{2}{\pi\hbar}}\,\mymat{u}^{(\mymat{K})}\left(\mathbbm{1}+\mymat{K}_{\mathrm{oo}}^{2}\right)^{-1/2}\,\,.
\end{align}
Instead of applying this conversion by a constant matrix directly,
one can equivalently apply it to the overlaps $c_{\multiindex{n}}^{k(\totenergy)}$
once they have been computed.
We shall return to this expansion later, when considering the
numerical solution of a scattering problem involving an initial
wavepacket.

\subsection{Discussion}

The starting point of this section was the set of coupled channel differential equations from the previous section.
By dividing space up into sections, we were able to impose physical boundary conditions on the collision
problem, culminating in a definition of the scattering cross section, which is a concrete observable
that can be compared to scattering experiments.
We have seen how this is directly related to the $\mymat{S}$-matrix, which connects to the differential
equation's asymptotic solutions, the spherical Bessel and Hankel functions.
We saw that with the help of the plane wave expansion and some clever bookkeeping, it is possible
to develop a `time-independent' picture of a dynamical collision process that nevertheless correctly
captures its outcome.

In anticipation of the numerical procedure we want to apply, we have assumed that the wavefunction
vanishes at short distances, on account of a strongly repulsive interaction barrier.
This might not universally be an adequate description, for instance when dealing with collisions
which could lead to an insertion reaction.
In such cases, the wavefunction does not vanish at short interparticle
distances and one requires coordinates capable of simultaneously representing different arrangements of the atoms
(and their channels), as in the case of so-called hyperspherical coordinates \cite{althorpe2003quantum}.
In addition, we have limited our discussion to interaction potentials which fall off quicker than 1/$R^2$.
Nevertheless, our approach can treat the important cases of both dipolar ($1/R^3$) and ion--neutral ($1/R^4$) collisions,
even though these may require large values of $R_{\max}$ and many partial waves to converge.
The case of a collision governed by pure Coulomb interaction, say an electron scattering off a proton,
requires a distinct mathematical treatment \cite{friedrich2013scattering}.

We have now introduced the conceptual tools to know what goes into a scattering simulation and more crucially
 what comes out (i.e. the cross section) and how to interpret it.
These concepts are concretely anchored to the set of coupled channel differential equations and the next
section will serve as a gentle introduction to the numerical algorithm their solution actually hinges on, for
the case of a single quantum channel.

\subsection{Section summary}

\begin{keyresults}

\kr{\label{kr:ch2-xsec}%
  The differential cross section between two internal states $\alpha\beta$ and $\alpha'\beta'$, expressed in terms of the scattering amplitude
  \begin{equation*}
    \frac{\infin{\sigma}}{\infin{\Omega}}
    = R^2\frac{|\myvec{j}_{\mathrm{scattered}}|}{|\myvec{j}_{\mathrm{incident}}|}
    = \frac{k_{\alpha'\beta'}}{k_{\alpha\beta}}\lvert f_{\alpha'\beta'\leftarrow\alpha\beta}(\theta,\varphi)\rvert^{2}\,\,.
  \end{equation*}
  \emph{Cf.}~\cref{eq:differential_cross_section}.}

\kr{\label{kr:ch2-smat}%
  In case of probability current-normalised wavefunctions, the $\mymat{S}$-matrix connects the incoming and outgoing asymptotic parts
  \begin{equation*}
    u_{\multiindex{n'}\multiindex{n}}(R\rightarrow\infty)
    = v_{\alpha'\beta'}^{-\frac{1}{2}} \left( \delta_{\multiindex{n'}\multiindex{n}}e^{-i(k_{\alpha'\beta'} R - \ell'\pi/2)}
    -  e^{i(k_{\alpha'\beta'} R - \ell'\pi/2)} S_{\multiindex{n'}\multiindex{n}} \right)\,\,.
  \end{equation*}
  \emph{Cf.}~\cref{eq:s_matrix_def}.}

\kr{\label{kr:ch2-ics}%
  The integral cross section between two internal states $\alpha\beta$ and $\alpha'\beta'$,
  expressed in terms of the $\mymat{S}$-matrix
  \begin{align*}
    \sigma_{\alpha'\beta'\leftarrow\alpha\beta} = \int \infin{\Omega}\,\frac{\infin{\sigma}}{\infin{\Omega}}
    = \frac{\pi}{k_{\alpha\beta}^2}\sum_{\ell'=0}^{\infty}\sum_{m'=-\ell'}^{\ell'}
    \left\lvert\sum_{\ell=0}^{\infty} i^{\ell}\sqrt{2\ell+1}\left(\delta_{\multiindex{n'}\multiindex{n}} - S_{\multiindex{n'}\multiindex{n}}\right)\right\rvert^2\,\,.
  \end{align*}
  \emph{Cf.}~\cref{eq:integral_cross_section}.
}

\kr{\label{kr:ch2-match}%
  When expressing asymptotic boundary conditions in terms of spherical Bessel
  functions of the first ($\mymat{F}$) and second ($\mymat{G}$) kind in matrix form, the matching depends on
  the reactance matrix $\mymat{K}$
  \begin{equation*}
    \mymat{u}^{(\mymat{K})}(R\gg R_{0}) = \mymat{F}(\matchannelwavenumber R) - \mymat{G}(\matchannelwavenumber R)\mymat{K}
  \end{equation*}
  \emph{Cf.}~\cref{eq:asymptotic_bessel_wavefunction}.}

\end{keyresults}

\section{Calculating single channel scattering cross sections}

\label{sect:single_channel_scattering}

Before delving into the treatment of inelastic
collision processes between atoms and molecules
in further detail, we will follow Lester's example
\cite{lester1976dynamics} of developing some
intuition by regarding the more simple case where
only a single channel is involved in the scattering
process.
Single channel scattering is the textbook example
\cite{cohen1986quantumv2} of
quantum scattering calculations that can be performed
analytically (to some degree).

In the single channel case, we can disregard
the internal structure of both particles entirely,
such that $\op{H}_{A,\mathrm{internal}}$ and
$\op{H}_{B,\mathrm{internal}}$ and the internal
coordinates $\Acoords$ and $\Bcoords$ need
not explicitly appear in \cref{eq:non_asym_SE_function}.
Additionally, we can drop subscripts with $\alpha$ and
$\beta$ for simplicity.

In this case we know that the
asymptotic solution of the time
independent Schr{\"o}dinger equation is composed of
a plane wave (which, without loss of generality, we can assume to be
incident along the $z$ direction)
and an outgoing spherical wave,
which in general can depend on
$\theta$ and $\varphi$
\begin{align}
    \psi(R,\theta,\varphi) \xrightarrow[R\rightarrow \infty]{} e^{ikz} + f(\theta,\varphi) e^{ikR}/R\,\,.
\end{align}
Similar to \cref{subsec:cross_sect},
the incoming and outgoing probability
current vectors are given by
\begin{align}
    \myvec{j}_{\mathrm{incident}} &= \frac{\hbar k\hat{z}}{\mu} \\
    \myvec{j}_{\mathrm{scattered}} &= |f(\theta,\varphi)|^{2} \frac{\hbar \myvec{k}}{\mu R^2}\,\,.
\end{align}
In this case the differential cross section
is given directly by the scattering amplitude
\begin{align}
    \frac{\infin{\sigma}}{\infin{\Omega}} = |f(\theta,\varphi)|^2\,\,.
\end{align}
If we further restrict ourselves to the case of a
central potential $V=V(R)$, the direction of the
incident wave vector becomes an axis of symmetry.
This translates into an azimuthal symmetry in the
probability amplitude and therefore the scattered
wavefunction.
The $R$ and $\theta$-dependent scattering state
can therefore be expanded in terms of Legendre
polynomials according to
\begin{align}
    \psi(R,\theta) = \sum_{\ell=0}^{\infty} R^{-1} u_{\ell}(R) P_{\ell}(\cos\theta)
\end{align}
and
\begin{align}
    f(\theta) = \sum_{\ell=0}^{\infty} f_{\ell} P_{\ell}(\cos\theta)\,\,.
\end{align}
The terms in the above sums correspond to
different partial waves, i.e. the orbital angular
momenta $\ell$.
Each $u_{\ell}(R)$ solves the radial single
channel differential equation
\begin{align}
    \partial_{R}^{2} u_{\ell}(R) + (k^2 - 2\mu V(R)/\hbar^2 - \ell(\ell+1)R^{-2})u_{\ell}(R) = 0\,\,.
    \label{eq:radial_single_channel}
\end{align}
Asymptotically, where we expect $V(R) \rightarrow 0$, this reduces to the Riccati--Bessel
differential equation.
As such, asymptotic solutions can be expressed as
arbitrary linear combinations of
$kRj_{\ell}(kR)$ and $kRy_{\ell}(kR)$.
Therefore
\begin{align}
    u_{\ell}(R) = kR(A_{\ell}j_{\ell}(kR) + B_{\ell}y_{\ell}(kR))\,\,,
\end{align}
for some partial wave-dependent constants $A_{\ell}$
and $B_{\ell}$.
From the asymptotic behaviour of the spherical
Hankel functions discussed in \cref{subsec:hankel}
the spherical Bessel functions of first and second
kind behave as
\begin{align}
    j_{\ell}(kR) &\rightarrow (kR)^{-1} \sin{(kR-\ell\pi/2)} \\
    y_{\ell}(kR) &\rightarrow -(kR)^{-1} \cos{(kR-\ell\pi/2)}\,\,.
\end{align}
The asymptotic radial wavefunction can be
expressed as a phase shifted sine function
\begin{align}
    u_{\ell}(R) \rightarrow &
    A_{\ell} \sin{(kR-\ell\pi/2)} - B_{\ell} \cos{(kR-\ell\pi/2)} \\
    = & C_{\ell} \sin{(kR-\ell\pi/2 + \eta_{\ell})} \\
    = & C_{\ell} \frac{e^{i(kR-\ell\pi/2 + \eta_{\ell})}-e^{-i(kR-\ell\pi/2 + \eta_{\ell})}}{i2}\,\,,
    \label{eq:single_channel_asymptotic}
\end{align}
with $C_{\ell} = (A_{\ell}^2+B_{\ell}^2)^{1/2}$, where $A_{\ell}$ and $B_{\ell}$ may share a common complex factor $\mathcal{N}_{\ell}$,
\begin{align}
    A_{\ell} = \mathcal{N}_{\ell}\,a_{\ell}\,\,,\qquad B_{\ell} = \mathcal{N}_{\ell}\,b_{\ell}\,\,,\qquad a_{\ell},b_{\ell}\in\mathbb{R}\,\,,~\mathcal{N}_{\ell}\in\mathbb{C}\,\,,
\end{align}
such that $\eta_{\ell}=\arctan{(-B_{\ell}/A_{\ell})}$ is real, whereas $C_{\ell}$ is in general complex; we choose the sign of the square root according to $C_{\ell}\cos\eta_{\ell}=A_{\ell}$.
Famously, $\eta_{\ell}$ is the so-called scattering
phase shift, i.e., the shift in the asymptotic radial wavefunction between the 
free solution (where the interparticle interaction is turned off) and the true 
solution accounting for the close range interaction of the two particles.
To see that the equality above holds, defining $x=kR-\ell\pi/2$, we note that
\begin{align*}
    A\frac{e^{ix}-e^{-ix}}{i2} - B\frac{e^{ix}+e^{-ix}}{2}
    &= \frac{A-iB}{i2} e^{ix} - \frac{A+iB}{i2}e^{-ix} \\
    &= \frac{1}{i2}\left(e^{i\gamma_1} e^{ix} - e^{-i\gamma_2} e^{-ix} \right) \\
    &= \frac{e^{i\gamma_3}}{i2} \left(e^{i(x+\gamma_1-\gamma_3)} - e^{-i(x+\gamma_2+\gamma_3)} \right) \\
    &= \frac{e^{i\gamma_3}}{i2} \left( e^{i(x+\gamma_4)} - e^{-i(x+\gamma_4)}\right) \\
    &= e^{i\gamma_3} \sin{(x+\gamma_4)}\,\,,
\end{align*}
where $\gamma_{1}=\ln{(A-iB)}/i$ and $\gamma_{2}=-\ln{(A+iB)}/i$.
Above, we have defined $\gamma_3$ such that $\gamma_4=\gamma_1-\gamma_3=\gamma_2+\gamma_3$,
therefore $\gamma_3 =  (\gamma_1-\gamma_2)/2$ and $\gamma_4=(\gamma_1+\gamma_2)/2$.
Finally
\begin{align*}
    e^{i\gamma_3} &= \exp{\left(i \frac{\ln((A-iB)(A+iB))}{i2}\right)} \\
    &= (A^2+B^2)^{1/2}
\end{align*}
and
\begin{align}
    \gamma_4 = \frac{\ln{(\frac{A-iB}{A+iB})}}{i2} = \frac{\ln{(\frac{i+\frac{B}{A}}{i-\frac{B}{A}})}}{i2} = \frac{i2 \arctan{(-B/A)}}{i2}\,\,.
\end{align}
Combining the plane wave expansion
\begin{align}
    e^{ikz} = e^{ikR\cos\theta} = \sum_{\ell=0}^{\infty} (2\ell+1)i^{\ell}j_{\ell}(kR)P_{\ell}(\cos\theta)
\end{align}
with the expansion of $f(\theta)$ yields
\begin{align}
    e^{ikz} + f(\theta)R^{-1} e^{ikR}
    = \sum_{\ell=0}^{\infty} \left( (2\ell+1) i^{\ell} j_{\ell}(kR) + f_{\ell}R^{-1}e^{ikR}\right) P_{\ell}(\cos\theta)\,\,.
    \label{eq:single_channel_expansion}
\end{align}
For asymptotically large $R$ we require equality
between \cref{eq:single_channel_asymptotic,eq:single_channel_expansion}
on a per partial wave basis
\begin{align}
    &R^{-1} C_{\ell}\frac{e^{i(kR-\ell\pi/2 + \eta_{\ell})}-e^{-i(kR-\ell\pi/2 + \eta_{\ell})}}{i2} \nonumber \\
    &= (2\ell+1) i^{\ell} (kR)^{-1} \sin{(kR-\ell\pi/2)} + f_{\ell}R^{-1}e^{ikR} \nonumber \\
    &= (2\ell+1) i^{\ell} (kR)^{-1} \frac{e^{i(kR-\ell\pi/2)}-e^{-i(kR-\ell\pi/2)}}{i2} + f_{\ell}R^{-1}e^{ikR} \nonumber \\
    &= R^{-1}\left((2\ell+1)\frac{i^{\ell}e^{-i\ell\pi/2}}{i2k} + f_{\ell} \right) e^{ikR}
    - R^{-1}(2\ell+1)\frac{i^{\ell}e^{i\ell\pi/2}}{i2k}e^{-ikR} \nonumber \\
    &= R^{-1}\left(\frac{2\ell+1}{i2k} + f_{\ell} \right) e^{ikR}
    - R^{-1}(2\ell+1)\frac{(-1)^{\ell}}{i2k}e^{-ikR}\,\,.
\end{align}
Comparing prefactors of the $R^{-1}e^{-ikR}$ terms
\begin{align}
    -C_{\ell}\frac{e^{-i(-\ell\pi/2+\eta_{\ell})}}{\cancel{i2}}
    &= -(2\ell+1) \frac{(-1)^{\ell}}{\cancel{i2}k} \\
    C_{\ell} &= (2\ell+1)\frac{i^{\ell}}{k} e^{i\eta_{\ell}}\,\,.
\end{align}
Similarly, for the $R^{-1}e^{ikR}$ terms we get
\begin{align}
    C_{\ell} \frac{e^{i(-\ell\pi/2+\eta_{\ell})}}{i2}
    &= \frac{2\ell+1}{i2k} + f_{\ell}
\end{align}
and, substituting in the found expression for $C_{\ell}$ and
solving for $f_{\ell}$,
\begin{align}
    f_{\ell}
    &= (2\ell+1)\frac{i^{\ell}}{k} e^{i\eta_{\ell}}\frac{e^{i(-\ell\pi/2+\eta_{\ell})}}{i2} - \frac{2\ell+1}{i2k} \nonumber \\
    &= \frac{2\ell+1}{i2k}e^{i\eta_{\ell}}
    \left(i^{\ell}e^{-i\ell\pi/2}e^{i\eta_{\ell}} - e^{-i\eta_{\ell}}\right) \nonumber \\
    &= (2\ell+1)\frac{e^{i\eta_{\ell}}}{k} \sin{\eta_{\ell}}\,\,.
\end{align}
Then, the expansion of the scattering probability amplitude
can be expressed as
\begin{align}
    f(\theta) = k^{-1} \sum_{\ell=0}^{\infty} (2\ell+1) e^{i\eta_{\ell}}\sin{\eta_{\ell}} P_{\ell}(\cos\theta)\,\,,
\end{align}
leading to the differential cross section
\begin{align}
    \frac{\infin{\sigma}}{\infin{\Omega}} = k^{-2}\sum_{\ell \ell'} (2\ell+1)(2\ell'+1) e^{i(\eta_{\ell}-\eta_{\ell'})} \sin{\eta_{\ell}}\sin{\eta_{\ell'}}P_{\ell}(\cos\theta)P_{\ell'}(\cos\theta)\,\,.
\end{align}
Integrating over the solid angle and making use of the
orthogonality of Legendre polynomials
\begin{align}
    \sigma &= k^{-2} \sum_{\ell\ell'} (2\ell+1)(2\ell'+1) e^{i(\eta_{\ell}-\eta_{\ell'})} \sin{\eta_{\ell}} \sin{\eta_{\ell'}} \int \infin{\Omega} P_{\ell}(\cos\theta)P_{\ell'}(\cos\theta) \nonumber \\
    &= k^{-2} \sum_{\ell\ell'} (2\ell+1)(2\ell'+1) e^{i(\eta_{\ell}-\eta_{\ell'})} \sin{\eta_{\ell}} \sin{\eta_{\ell'}} \delta_{\ell\ell'} \frac{2}{2\ell+1} 2\pi \nonumber \\
    &= 4\pi k^{-2} \sum_{\ell} (2\ell+1) \sin^{2}\eta_{\ell}\,\,.
\end{align}
In the following, we will demonstrate that we can
portray the scattering state in different ways
by choosing $A_{\ell}$ and $B_{\ell}$ (consistent with
their relation to the scattering phases $\eta_{\ell}$).
This way, we can express the in- and outgoing terms in the
single channel scattering wavefunction in terms of
scalar quantities corresponding to
the `reactance', `scattering' and `transition' collision matrices
$\mymat{K},\,\mymat{S}$ and $\mymat{T}$.

\subsection{$K_{\ell}$ amplitude}

One possible choice of $A_{\ell}$ and $B_{\ell}$ that
satisfies $\eta_{\ell}=\arctan{(-B_{\ell}/A_{\ell})}$
is $A_{\ell}=e^{i\eta_{\ell}}\cos\eta_{\ell}$ and
$B_{\ell}=-e^{i\eta_{\ell}}\sin\eta_{\ell}$ leading to
$C_{\ell}=e^{i\eta_{\ell}}$ and
\begin{align}
    u_{\ell}(R)
    &= kR e^{i\eta_{\ell}} \left(\cos\eta_{\ell}j_{\ell}(kR) - \sin\eta_{\ell}y_{\ell}(kR) \right) \nonumber \\
    &\rightarrow e^{i\eta_{\ell}} \big(\cos\eta_{\ell}\sin{(kR-\ell\pi/2)} + \sin\eta_{\ell}\cos{(kR-\ell\pi/2)} \big) \nonumber \\
    &=e^{i\eta_{\ell}}\cos\eta_{\ell} \big(\sin{(kR-\ell\pi/2)} + \tan\eta_{\ell}\cos{(kR-\ell\pi/2)} \big)
\end{align}
Following \cite{lester1976dynamics}, in this case
we can denote the amplitude of the irregular (asymptotically cosine)
standing wave relative to the regular (asymptotically sine) one by
\begin{align}
    K_{\ell} = \tan{\eta_{\ell}}\,\,.
    \label{eq:k_ell}
\end{align}

\subsection{$T_{\ell}$ amplitude}

Another valid choice for $A_{\ell}$ and $B_{\ell}$ is
$A_{\ell}=1+ie^{i\eta_{\ell}}\sin\eta_{\ell}$
and $B_{\ell}=-e^{i\eta_{\ell}}\sin\eta_{\ell}$
in which case we can represent the asymptotic
radial wavefunction as follows
\begin{align}
    u_{\ell}(R)
    &= kR\left( j_{\ell}(kR) - e^{i\eta_{\ell}}\sin\eta_{\ell} (y_{\ell}(kR) - i j_{\ell}(kR) ) \right) \nonumber \\
    &\rightarrow \sin{(kR-\ell\pi/2)} + e^{i\eta_{\ell}}\sin\eta_{\ell} (\cos{(kR-\ell\pi/2)}+i\sin{(kR-\ell\pi/2)}) \nonumber \\
    &= \sin{(kR-\ell\pi/2)} + e^{i\eta_{\ell}}\sin\eta_{\ell} e^{i(kR-\ell\pi/2)}
\end{align}
Please note that, in contrast to what is stated in \cite{lester1976dynamics},
we find that the two terms in $A_{\ell}$ have
the same sign, since
\begin{align}
    \tan{\eta_{\ell}} &= \frac{2}{i2} \frac{e^{i\eta_{\ell}}-e^{-i\eta_{\ell}}}{e^{i\eta_{\ell}}+e^{-i\eta_{\ell}}} \nonumber \\
    -\frac{B_{\ell}}{A_{\ell}}
    &= - \frac{-e^{i\eta_{\ell}}\sin\eta_{\ell}}{1+ie^{i\eta_{\ell}}\sin\eta_{\ell}} \nonumber \\
    &= - \frac{-e^{i\eta_{\ell}}(e^{i\eta_{\ell}}-e^{-i\eta_{\ell}})/(i2)}{1+ie^{i\eta_{\ell}}(e^{i\eta_{\ell}}-e^{-i\eta_{\ell}})/(i2)} \nonumber \\
    &= - \frac{-e^{i\eta_{\ell}}(e^{i\eta_{\ell}}-e^{-i\eta_{\ell}})}{i2e^{i\eta_{\ell}}e^{-i\eta_{\ell}}+ie^{i\eta_{\ell}}(e^{i\eta_{\ell}}-e^{-i\eta_{\ell}})} \nonumber \\
    &= \frac{e^{i\eta_{\ell}}-e^{-i\eta_{\ell}}}{i(e^{i\eta_{\ell}}+e^{-i\eta_{\ell}})}
\end{align}
Consequently, we denote the amplitude of the
outgoing wave by \cite{lester1976dynamics}
\begin{align}
    T_{\ell} = e^{i\eta_{\ell}}\sin\eta_{\ell}\,\,.
    \label{eq:t_ell}
\end{align}

\subsection{$S_{\ell}$ amplitude}

Yet another choice is to set
$A_{\ell}=\frac{i}{i2}(e^{i2\eta_{\ell}}+1)$
and
$B_{\ell}=-\frac{1}{i2}(e^{i2\eta_{\ell}}-1)$.
In this case, the asymptotic radial wavefunction takes the form
\begin{align}
    u_{\ell}(R) =& kR\left( \frac{i}{i2}(e^{i2\eta_{\ell}}+1) j_{\ell}(kR) - \frac{1}{i2}(e^{i2\eta_{\ell}}-1)y_{\ell}(kR) \right) \nonumber \\
    =& \frac{kR}{i2} \Big( (y_{\ell}(kR) + ij_{\ell}(kR)) - e^{i2\eta_{\ell}}(y_{\ell}(kR)-ij_{\ell}(kR)) \Big) \nonumber \\
    \rightarrow&  \frac{1}{i2} \Big( -\cos{(kR-\ell\pi/2)}+i\sin{(kR-\ell\pi/2)} \nonumber \\
    &- e^{i2\eta_{\ell}} \left(-\cos{(kR-\ell\pi/2)} -i \sin{(kR-\ell\pi/2)}\right) \Big) \nonumber \\
    =& \frac{1}{i2} \left( -e^{-i(kR-\ell\pi/2)} + e^{i2\eta_{\ell}} e^{i(kR-\ell\pi/2)}  \right)\,\,.
\end{align}
Here too, to ensure
consistency with our definition of
$\eta_{\ell}$, we obtain different signs than in \cite{lester1976dynamics}, namely
\begin{align}
    \tan{\eta_{\ell}} &= \frac{2}{i2} \frac{e^{i\eta_{\ell}}-e^{-i\eta_{\ell}}}{e^{i\eta_{\ell}}+e^{-i\eta_{\ell}}} \nonumber \\
    -\frac{B_{\ell}}{A_{\ell}}
    &= -\frac{\frac{1}{i2}(e^{i2\eta_{\ell}}-1)}{-\frac{i}{i2}(e^{i2\eta_{\ell}}+1)} \nonumber \\
    &= \frac{1}{i} \frac{e^{i\eta_{\ell}}}{e^{i\eta_{\ell}}} \frac{e^{i\eta_{\ell}}-e^{-i\eta_{\ell}}}{e^{i\eta_{\ell}}+e^{-i\eta_{\ell}}} \,\,.
\end{align}
This final relative amplitude is denoted by
\begin{align}
    S_{\ell} = e^{i2\eta_{\ell}}\,\,.
    \label{eq:s_ell}
\end{align}

\subsection{Single channel Numerov's method}

\label{subsec:numerov_sc}

Our ultimate goal is to solve a differential equation of the form
\begin{align}
    \partial_{x}^2 y(x) = -\numlinsc(x)y(x) + g(x)\,\,,
    \label{eq:numerov_diff_eq}
\end{align}
not for all $x \in \mathbb{R}$ but for certain, specific values
on a spatial grid $x \in \left\{x_{\min},\ldots,x_{\max}\right\}$.
In order to actually perform a numerical calculation of the
wavefunction, we have to choose this (in our case uniform) finite grid,
which can be parameterised by three quantities
\begin{enumerate}
    \item $x_{\min}$ - the closest inter-particle separation, has to be chosen sufficiently small, that the wavefunction vanishes in the strongly repulsive area of the potential
    \item $x_{\max}$ - the largest inter-particle separation, at which asymptotic boundary conditions should hold (for potentials that fall off faster than $1/x^2$ we recall that we first encounter `spherical Bessel' and, going even further, the `plane wave' behaviour of non-interacting particles)
    \item $\Delta x$ - the grid point separation, determines the numerical error in the evaluation of the eigenstates
      calculated with Numerov's method.
\end{enumerate}
For our collision problem, we identify the physical quantities in \cref{eq:radial_single_channel}
with the following variables in \cref{eq:numerov_diff_eq}:
$x=R$, $y=u_{\ell}$ (and $g=0$, though we will perform the derivation for the general
case and specialise later) and
\begin{align}
    \numlinsc(R) = k^2 - \frac{2\mu}{\hbar^2}V(R) - \frac{\ell(\ell+1)}{R^2}\,\,.
    \label{eq:s_single_channel}
\end{align}

In the context of scattering, just being able to evaluate $y(x)$ at the final two
grid points would typically already give enough information to obtain the collision cross section,
since that is what is required to determine the scattering phases
introduced in \cref{sect:single_channel_scattering}.
First, we will consider the first 6 terms of a Taylor expansion
of our solution function $y(x)$ around some point $x_{0}$
\begin{align}
    y(x) = \sum_{k=0}^{5} \frac{\partial_{x}^{k}y(x)|_{x=x_{0}}}{k!} (x-x_{0})^{k} + \mathcal{O}((x-x_{0})^{6})\,\,.
\end{align}
Defining $h=x-x_{0}$ to be a small distance between $x$ and $x_{0}$ (e.g. the
spacing between $x_{i}$ and $x_{i+1}$, the $i^{\mathrm{th}}$ and $(i+1)^{\mathrm{st}}$ points on an equidistant numerical grid) it follows that
\begin{align}
    y(x_{0}+h) = \sum_{k=0}^{5} \frac{\partial_{x}^{k}y(x)|_{x=x_{0}}}{k!} h^{k} + \mathcal{O}(h^{6})\,\,,
\end{align}
or, going either forwards or backwards on our grid,
\begin{align}
    y(x_{i}+h) &= \sum_{k=0}^{5} \frac{\partial_{x}^{k}y(x)|_{x=x_{i}}}{k!} h^{k} + \mathcal{O}(h^{6}) \\
    y(x_{i}-h) &= \sum_{k=0}^{5} \frac{\partial_{x}^{k}y(x)|_{x=x_{i}}}{k!} (-h)^{k} + \mathcal{O}(h^{6})\,\,.
\end{align}
Due to the fact that $h^{k}+ (-1)^{k}h^k=0$ when $k$ is odd
\begin{align}
    y(x_{i+1}) + y(x_{i-1}) = 2y(x_{i}) + \partial_{x}^{2}y(x)|_{x=x_{i}}h^2 + \partial_{x}^4y(x)|_{x=x_{i}}\frac{h^4}{12} + \mathcal{O}(h^6)\,\,.
\end{align}
We can already see that this has the makings of a recursion relation - and
Numerov's trick was now to define a second function, namely
$w(x)=\partial_{x}^{2}y(x)$, and substitute that into the equation above
\begin{align}
    y(x_{i+1}) + y(x_{i-1}) -2y(x_i) = w(x)|_{x=x_{i}}h^{2} + \partial_{x}^{2} w(x)|_{x=x_{i}}\frac{h^4}{12} + \mathcal{O}(h^6)\,\,.
    \label{eq:numerov_yw}
\end{align}
Now we can note that the exact same Taylor expansion can also be applied to $w(x)$, which yields
\begin{align}
    \frac{h^2}{12} \left(w(x_{i+1}) + w(x_{i-1}) - 2w(x_{i})\right) = \frac{h^2}{12} \left(\partial_{x}^{2}w(x)|_{x=x_{i}}h^2 + \partial_{x}^{4}w(x)|_{x=x_{i}}\frac{h^4}{12} + \mathcal{O}(h^6)\right) \,\,.
\end{align}
The additional prefactor of $h^2$ pushes the fourth derivative term above the
$\mathcal{O}(h^6)$ threshold, allowing it to be dropped
\begin{align}
    \frac{h^4}{12}\partial_{x}^{2}w(x)|_{x=x_{i}}= \frac{h^2}{12} \left(w(x_{i+1}) + w(x_{i-1}) - 2w(x_{i})\right) + \mathcal{O}(h^6)\,\,.
\end{align}
At this point we are able to substitute in the original differential equation
$w(x)=-\numlinsc(x)y(x)+g(x)$
\begin{align}
    \frac{h^4}{12}\partial_{x}^{2}w(x)|_{x=x_{i}}= \frac{h^2}{12} (&-\numlinsc(x_{i+1})y(x_{i+1})+g(x_{i+1}) \nonumber \\
     &-\numlinsc(x_{i-1})y(x_{i-1})
    +g(x_{i-1}) + 2\numlinsc(x_{i})y(x_{i}) - 2g(x_{i})) + \mathcal{O}(h^6)\,\,.
\end{align}
Inserting the expressions for $w(x)$ and $\frac{h^4}{12}\partial_{x}^{2}w(x)|_{x=x_{i}}$ into \cref{eq:numerov_yw}
finally yields an equation up to sixth order in $h$, which depends
only on the grid points of $\numlinsc$, $g$ and our unknown solution $y$
without any leftover derivative terms
\begin{align}
    y(x_{i+1}) + y(x_{i-1}) -2y(x_i) = \left(-\numlinsc(x_{i})y(x_{i})+g(x_{i}) \right) h^{2}
    + \frac{h^2}{12} \big(-\numlinsc(x_{i+1})y(x_{i+1})+g(x_{i+1}) \nonumber \\
    -\numlinsc(x_{i-1})y(x_{i-1})
    +g(x_{i-1}) + 2\numlinsc(x_{i})y(x_{i}) - 2g(x_{i})\big) + \mathcal{O}(h^6)\,\,.
\end{align}
This recursion relation means that given two neighbouring points, say
$y(x_{0})$ and $y(x_{1})$, in principle all further points can be
evaluated according to
\begin{align}
    y(x_{i+1}) \approx \frac{\Big( \frac{h^2}{12}(g(x_{i+1})+10g(x_{i})+g(x_{i-1})) + 2y(x_{i}) (1- \frac{5h^2}{12}\numlinsc(x_{i})) - y(x_{i-1}) (1+\frac{h^2}{12}\numlinsc(x_{i-1})) \Big)}{1+\frac{h^2}{12}\numlinsc(x_{i+1})}\,\,,
\end{align}
if $h$ is sufficiently small.
In the homogeneous case ($g(x)=0$), which is of interest for the
solution of the coupled channel problem formulated above, the equation
simplifies to
\begin{align}
    y(x_{i+1}) \approx \frac{\Big(2y(x_{i}) (1- \frac{5h^2}{12}\numlinsc(x_{i})) - y(x_{i-1}) (1+\frac{h^2}{12}\numlinsc(x_{i-1})) \Big)}{1+\frac{h^2}{12}\numlinsc(x_{i+1})}\,\,.
    \label{eq:numerov_homog}
\end{align}
In practice, the base implementation of Numerov's method, as shown
here, suffers from some drawbacks, which can be remedied by renormalisation,
which we will address next before generalising Numerov's method
to the multi-channel case.

\subsection{Renormalised Numerov's method}

In the base formulation of Numerov's method, numerical instability
can arise when propagating through classically forbidden regions
(i.e., spatial regions where the system's potential energy would exceed
the total amount of available energy) and
can manifest itself in terms of both numerical overflow and the linear
dependence of solutions (in the coupled channels case) even though they
should actually be mutually linearly independent
\cite{karman2014renormalized}.
The renormalised version of Numerov's method \cite{johnson1977new} addresses these issues by
attaining solutions to the ordinary differential equation
in a slightly more indirect way, by stepping through the ratio of solutions
between two neighbouring points, as opposed to the solution itself directly.
Let $q(x_{i})$ be the ratio between the solution in the previous
$(i-1)^{\mathrm{st}}$ and current $i^{\mathrm{th}}$ grid point
\begin{align}
    y(x_{i-1}) = q(x_{i}) y(x_{i})\,\,.
    \label{eq:renormalised_recursion_sc}
\end{align}
Similarly, going from the current to the next grid point
\begin{align}
    y(x_{i+1}) = q(x_{i+1})^{-1} y(x_{i})\,\,.
\end{align}
Under the assumption that $y(x_{i}) \neq 0$,
both expressions can be directly substituted into
\cref{eq:numerov_homog}, leading to
\begin{align}
    q(x_{i+1})^{-1} \cancel{y(x_{i})} \approx \frac{\Big(2\cancel{y(x_{i})} (1- \frac{5h^2}{12}\numlinsc(x_{i})) - \numratiosc(x_{i}) \cancel{y(x_{i})} (1+\frac{h^2}{12}\numlinsc(x_{i-1})) \Big)}{1+\frac{h^2}{12}\numlinsc(x_{i+1})}
    \label{eq:ren_numerov_homog}
\end{align}
which can be inverted for a single-term recursion relation
for the ratio function $q(x_{i})$.
The propagation is started in the strongly repulsive region with $y(x_{1})=0$ and an arbitrary $y(x_{2})\neq 0$, i.e.\ $q(x_{2})=0$.

Determining $q(x)$ at each grid point
contains the information needed to
obtain the solution $y(x)$ but in the
context of scattering, the cross section
(or the corresponding scattering phase)
can be determined directly from $q(x)$
in the final two grid points
(i.e., $x_{\max}-h$ and $x_{\max}$ are both
assumed to be in the asymptotic region of the
inter-particle interaction),
after fixing the boundary conditions at the end of
the spatial grid.
Matching $u_{\ell}=F_{\ell}-G_{\ell}K_{\ell}$, with $F_{\ell}(x)=kx\,j_{\ell}(kx)$ and $G_{\ell}(x)=kx\,y_{\ell}(kx)$, at the last two grid points $x_{n-1}$ and $x_{n}=x_{\max}$ gives
\begin{align}
    K_{\ell} = \tan\eta_{\ell} = \frac{F_{\ell}(x_{n-1})-q(x_{n})F_{\ell}(x_{n})}{G_{\ell}(x_{n-1})-q(x_{n})G_{\ell}(x_{n})}\,\,,
    \label{eq:K_ell_from_q}
\end{align}
the single-channel version of \cref{eq:K_from_Q}.
In this tutorial we will also be interested
in explicitly obtaining solutions, i.e.
wavefunctions in the context of quantum
scattering, which requires us to store $q(x)$
at each grid point and then iterate from the
end of the grid (where the asymptotic
boundary conditions have been fixed)
inwards by the recursion relation
given by \cref{eq:renormalised_recursion_sc}.

\subsection{Discussion}

In this section, we gained a concrete prescription for taking a special case of the boundary value problem formulated in the previous
two sections, which is not analytically solvable for arbitrary interaction potentials, and evaluating it
algorithmically using Numerov's method.
We saw how we can make this approach more robust by propagating the ratio of the wavefunction in two neighbouring points
as opposed to the wavefunction itself directly, which we refer to as the `renormalised' Numerov's method.
In addition, we have seen how the single channel asymptotic wavefunction can be portrayed in different ways in terms of the scattering
phase, using `reactance', `scattering' and `transition' collision amplitudes.

The beauty of this perspective on collision simulations is that apart from the Born-Oppenheimer approximation (which relies
on a well-isolated potential energy surface), the wavefunctions we obtain can be made essentially exact.
As discussed at the end of the previous section, there are restrictions on which interaction potentials are valid for this
approach, namely that they fall off faster than 1/$R^2$ and possess a strong repulsive wall at short distances where the
wavefunction is expected to vanish.
Numerov's method, however, is based on Taylor expansion of the wavefunction and is therefore a local method of solving a differential
equation and leads to a numerical error.
As a consequence, small errors in one point of the propagated solution are carried over to the next, making it crucial to carefully ensure
convergence by the choice of an adequate spatial grid.
Depending on the interaction potential, this grid can become very dense, increasing the numerical complexity of the problem.
This issue can be addressed by modifying Numerov's method to account for non-uniform grids \cite{vigo2005variable}, or similar propagator
methods which can, for instance, follow the strength of the
potential in different regions of space \cite{karman2014renormalized}.
One commonly used alternative, the log-derivative method \cite{johnson1973multichannel,manolopoulos1986improved}, 
similarly generates solutions to the coupled channel equations.
As the name implies, this is accomplished by propagating the logarithmic derivative of the wavefunction
between neighbouring spatial grid points.
Special care should also be taken to ensure that the first and last grid points are chosen near and far enough such that
the wavefunction does truly vanish at the start of propagation and the asymptotic solution actually is independent of the potential energy
surface's influence on the collision.

We now have an intuitive understanding of how to treat a problem consisting of just a single collision channel,
suitable for the collision between two atoms, for example, in regimes where there is no coupling between their internal states.
This grasp provides a good springboard to generalise to the multichannel version of Numerov's method, with which we can actually
solve sets of coupled differential equations, taking internal structures of the colliding particles into account.
The next section will introduce the machinery and make the treatment concrete for the case of a (structureless) atom colliding with
a diatomic molecule which can both rotate and vibrate.

\subsection{Section summary}

\begin{keyresults}

\kr{\label{kr:ch3-sell}%
  In the single channel case, reactance ($K_{\ell}$), transition ($T_{\ell}$) and scattering ($S_{\ell}$) amplitudes
  can be expressed in terms of a single scattering phase $\eta_{\ell}$
  \begin{equation*}
    K_{\ell} = \tan\eta_{\ell}  \quad T_{\ell} = e^{i\eta_{\ell}}\sin\eta_{\ell}  \quad S_{\ell} = e^{i2\eta_{\ell}}\,\,.
  \end{equation*}
  \emph{Cf.}~\cref{eq:k_ell}, \cref{eq:t_ell} and \cref{eq:s_ell}.}

\kr{\label{kr:ch3-numerov}%
  The single-channel Numerov recursion relation for $\partial_{x}^2 y(x) = -\numlinsc(x)y(x)$ discretised on a grid with uniform spacing $x_{i+1}=x_{i}+h$ is given by
  \begin{equation*}
    y(x_{i+1}) \approx \frac{\Big(2y(x_{i}) (1- \frac{5h^2}{12}\numlinsc(x_{i})) - y(x_{i-1}) (1+\frac{h^2}{12}\numlinsc(x_{i-1})) \Big)}{1+\frac{h^2}{12}\numlinsc(x_{i+1})} .
  \end{equation*}
  \emph{Cf.}~\cref{eq:numerov_homog}.}

\kr{\label{kr:ch3-ren}
The single-term recursion relation for the ratio $q(x_i) = y(x_{i-1})/y(x_i)$ leads to a more stable propagation (referred to as renormalised Numerov's method)
\begin{equation*}
    q(x_{i+1})^{-1} \approx \frac{\Big(2(1- \frac{5h^2}{12}\numlinsc(x_{i})) - \numratiosc(x_{i}) (1+\frac{h^2}{12}\numlinsc(x_{i-1})) \Big)}{1+\frac{h^2}{12}\numlinsc(x_{i+1})}
\end{equation*}
\emph{Cf.}~\cref{eq:ren_numerov_homog}.}

\end{keyresults}

\section{Calculating cross sections of inelastic atom-diatom collisions}

\label{sect:atom_diatom_collisions}

The following workflow provides an overview of how the individual pieces of this section combine into a scattering calculation
that can be used to describe atom-diatom collisions.
\begin{tcolorbox}[enhanced, colback=blue!3, colframe=blue!45!black, boxrule=0.6pt, arc=2pt, fonttitle=\bfseries\sffamily, coltitle=white, title={Workflow of an atom--diatom scattering calculation}]
\begin{enumerate}
  \item Obtain the potential energy surface $V(R,r,\theta)$ (cf.\ \cref{sect:formalism}) and the diatomic potential $V_{BC}(r)$.
  \item Use the sinc-DVR to solve the radial diatomic eigenproblem (\cref{subsec:dvr}), yielding $\chi_{vj}(r)$ and $\intenergy_{vj}$ for all $v$ at once, with one diagonalisation per $j$.
  \item Perform a Legendre expansion of the interaction potential (\cref{eq:legendre_theta_potential_expansion}) to obtain its expansion coefficients.
  \item Divide the problem into blocks of fixed total angular momentum $J$ and parity $(-1)^{j+\ell}$. 
  \item For each block $J$, $(-1)^{j+\ell}$, set up the channels $\multiindex{n}=(v,j,\ell)$ and gather the corresponding coupling matrix elements $\bra{\multiindex{n'}}\op{V}\ket{\multiindex{n}}(R)$ (\cref{eq:percival_seaton,eq:final_V_matrix_element}) to evaluate $\matnumlinsc(R)$ (\cref{eq:s_multichannel}).
  \item For each collision energy $\totenergy$, use renormalised Numerov's method to propagate the ratio matrix $\matnumratiosc$ from $R_{\min}$, where $\matnumratiosc=0$, to $R_{\max}$ (\cref{eq:multichan_ren_recursion}).
  \item Insert the final ratio matrix $\matnumratiosc$ into the asymptotic boundary matching formula (\cref{eq:K_from_Q}) to obtain $\mymat{K}$, and convert its open-open block $\mymat{K}_{\mathrm{oo}}$ into $\mymat{S}_{\mathrm{oo}}$ (\cref{eq:S_from_Koo}).
  \item Accumulate the integral (\cref{eq:arthurs_dalgarno}) and differential (\cref{eq:dcs_atom_diatom}) cross sections over $J$ and parity, and check convergence (\cref{subsec:convergence}).
\end{enumerate}
\end{tcolorbox}

\subsection{Jacobi-coordinate system}

\label{subsec:jacobi_coord_system}
\begin{figure}
    \centering
    \resizebox{0.7\textwidth}{!}{
    \begin{tikzpicture}
        \node (Htop) at (0,2cm) {};
        \node [] at (Htop) {$B$};
        \node (Hbottom) at (0,-2cm) {};
        \node [] at (Hbottom) {$C$};
        \node (HCOM) at (0,0) {};
        \node (Ne) at (-5cm,3cm) {$A$};
        \draw (Htop) circle (0.3cm);
        \draw (Hbottom) circle (0.3cm);
        \draw (HCOM) circle (0.1cm) node [below] {COM};
        \draw (Ne) circle (0.6cm);
        \draw [] (Ne) -- (HCOM) node [pos=0.5,below] {$\norm{\myvec{R}}$};
        \draw [] (Htop) -- (Hbottom);
        \draw [decorate,decoration={brace,amplitude=5pt,raise=15.5pt},xshift=-100pt] (Htop) -- (Hbottom) node [xshift=20,pos=0.5,right] {$\norm{\myvec{r}}$};
        \draw [dashed] (0,0) -- (-5cm,0) node [midway,below] {$x$};
        \draw [dashed] (Ne) -- (-5cm,0) node [midway,left] {$z_B+s$};
        \draw (HCOM) ++( 149 : 1 ) arc ( 149:90:1 ) node [midway,below] {$\theta$};

        \node (rHCOM) at (0,\jcsshift) {};
        \node (rHtop) at (0,\jcsshift+2cm) {};
        \node [] at (rHtop) {$B$};
        \node (rHbottom) at (0,\jcsshift-2cm) {};
        \node [] at (rHbottom) {$C$};
        \node (rNe) at (-5cm,\jcsshift+3cm) {$A$};
        \draw (rHtop) circle (0.3cm);
        \draw (rHbottom) circle (0.3cm);
        \draw (rNe) circle (0.6cm);
        \draw [decorate,decoration={brace,amplitude=5pt,raise=15.5pt},xshift=-100pt] (rHbottom) -- (rNe) node [pos=0.5,yshift=-20,xshift=-20] {$R_{AC}$};
        \draw (rNe) -- (rHtop) node [pos=0.5,above] {$R_{AB}$};
        \draw (rNe) -- (rHbottom);%
        \draw (rHtop) -- (rHbottom) node [pos=0.5,right] {$R_{BC}$};
        \draw [dashed] (rNe) -- (0,\jcsshift+3cm) node [midway,above] {$x$};
        \draw [dashed] (0,\jcsshift+3cm) -- (rHtop) node [midway,right] {$s$};
        \draw [gray!30] (rNe) -- (rHCOM.center) node [pos=0.5,below] {$\norm{\myvec{R}}$};
        \draw [gray!30] (rHCOM) ++( 149 : 1 ) arc ( 149:90:1 ) node [midway,below] {$\theta$};
    \end{tikzpicture}
    }
    \caption{Two equivalent coordinate systems for the three-body system formed by $A,B$ and $C$.
    Above, the configuration is defined by
    $\norm{\myvec{r}}$, $\norm{\myvec{R}}$ and $\theta$.
    Below, we express the configuration in terms of the
    relative distances $R_{AB},R_{AC}$ and $R_{BC}$.
    For convenience, additional points defining the quantities
    $x$ and $s$, forming various right triangles have been included to facilitate the conversion between both sets of coordinates.}
    \label{fig:coordinates}
\end{figure}
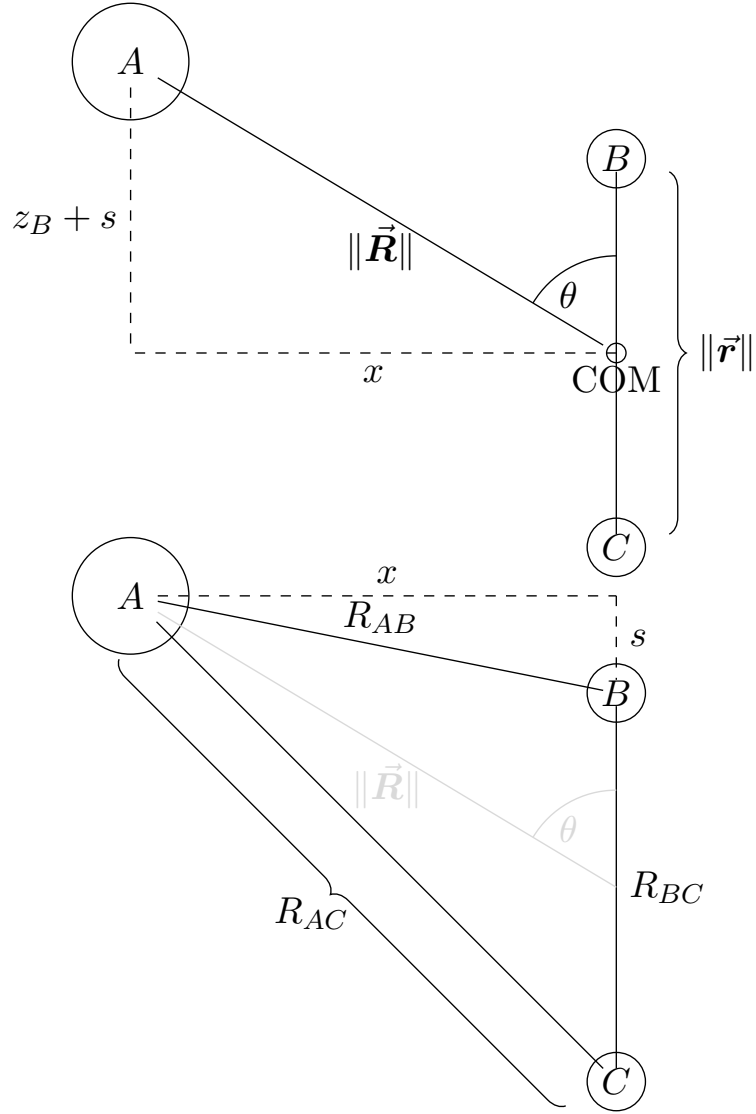

Taking the overall translational and
rotational symmetry into account,
the system's spatial configuration can
fully be described using just three coordinates.
This becomes apparent, by considering the distances between
each pair of bodies ($R_{AB}, R_{AC}$ and $R_{BC}$)
as the three sides of a triangle, and is
therefore uniquely defined by the three distances.
Whilst working with these coordinates might be a more natural
fit in the context of a three-body problem where all the
particles are treated in the same way, we would like to
make use of the fact that we have cast our scattering
problem in the form of a set of coupled one-dimensional
differential equations along the atom-diatom axis.

The Jacobi coordinates of a system comprised of three bodies
is defined by the length of $\myvec{R}$, the vector pointing
from the $BC$ centre of mass to $A$, as well as the length of
$\myvec{r}$, the vector pointing from $C$ to $B$, and finally
$\theta$, the angle between $\myvec{R}$ and $\myvec{r}$.

Converting back and forth between the two sets of
coordinates is relatively straight forward trigonometry.
For this purpose, we first introduce a coordinate system with
the $BC$ centre of mass in the origin
\begin{align}
    \myvec{X}_{BC}=\frac{m_{B}\myvec{X}_{B}+m_{C}\myvec{X}_{C}}{m_{B}+m_{C}}=\myvec{0}\,\,,
\end{align}
where $\myvec{X}_{B}=(0,0,z_{B})^{T}$ and $\myvec{X}_{C}=(0,0,z_{C})^{T}$
are the vectors pointing from the origin to $B$ and $C$,
and we were free to orient our coordinate system such that
both points lie along the $z$-axis.
In that case, since $\norm{\myvec{r}}=R_{BC}=|z_{B}|+|z_{C}|=z_B-z_C$, we have
$m_{B}(R_{BC}+z_{C}) + m_{C}z_{C}=0$ and
\begin{align}
    \myvec{X}_{B} &= \begin{pmatrix}
        0 \\ 0 \\ \frac{m_{C}R_{BC}}{m_{B}+m_{C}}
    \end{pmatrix} \\
    \myvec{X}_{C} &= \begin{pmatrix}
        0 \\ 0 \\ -\frac{m_{B}R_{BC}}{m_{B}+m_{C}}
    \end{pmatrix}
\end{align}
Without loss of generality, we can choose $A$ to lie in the
$xz$-plane, such that in spherical polar coordinates
\begin{align}
    \myvec{X}_{A}= \begin{pmatrix}
        R\sin\theta \\ 0 \\ R\cos\theta
    \end{pmatrix}\,\,.
\end{align}
Consequently, we have
\begin{align}
    \label{eq:R_AB_from_Jacobi}
    R_{AB} &= \norm{\myvec{X}_{A}-\myvec{X_{B}}} = \sqrt{R^{2}(\sin\theta)^2 + \left(R\cos\theta-\frac{m_{C}R_{BC}}{m_B+m_C}\right)^2}\,\,, \\
    R_{AC} &= \norm{\myvec{X}_{A}-\myvec{X_{C}}} = \sqrt{R^{2}(\sin\theta)^2 + \left(R\cos\theta+\frac{m_{B}R_{BC}}{m_B+m_C}\right)^2}\,\,.
    \label{eq:R_AC_from_Jacobi}
\end{align}
Otherwise, to obtain the Jacobi-coordinates from relative
distances (and individual masses), we first
notice that by Pythagoras' rule
\begin{align}
    x^2+(R_{BC}+s)^2 = R_{AC}^2 \\
    x^2+s^2 = R_{AB}^2
\end{align}
which by eliminating the (as of yet) unknown variable $x=R\sin\theta$, can be
solved for $s$
\begin{align}
    s = \frac{R_{AC}^2-R_{AB}^2-R_{BC}^2}{2R_{BC}}\,\,.
\end{align}
Consequently, we can obtain $R$ by Pythagoras' theorem as well
\begin{align}
    R = \sqrt{x^2 + (s+z_B)^2} = \sqrt{R_{AB}^2-s^2 + (s+z_B)^2} =
    \sqrt{R_{AB}^2+2sz_B +z_B^2}\,\,.
\end{align}
Finally, knowing both $R$ and that $z_B+s=R\cos\theta$,
which can be solved for $\theta$
\begin{align}
    \theta = \arccos{\frac{z_B+s}{R}}
\end{align}
gives us the final piece of information needed.
Under substitution of one of the constituents of the
diatomic molecule, the centre of mass along the
diatomic axis is shifted accordingly, leading to
a modification in the $R$ and $\theta$ coordinates as
well.
In \cite{liu1978effect}, a method for transforming the
Legendre expansion of the potential along with a
corresponding quadrature procedure are discussed.

\subsection{Atom-diatom Hamiltonian}

\label{subsec:atom_diatom_hamiltonian}

Having gotten most of the preliminaries out of the way in the prior discussion
of the abstract two particle case, it is a straightforward matter to write
down the centre of mass Hamiltonian for the case of an atom-diatom system
\begin{align}
    \label{eq:atom_diatom_hamiltonian}
    \op{H} &= -\frac{\hbar^2}{2\mu} \Delta_{\myvec{R}} + \op{H}_{BC}(\myvec{r}) + \op{V}(\myvec{R},\myvec{r}) \\
    \label{eq:diatom_hamiltonian}
    \op{H}_{BC} &= -\frac{\hbar^2}{2\mu_{BC}} \Delta_{\myvec{r}} + \op{V}_{BC}(\myvec{r})\,\,.
\end{align}
Here, we have introduced the reduced mass of the diatomic molecule $\mu_{BC}=m_B m_C/(m_B+m_C)$
and, with the combined mass $m_{BC}=m_B+m_C$, $\mu=m_A m_{BC}/(m_A+m_{BC})$ the reduced mass
of the relative motion of the atom-diatom system.
Recalling \cref{eq:spherical_laplace}, the Laplace operator can be expressed in
terms of an angular momentum operator, we can express the rotational parts of each kinetic
energy terms in the total Hamiltonian with individual angular momenta $\opell$
(the rotation of the entire complex around $A-BC$ centre of mass) and $\op{j}$
(the rotation of the $BC$ particle around its own centre of mass)
\begin{align}
    \op{H} = -\frac{\hbar^2}{2\mu} R^{-1}\partial_{R}^{2} R +\frac{\opell^2}{2\mu R^2} + \op{V}(\myvec{R},\myvec{r})
    -\frac{\hbar^2}{2\mu_{BC}}r^{-1}\partial_{r}^{2} r +\frac{\op{j}^2}{2\mu_{BC}r^2} + \op{V}_{BC}(\myvec{r})\,\,.
\end{align}

\subsubsection*{Diatomic molecule eigenstates}

As a molecule, we consider the (bound) states of $BC$, given by the Schr{\"o}dinger equation
\begin{align}
    \op{H}_{BC} \Phi_{BC}(\myvec{r}) = \intenergy_{vj} \Phi_{BC}(\myvec{r})
\end{align}
where each eigenstate is characterised by the quantum numbers $v$ (corresponding
to the bond length degree of freedom) and $j$ (corresponding to the diatomic
rotational degree of freedom) and $\intenergy_{vj}$ are the corresponding eigenvalues.
The $\Phi_{BC}(\myvec{r})$ can be written as a product of one dimensional (radial)
eigenfunctions and spherical Harmonics in $j$, $m$
\begin{align}
    \Phi_{BC}(\myvec{r}) = r^{-1} \chi_{vj}(r) Y_{jm}(\myvec{r}/\norm{\myvec{r}})\,\,.
\end{align}
Here, as we did above \cref{eq:cc}, we can again exploit
the fact that introducing the prefactor $r^{-1}$
to the radial part of the wavefunction leads to a simplification
of the application of the differential operator.
Whilst the spherical harmonics and their various properties
are well-known, the bound radial eigenfunctions
$\chi_{vj}(r)$ can be obtained for most simple diatomic molecules
by solving the eigensystem (either by diagonalisation or by more sophisticated methods
such as using a discrete variable representation
(DVR) discussed in \cref{subsec:dvr}).

\subsection{Parity and its conservation}

To fully describe a molecular state of a closed quantum system
one requires a complete set of commuting variables, generated,
for instance, by finding all operators which commute with the
Hamiltonian $\op{H}$.
Amongst these
is the parity operator $\op{P}$, which
inverts a space-fixed coordinate system
leading to the transformation $\myvec{r}\rightarrow -\myvec{r}$.
In spherical polar coordinates this corresponds
to the transformation $\theta \rightarrow \pi -\theta$
and $\varphi \rightarrow \varphi + \pi$.
Seeing how the parity operator affects the entire
molecular state can be a somewhat involved process,
which is best accomplished by changing from a space-fixed
to a molecule-fixed (or body-fixed) coordinate system for
the electronic part of the wavefunction
\cite{krems2018molecules}.

Eigenstates $\ket{\phi}$ of the parity operator
fulfil the eigenvalue equation
\begin{align}
  \op{P}\ket{\phi} = p \ket{\phi}
\end{align}
for some $p~\in\mathbb{C}$,
with the additional property that a second
inversion reverts the coordinate system back
to its original state, i.e.
\begin{align}
  \underbrace{\op{P} \left(\op{P}\ket{\phi}\right)}_{ = p^2 \ket{\phi}} =
  \underbrace{\mathbbm{1} \ket{\phi}}_{=\ket{\phi}}  \,\,.
\end{align}
Consequently, the above equation admits the eigenvalues $p=\pm 1$.

Now let's see how parity affects angular momenta.
The explicit form of the spherical harmonics
is given by
\begin{align}
  Y_{\ell m}(\theta,\varphi) = \frac{(-1)^m}{2^{\ell}\ell !} \sqrt{\frac{(2\ell + 1)}{4\pi}\frac{(\ell-m)!}{(\ell+m)!}}e^{im\varphi}(1-z^2)^{m/2} \left(\frac{d}{dz}\right)^{\ell+m} (z^2-1)^{\ell}\,\,,
\end{align}
where, as usual $z=\cos\theta$, leading to the substitution
$z \rightarrow -z$ under parity.
Therefore, applying the parity transformation
\begin{align}
  Y_{\ell m}(\theta,\varphi) \rightarrow Y_{\ell m}(\pi-\theta,\pi+\varphi)
  = (-1)^{m} (-1)^{\ell+m} Y_{\ell m}(\theta,\varphi)= (-1)^{\ell} Y_{\ell m}(\theta,\varphi)\,\,,
\end{align}
introduces a prefactor of $(-1)^{\ell}$.

The vibrational part of the wavefunction is
invariant under parity, since it only depends
on the scalar distances between particles
which are conserved under parity.
The treatment of the electronic part of the
wavefunction under parity, however,
is more involved - since we are mainly interested
in the nuclear motion we shall therefore refrain
from discussing it in detail here,
the interested reader is referred
to \cite{krems2018molecules} for a detailed
explanation.

For two particles rotating around their common
centre of mass with the orbital quantum number
$\ell$, the total parity is comprised
of the product of each particle's parity
multiplied by a parity accounting for the orbital rotation, $(-1)^{\ell}$
\cite{krems2018molecules,landau1991quantum}
\begin{align}
  \op{P}_{\mathrm{tot}} = (-1)^{\ell}\op{P}_{1} \op{P}_{2}\,\,.
\end{align}
Above, $\op{P}_{1}$ and $\op{P}_{2}$ are the
parities of the first and second particles, respectively.
In the specific case of a diatomic molecule with
a rotational quantum number $j$ colliding with an
atom whose internal structure can be neglected, the
total parity may be expressed as
\begin{align}
  \op{P}_{\mathrm{tot}} = (-1)^{j+\ell}\,\,.
\end{align}
Finally, whether or not channels of opposite
parity are allowed to couple, depends on
the operator for which we are
attempting to evaluate matrix elements, leading to the so-called
parity selection rules.
True scalars always connect even-parity states
with other even parity states and odd parity
states with other odd ones, whilst `pseudoscalars'
(e.g. the scalar products between polar and axial
vectors) connect states of opposing parity.
In fact, it can be shown that polar and axial vectors
connect states of opposing and equal parity
respectively \cite{landau1991quantum}.
Consequently, the atom-diatom interaction potential, which only depends on $R$, $r$ and $\theta$ and is therefore invariant under parity,
only couples channels of equal parity,
such that $(-1)^{j+\ell}=(-1)^{j'+\ell'}$.
Below, we will derive a closed expression for the angular part of the 
potential matrix elements in terms of 3j symbols, where selection 
rules enforce this parity condition.
The advantage we gain from this separation is that blocks of fixed total 
angular momentum $J$ are further split into blocks of opposite parity 
which can be solved separately.
For homonuclear diatomic molecules, we will see that symmetry leads
to a decoupling of even and odd $j$, effectively halving the necessary size of our channel basis in each case.

\subsection{Discrete variable representation}
\label{subsec:dvr}

The following brief introduction to the
topic of the discrete variable representation is
mainly based on two sources
\cite{groenenboom1993combining,groenenboom2001discrete},
with a particular focus on the so-called
`sinc-DVR' which is the main type of discrete
variable representation utilised in this work.
As such, we shall sacrifice generality for the sake of
describing the succinct solution to the
problem of obtaining the diatomic eigenstates
required in \cref{subsec:atom_diatom_hamiltonian}.
We shall be content in noting that
other approaches and interpretations of
discrete variable representations
\cite{lill1982discrete,colbert1992novel} exist
and that there are comprehensive overviews
on the former and their relation to a number
of closely connected topics \cite{meyer2017numerical}.

The conventional application for a discrete variable
representation is to obtain numerical solutions to
time independent Schr{\"o}dinger equations of the form
\begin{align}
    \op{H}\Psi = (\op{H}_{0} + \op{V})\Psi= \totenergy\Psi\,\,,
\end{align}
where $\op{H}$ might for example be given by \cref{eq:diatom_hamiltonian},
as is the case for a diatomic molecule.
In spherical coordinates, the Laplace operator
can be written as a sum of a chain of derivatives
in $r$ and an angular term expressed with the angular
momentum operator $\op{j}^2/r^2$.
For the radial functions $\chi_{vj}(r)$ of the diatomic molecule, this angular term reduces to the centrifugal term $\hbar^2 j(j+1)/(2\mu_{BC}r^2)$, which is simply added to $V_{BC}(r)$, so that a separate radial eigenproblem is solved for each $j$.
Ultimately, we would like to break down the above
equation to a linear algebra problem, representing the
eigenstate $\Psi=\sum_{i=0}^{n}c_{i}\psi_{i}$ as
a vector of expansion coefficients for some given
basis set and the total Hamiltonian, $\op{H}$ by
a matrix in the `best possible' way.
The critical insight is to find a `pseudo spectral'
representation, consisting of a global basis set
and finite set of grid points,
for which the potential matrix,
whose elements are given by
$\bra{i}\op{V}\ket{j}$,
which can be evaluated as a quadrature integral \cite{abramowitz1988handbook},
becomes diagonal.
At the same time, we would like to be able to
evaluate quadrature integrals of $k^{\mathrm{th}}$-derivative
terms $\bra{i}\partial_{x}^{k}\ket{j}$, such as they appear
in the kinetic energy operator,
in closed form.

We begin by considering an orthonormal
basis set
\begin{align}
    \mathcal{F} = \left\{ \phi_{i}(x)=\braket{x|i},~i=1,...,n \right\}
\end{align}
and a quadrature, consisting of pairs
of points and weights
\begin{align}
    \mathcal{Q} = \left\{ (x_k,w_k),~k=1,...,n \right\}\,\,.
\end{align}
We require that our basis functions
fulfill the property
\begin{align}
    \phi_{i}(x_k) = w_k^{-1/2} \delta_{ik}\,\,,
    \label{eq:dvr_property}
\end{align}
which means that, sampled at each quadrature
point, the $i^{\mathrm{th}}$ basis
function always vanishes except for at its matching point $x_i$.
In this case, performing a quadrature
integration to obtain the potential
matrix elements
\begin{align}
    \bra{i}\op{V}\ket{j}
    =& \int \infin{x} \phi_i(x) V(x) \phi_j(x) \nonumber \\
    \approx& \sum_{k=1}^{n} w_k \phi_i(x_k) V(x_k) \phi_j(x_k) \nonumber \\
    =& \sum_{k=1}^{n} \delta_{ik} V(x_k) \delta_{jk} = V(x_i) \delta_{ij}
\end{align}
yields a diagonal matrix $\mymat{V}$.
Above, we have exploited the
required property in \cref{eq:dvr_property}.

\subsubsection*{Sinc-function discrete variable representation}

There are several examples of
different functions and
weights that are appropriate
for constructing a discrete variable representation
as introduced above
but we will discuss the so-called sinc-DVR,
since this is the one utilised throughout
this tutorial.
We shall begin by requiring that our
sought quadrature should be
uniformly spaced and with uniform weights.
In that case, the set of functions
\begin{align}
  \mathcal{F}_{\sinc} = \left\{ \phi_{i}(x) = \Delta^{-1/2} \sinc{\left(\pi\left(\frac{x}{\Delta}-i\right)\right)},~i\in\mathbb{Z}\right\}\,\,,%
    \label{eq:sinc_dvr_basis}
\end{align}
with $\sinc(x)=\sin(x)/x$, and $\Delta$ the distance between
grid points along $x$, can be shown to
be a good candidate for a discrete variable
representation \cite{colbert1992novel,groenenboom2001discrete}.
Recalling that, by L'Hopital's rule,
$\lim_{x\rightarrow 0} \frac{\sin(x)}{x} = \lim_{x\rightarrow 0} \frac{\cos(x)}{1} = 1$.
Furthermore, the numerator of $\phi_i(x)$
vanishes whenever $x/\Delta -i$ is a nonzero
integer.
Consequently, \cref{eq:dvr_property} is fulfilled
with the weights $w_{k}=\Delta$, therefore $\Delta$
appears both in the points and weights of the
quadrature
\begin{align}
  \mathcal{Q}_{\sinc} = \left\{(k\Delta,\Delta),~k\in\mathbb{Z}\right\}\,\,.
    \label{eq:sinc_dvr_quadrature}
\end{align}
We will begin by showing that the band-limited
set of functions
\begin{align}
    \tilde{\phi}_n(p) =
    \begin{cases}
        (2p_{\max})^{-1/2} e^{in \pi p/p_{\max}}, &p \leq p_{\max} \\
        0, &\mathrm{else}
    \end{cases}
\end{align}
is the  Fourier-transformed set of functions corresponding
to the $\sinc$-basis defined in real space above and
which form a basis in momentum space.
Evaluating the Fourier transform of these
momentum functions explicitly,
we see that
\begin{align}
    \phi_{n}(x)
    =& (2\pi)^{-1/2} \int_{\mathbb{R}} \infin{p}~e^{-ipx} \tilde{\phi}_n(p) \nonumber \\
    =& (2\pi)^{-1/2} \int_{-p_{\max}}^{p_{\max}} \infin{p}~e^{-ipx} (2p_{\max})^{-1/2} e^{in\pi p/p_{\max}} \nonumber \\
    =& (4\pi p_{\max})^{-1/2} \left[\frac{e^{-ip(x-\frac{n\pi}{p_{\max}})}}{-i(x-\frac{n\pi}{p_{\max}})}\right]_{-p_{\max}}^{p_{\max}} \nonumber \\
    =& (4\pi p_{\max})^{-1/2} \frac{1}{-i(x-\frac{n\pi}{p_{\max}})} \underbrace{\left(e^{-ip_{\max}(x-\frac{n\pi}{p_{\max}})}-e^{ip_{\max}(x-\frac{n\pi}{p_{\max}})}\right)}_{=-i2 \sin{\left(p_{\max}(x-\frac{n\pi}{p_{\max}})\right)}} \nonumber \\
    =& \frac{\sqrt{p_{\max}}}{\sqrt{\pi}} \sinc{\left(p_{\max}(x-\frac{n\pi}{p_{\max}})\right)}\,\,.
\end{align}
Finally, identifying $p_{\max}=\pi/\Delta$ yields the set of functions in \cref{eq:sinc_dvr_basis}.
The band limitation is important, because it
tells us that our basis spans all functions
whose highest momentum component is less than
$p_{\max}$.
For overlap matrices, including those involving
functions whose highest frequency components are
less than $2p_{\max}$, the quadrature in
\cref{eq:sinc_dvr_quadrature} is exact
\begin{align}
    \braket{n|m} =& \int \infin{x} \phi_{n}(x)\phi_{m}(x) \nonumber \\
    =& \sum_{i=-\infty}^{\infty} \Delta \frac{\sinc{(\pi(i-n))}}{\sqrt{\Delta}} \frac{\sinc{(\pi(i-m))}}{\sqrt{\Delta}} \nonumber \\
    =& \sum_{i=-\infty}^{\infty} \delta_{i,n}\delta_{i,m} = \delta_{n,m}
\end{align}
In particular, we are interested in
powers of $x$-derivatives, for
which we can exploit the fact that
derivatives of band limited functions
are also band limited
\begin{align}
    \partial_x^k %
    \phi_{n}(x)
    =& (2\pi)^{-1/2}\int_{-\infty}^{\infty} \infin{p}~\underbrace{\tilde{\phi}_n(p)}_{=0~\mathrm{for}~|p| > p_{\max}}(-ip)^{k}e^{-ipx} \\
    =& (2\pi)^{-1/2}\int_{-p_{\max}}^{p_{\max}} \infin{p}~\tilde{\phi}_n(p)(-ip)^{k}e^{-ipx}\,\,.
\end{align}
By using the quadrature we can obtain
a simplified expression for the matrix
elements of all derivatives
\begin{align}
    \bra{n} \partial_x^k%
    \ket{m}
    &= \int_{-\infty}^{\infty} \infin{x} \frac{1}{\Delta} \sinc{(\pi(\frac{x}{\Delta}-n))}
    \partial_x^{k}
    \sinc{(\pi(\frac{x}{\Delta}-m))} \\
    &= \sum_{i=-\infty}^{\infty} \sinc{(\pi(i-n))}
    \partial_x^{k}
    \sinc{(\pi(\frac{x}{\Delta}-m))}\Big|_{x=i\Delta} \\
    &=
    \partial_x^{k}
    \sinc{(\pi\frac{x}{\Delta})}\Big|_{x=(n-m)\Delta}\,\,.
\end{align}
Evaluating the explicit forms of the first two
derivatives gives
\begin{align}
    \bra{n}
    \partial_x
    \ket{m} = \frac{\cos{(\frac{\pi}{\Delta}x)}}{x}
    -\frac{\sin{(\frac{\pi}{\Delta}x)}}{\frac{\pi}{\Delta}x^2} \Bigg|_{x=(n-m)\Delta}
    = \begin{cases}
        0, &n=m \\
        \frac{1}{\Delta} \frac{(-1)^{n-m}}{n-m}, &n \neq m
    \end{cases}
\end{align}
and
\begin{align}
    \bra{n}
    \partial_x^{2}
    \ket{m} =&
    -\frac{\frac{\pi}{\Delta}\sin(\frac{\pi}{\Delta}x)}{x}
    -\frac{\cos{(\frac{\pi}{\Delta}x)}}{x^2}
    -\frac{\cos{(\frac{\pi}{\Delta}x)}}{x^2}
    +\frac{2\sin{(\frac{\pi}{\Delta}x)}}{\frac{\pi}{\Delta}x^3}\Bigg|_{x=(n-m)\Delta} \nonumber \\
    =& \begin{cases}
        -\frac{1}{3}\frac{\pi^2}{\Delta^2}, &n=m \\
        -\frac{2}{\Delta^2}\frac{(-1)^{n-m}}{(n-m)^{2}}, &n \neq m\,\,.
    \end{cases}
    \label{eq:dvr_second_derivative}
\end{align}
Above, we can use
Taylor expansion of $\sin$ and $\cos$ to
evaluate the various
limits when $n=m$.

\subsection{Potential energy surface expansion}

In the absence of external fields, the total angular momentum of the
entire complex must be conserved.
In the atom-diatom case, we obtain the total angular momentum $\myvec{J}=\myvec{j}+\myvec{\ell}$ by coupling $\op{j}$ and $\opell$.
In terms of spherical harmonics
\begin{align}
    \mathcal{Y}_{j\ell}^{JM}\left(\frac{\myvec{R}}{\norm{\myvec{R}}},\frac{\myvec{r}}{\norm{\myvec{r}}}\right) = \sum_{m_{j}=-j}^{j} \sum_{m_{\ell}=-\ell}^{\ell} C_{jm_{j}\ell m_{\ell}}^{JM}
    Y_{jm_{j}}\left(\frac{\myvec{r}}{\norm{\myvec{r}}}\right)
    Y_{\ell m_{\ell}}\left(\frac{\myvec{R}}{\norm{\myvec{R}}}\right)
\end{align}
where $C_{jm_j \ell m_{\ell}}^{JM} = \braket{jm_j \ell m_{\ell}|JM}$ is the Clebsch-Gordan coefficient
for transforming between the uncoupled and coupled
angular momentum states.
Recalling \cref{eq:non_asymptotic_expansion}, we are already familiar with the concept of
expanding the eigenstates of the full (i.e. non-asymptotic)
Hamiltonian in coefficients dependent on $R$.
The $\myvec{R}$ and $\myvec{r}$-dependent eigenstates
can therefore be written in two ways, the uncoupled basis,
where
\begin{align}
    \Psi_{\multiindex{n}}(\myvec{R},\myvec{r})
    =& \sum_{\multiindex{n'}} R^{-1} u_{\multiindex{n'}\multiindex{n}}^{\totenergy}(R) \Phi_{\multiindex{n'}}\left(\frac{\myvec{R}}{\norm{\myvec{R}}},\myvec{r}\right)
\end{align}
with
\begin{align}
  \Phi_{\multiindex{n}}\left(\frac{\myvec{R}}{\norm{\myvec{R}}},\myvec{r}\right) =&\,r^{-1} \chi_{vj}(r) Y_{jm_j}\left(\frac{\myvec{r}}{\norm{\myvec{r}}}\right)
    Y_{\ell m_{\ell}}\left(\frac{\myvec{R}}{\norm{\myvec{R}}}\right)
\end{align}
or, using the above definition to express the wavefunction
for a fixed total angular momentum and projection $J,M$
\begin{align}
    \Psi_{\multiindex{n}}^{JM}(\myvec{R},\myvec{r}) =& \sum_{\multiindex{n'}} R^{-1} u_{\multiindex{n'}\multiindex{n}}^{\totenergy}(R) \Phi_{\multiindex{n'}}^{JM}\left(\frac{\myvec{R}}{\norm{\myvec{R}}},\myvec{r}\right)
\end{align}
with
\begin{align}
    \Phi_{\multiindex{n}}^{JM}\left(\frac{\myvec{R}}{\norm{\myvec{R}}},\myvec{r}\right) =&\,r^{-1} \chi_{vj}(r)
    \mathcal{Y}_{j\ell}^{JM}\left(\frac{\myvec{R}}{\norm{\myvec{R}}},\frac{\myvec{r}}{\norm{\myvec{r}}}\right)\,\,.
\end{align}
Depending on which basis we are using, the (multi)-index is comprised of the quantum numbers
$\multiindex{n}=(v,j,\ell,m_j,m_{\ell})$ in the uncoupled case, whilst
in the coupled case the multiindex is restricted to $\multiindex{n}=(v,j,\ell)$.
With the above representation, we can now turn our attention to
evaluating $R$-dependent matrix elements of the potential
energy surface in the channel basis
\begin{align}
  \bra{\multiindex{n'}}\op{V}\ket{\multiindex{n}}(R) =& \int \infin{\myvec{r}} \int \infin{\Omega_{\myvec{R}}}
    {\Phi_{\multiindex{n'}}^{JM}}^{*}\left(\frac{\myvec{R}}{\norm{\myvec{R}}},\myvec{r}\right)
    V(\myvec{R},\myvec{r})
    \Phi_{\multiindex{n}}^{JM}\left(\frac{\myvec{R}}{\norm{\myvec{R}}},\myvec{r}\right) \nonumber \\
    =& \int \infin{r} \int \infin{\Omega_{\myvec{r}}} \int \infin{\Omega_{\myvec{R}}} \cancel{\frac{r^{2}}{r^2}}
    {\chi_{v'j'}(r)}^{*}{\mathcal{Y}_{j'\ell'}^{JM}}^{*}\left(\frac{\myvec{R}}{\norm{\myvec{R}}},\frac{\myvec{r}}{\norm{\myvec{r}}}\right) \nonumber \\
    &\times V(\myvec{R},\myvec{r})
    {\chi_{vj}(r)}{\mathcal{Y}_{j\ell}^{JM}}\left(\frac{\myvec{R}}{\norm{\myvec{R}}},\frac{\myvec{r}}{\norm{\myvec{r}}}\right)\,\,.
    \label{eq:potential_matrix_element_full}
\end{align}
From here, we proceed by expanding the
(in general) $\myvec{R}$ and $\myvec{r}$-dependent
potential energy surface $V(\myvec{R},\myvec{r})=V(R,r,\theta)$  in Legendre polynomials
in the $\theta$ (angle between $\myvec{R}$ and $\myvec{r}$)
degree of freedom
\begin{align}
    V(\myvec{R},\myvec{r}) = \sum_{\lambda=0}^{\infty} V_{\lambda}(R,r)P_{\lambda}(\cos\theta)\,\,.
    \label{eq:legendre_theta_potential_expansion}
\end{align}
Note that $V(R,r,\theta)$ is symmetric under $\theta\rightarrow\pi-\theta$ only for homonuclear diatomic molecules; otherwise, attention should be paid to which orientation of $\myvec{r}$ the angle $\theta$ is defined with respect to.
By introducing a sum over the order of the Legendre
expansion, we can factorise the integral
into separate parts for angular and radial (in $r$) integration
\begin{align}
  \bra{\multiindex{n'}}\op{V}\ket{\multiindex{n}}(R) =&\sum_{\lambda=0}^{\infty}\int \infin{r}
    {\chi_{v'j'}(r)}^{*}V_{\lambda}(R,r){\chi_{vj}(r)} \nonumber \\
    &\times
    \underbrace{\int \infin{\Omega_{\myvec{r}}}  \int  \infin{\Omega_{\myvec{R}}}{\mathcal{Y}_{j'\ell'}^{JM}}^{*}\left(\frac{\myvec{R}}{\norm{\myvec{R}}},\frac{\myvec{r}}{\norm{\myvec{r}}}\right)
    P_{\lambda}(\cos\theta)
    {\mathcal{Y}_{j\ell}^{JM}}\left(\frac{\myvec{R}}{\norm{\myvec{R}}},\frac{\myvec{r}}{\norm{\myvec{r}}}\right)}_{=\bra{j'\ell'J}P_{\lambda}(\cos\theta)\ket{j\ell J}}\,\,.
    \label{eq:potential_matrix_element_expanded}
\end{align}
Bringing the full machinery of angular momentum algebra
to bear on the second factor (conveniently shortened using bra-ket notation) we arrive at a closed
expression dependent on Wigner 3j and 6j symbols
\begin{align}
    \bra{j'\ell'J}P_{\lambda}(\cos\theta)\ket{j\ell J} =&
    (-1)^{j+j'-J}
     \sqrt{(2j+1)(2j'+1)(2\ell+1)(2\ell'+1)} \nonumber \\
    & \times
    \begin{Bmatrix}
        J & \ell' & j' \\ \lambda & j & \ell
    \end{Bmatrix}
    \begin{pmatrix}
        j & \lambda & j' \\ 0 & 0 & 0
    \end{pmatrix}
    \begin{pmatrix}
        \ell & \lambda & \ell' \\ 0 & 0 & 0
    \end{pmatrix}\,\,,
    \label{eq:percival_seaton}
\end{align}
which is known in the literature as the Percival-Seaton
coefficient \cite{percival1957partial,wormer2006angular}.
By rescaling the $r$ axis (which in theory includes all of
$\mathbb{R}_{+}$ but in practice, i.e. for numerical
purposes, is a numerical grid going
from some $r_{\min}$ to some $r_{\max}$) to the range
$\tilde{r}\in \left[-1,1\right]$, we can play
the Legendre expansion trick a second time
\begin{align}
    V(\myvec{R},\myvec{r}) = \sum_{\lambda=0}^{\infty} \sum_{\eta=0}^{\infty} V_{\lambda\eta}(R)P_{\lambda}(\cos\theta)P_{\eta}(\tilde{r})\,\,.
\end{align}
$\tilde{r}$ can be expressed as a function of $r \in [r_{\min},r_{\max}]$ as follows
\begin{align}
    \tilde{r}(r) = 2\left(\frac{r-r_{\min}}{r_{\max}-r_{\min}} -\frac{1}{2}\right)\,\,,
\end{align}
which can easily be inverted
\begin{align}
    r(\tilde{r}) = r_{\min} + (r_{\max}-r_{\min})\left(\frac{\tilde{r}}{2}+\frac{1}{2}\right)\,\,.
\end{align}
The actual $R$-dependent coefficients of our potential
energy surface are
obtained by using the orthogonality property of
Legendre polynomials
\begin{align}
    \int_{-1}^{1} \infin{x} P_{m}(x) P_{n}(x) = \frac{2}{2n+1} \delta_{nm}
\end{align}
i.e.
\begin{align}
    V_{\lambda\eta} =& \frac{2\eta+1}{2} \int_{-1}^{1} \infin{\tilde{r}} P_{\eta}(\tilde{r}) V_{\lambda}(R,r(\tilde{r})) \nonumber \\
    =& \frac{2\eta+1}{2} \int_{-1}^{1} \infin{\tilde{r}} P_{\eta}(\tilde{r}) \left(\frac{2\lambda+1}{2} \int_{-1}^{1} \infin{\cos\theta} V(R,r(\tilde{r}),\theta) P_{\lambda}(\cos\theta)\right)\,\,.
\end{align}
Since $V(R,r,\theta)$ is only known numerically, we cannot
solve these integrals analytically.
For fast and accurate numerical evaluation,
it is convenient to approximate the
integrals using Gauss-Legendre quadratures
\begin{align}
    \int_{-1}^{1} \infin{x} f(x) \approx \sum_{i=1}^{n}w_{i}f(x_{i})
\end{align}
where $w_{i}=\frac{2}{(1-x_{i}^2)(P_{n}'(x_{i}))^2}$ are quadrature weights and $x_{i}$ are roots
of the $n^{\mathrm{th}}$ Legendre polynomial.
In terms of quadratures, the evaluation
of the expansion coefficients becomes
\begin{align}
    V_{\lambda\eta}(R) = \frac{(2\lambda+1)(2\eta+1)}{4} \sum_{p=1}^{n} w_{p}P_{\lambda}(x_{p})
    \sum_{q=1}^{n} w_q P_{\eta}(x_q) V(R,r(x_q),\theta(x_p))\,\,,
\end{align}
where the angle is evaluated at $\theta(x_p)=\arccos(x_p)$.
Turning back to \cref{eq:potential_matrix_element_full}
we now have
\begin{align}
  \bra{\multiindex{n'}}\op{V}\ket{\multiindex{n}}(R)
  =& \frac{r_{\max}-r_{\min}}{2} \sum_{\lambda\eta} V_{\lambda\eta}(R) \int \infin{\tilde{r}} \chi_{v'j'}^{*}(r(\tilde{r}))P_{\eta}(\tilde{r})\chi_{vj}(r(\tilde{r})) \nonumber \\
    &\times \bra{j'\ell'J}P_{\lambda}(\cos\theta)\ket{j\ell J}\,\,.
    \label{eq:final_V_matrix_element}
\end{align}
With this expression for the potential energy surface
matrix elements in place, we are now fully equipped with the necessary
background information to cast the collision problem as
a numerical one.

\subsection{Multi-channel Numerov's method}

To match the coupled channels differential
equations formulated above, we go from
a single, one dimensional homogeneous
second order ordinary differential
equation, as treated in \cref{subsec:numerov_sc}, to a system
of differential equations, such as
\begin{align}
    \partial_x^2 y_{1}(x) =& -\sum_{k=1}^{N} \numlinsc_{1k}(x) y_{k}(x) \nonumber \\
    &\vdots \nonumber \\
    \partial_x^2 y_{j}(x) =& -\sum_{k=1}^{N} \numlinsc_{jk}(x) y_{k}(x) \nonumber \\
    &\vdots \nonumber \\
    \partial_x^2 y_{N}(x) =& -
    \sum_{k=1}^{N} \numlinsc_{Nk}(x) y_{k}(x)\,\,.
    \label{eq:multichannel_solution}
\end{align}
The set of `one dimensional' solutions $\left\{y_{j}(x)\right\}$ to
the set of differential equations above
combine to form a single total solution to the coupled channel
differential equations.
We should recall, however,  that a single second
order differential equation has two
linearly independent solutions, and that
a system of $N$ coupled differential equations has a total
of $2N$ linearly independent solutions, with each solution
given as a set consisting of $N$ elements, corresponding to
each entry in \cref{eq:multichannel_solution}.
By imposing $y_{i}(x_{\min})=0$ in the strongly repulsive region,
the regular solutions are selected and propagated.
We therefore label
the $j^{\mathrm{th}}$ combination of these regular solutions
$y_{i}^{(j)}(x)$, or arrange them as
the coefficients of an $x$-dependent
matrix $\matnumvar(x)$ i.e. $y_{i}^{(j)}(x)=y_{ij}(x)$ and similarly for $\matnumlinsc$ comprised of $\numlinsc_{ij}$.
\begin{align}
    \partial_{x}^{2} \matnumvar(x) = -\matnumlinsc(x) \matnumvar(x)
\end{align}
Here, we denote the matrix form of the single-channel coupling $\numlinsc(x)$ from
\cref{subsec:numerov_sc} as $\matnumlinsc(x)$, which should not be confused with the
scattering $\mymat{S}$-matrix introduced in \cref{subsec:K_and_S_matrices}.
As in the single channel case, we identify the coupled channels formulation of our problem
with the relations $x=R$, $\matnumvar=\mymat{u}$
(where the columns of $\mymat{u}$ correspond to different linearly independent solutions).
By multiplying \cref{eq:cc} by $-2\mu/\hbar^2$ we see that
\begin{align}
    \numlinsc_{\multiindex{n''}\multiindex{n'}}(R) = \left(k_{\multiindex{n''}}^2 - \frac{\ell''(\ell''+1)}{R^2}\right)\delta_{\multiindex{n''}\multiindex{n'}} - \frac{2\mu}{\hbar^2}\bra{\multiindex{n''}}\op{V}\ket{\multiindex{n'}}\,\,,
    \label{eq:s_multichannel}
\end{align}
where $k_{\multiindex{n''}}^2 = 2\mu(\totenergy-\intenergy_{\multiindex{n''}})/\hbar^2$.
On a coefficient-wise basis, we can perform the same Taylor expansion as above
\begin{align}
    y_{ij}(x) = \sum_{k=0}^{5} \frac{\partial_{x}^{k}y_{ij}(x)|_{x=x_{0}}}{k!}(x-x_{0})^{k} + \mathcal{O}\left((x-x_{0})^{6}\right)
\end{align}
which we will write in matrix form as
\begin{align}
    \matnumvar(x) = \sum_{k=0}^{5} \frac{\partial_{x}^{k}\matnumvar(x)|_{x=x_{0}}}{k!}(x-x_{0})^{k} + \mathcal{O}\left((x-x_{0})^{6}\right)\,\,.
\end{align}
Following the same steps as above, the single channel formulation can be generalised to the
multi-channel case
\begin{align}
    \matnumvar(x_{i+1}) + \matnumvar(x_{i-1}) -2\matnumvar(x_{i}) =&  \partial_{x}^{2}\matnumvar(x)|_{x=x_{i}}h^2 + \partial_{x}^4\matnumvar(x)|_{x=x_{i}}\frac{h^4}{12} + \mathcal{O}(h^6) \\
    \frac{h^2}{12} \left(\matnumdiffeq(x_{i+1}) + \matnumdiffeq(x_{i-1}) - 2\matnumdiffeq(x_{i})\right) =& \frac{h^2}{12} \left(\partial_{x}^{2}\matnumdiffeq(x)|_{x=x_{i}}h^2 + \partial_{x}^{4}\matnumdiffeq(x)|_{x=x_{i}}\frac{h^4}{12} + \mathcal{O}(h^6)\right)
\end{align}
where, similarly to how we defined $w(x)$ above,
$\matnumdiffeq(x) =\partial_{x}^{2}\matnumvar(x)=-\matnumlinsc(x)\matnumvar(x)$ (restricted to the homogeneous case).
\begin{align}
    \matnumvar(x_{i+1}) + \matnumvar(x_{i-1}) -2\matnumvar(x_i) =& -\matnumlinsc(x_{i})\matnumvar(x_{i}) h^{2}
    + \frac{h^2}{12} \big(-\matnumlinsc(x_{i+1})\matnumvar(x_{i+1}) \nonumber \\
    &-\matnumlinsc(x_{i-1})\matnumvar(x_{i-1})
     + 2\matnumlinsc(x_{i})\matnumvar(x_{i})\big) + \mathcal{O}(h^6)\,\,,
\end{align}
which can again be solved for $\matnumvar(x_{i+1})$,
provided that $\mathbbm{1}+\frac{h^2}{12}\matnumlinsc(x_{i+1})$ is invertible
\begin{align}
    \matnumvar(x_{i+1}) \approx&  \left(\mathbbm{1}+\frac{h^2}{12}\matnumlinsc(x_{i+1})\right)^{-1} \nonumber \\
     & \times \left(2 \left(\mathbbm{1}- \frac{5h^2}{12}\matnumlinsc(x_{i})\right)\matnumvar(x_{i}) -  \left(\mathbbm{1}+\frac{h^2}{12}\matnumlinsc(x_{i-1})\right)\matnumvar(x_{i-1}) \right)
     \label{eq:multichan_numerov_homog}
\end{align}
Similarly, generalisation to the renormalised
multi-channel Numerov's method is also straight
forward, since the solution ratios $\matnumratiosc(x)$ can accordingly
be replaced with their matrix counterparts
\begin{align}
    \label{eq:Q_matrix_ratio}
    \matnumvar(x_{i-1}) =& \matnumratiosc(x_{i})\matnumvar(x_{i}) \\
    \matnumvar(x_{i+1}) =& \matnumratiosc(x_{i+1})^{-1}\matnumvar(x_{i})
\end{align}
leading to the single-term recursion relation
\begin{align}
    \matnumratiosc(x_{i+1})^{-1} \approx &  \left(\mathbbm{1}+\frac{h^2}{12}\matnumlinsc(x_{i+1})\right)^{-1} \nonumber \\
    & \times \Big(2(\mathbbm{1}- \frac{5h^2}{12}\matnumlinsc(x_{i})) -  (\mathbbm{1}+\frac{h^2}{12}\matnumlinsc(x_{i-1})) \matnumratiosc(x_{i}) \Big)
    \label{eq:multichan_ren_recursion}
\end{align}
under the assumption that $\matnumvar(x_{i})$
is invertible.

\subsection{Obtaining $\mymat{K}$ from $\matnumratiosc$ after Numerov propagation}

\label{subsec:Q_to_K}

Recalling the expression for the asymptotic behaviour
of the wavefunction discussed in \cref{subsec:K_and_S_matrices},
specifically \cref{eq:asymptotic_boundary_matrix},
we are able to obtain an expression for $\mymat{K}$ from
the final $\matnumratiosc$-matrix \cref{eq:Q_matrix_ratio}
\begin{align}
     \Big(\mymat{F}(\matchannelwavenumber R_{i-1})-\mymat{G}(\matchannelwavenumber R_{i-1}) \mymat{K}\Big)
     = \matnumratiosc(R_i) \Big(\mymat{F}(\matchannelwavenumber R_{i})-\mymat{G}(\matchannelwavenumber R_{i}) \mymat{K}\Big) \,\,,
\end{align}
where, as before (\cref{eq:F_matrix,eq:G_matrix} for
the open channels) the $\mymat{F}$
and $\mymat{G}$ matrices are diagonal matrices proportional
to spherical Bessel functions of the first and second kinds
respectively.
Solving for $\mymat{K}$, we obtain
\begin{align}
    \mymat{K} =
        \Big(\mymat{G}(\matchannelwavenumber R_{i-1})-\matnumratiosc(R_{i})\mymat{G}(\matchannelwavenumber R_i)\Big)^{-1}
        \Big(\mymat{F}(\matchannelwavenumber R_{i-1})-\matnumratiosc(R_{i})\mymat{F}(\matchannelwavenumber R_i)\Big)\,\,.
        \label{eq:K_from_Q}
\end{align}
In principle, by knowing $\mymat{K}$, the scattering wavefunction
can be easily evaluated at any inter-particle separation greater
than $R_{\max}$ (assuming $R_{\max}$ has been chosen large enough
to be in the asymptotic region of the potential)
by constructing the appropriate $\mymat{F}$ and
$\mymat{G}$ matrices, without having to use Numerov's method, which
involves performing computationally expensive linear algebra
operations such as matrix inversion.
\Cref{eq:K_from_Q} yields the augmented reactance matrix including
blocks coupling open and closed ($\mathrm{oc}$), closed and open ($\mathrm{co}$)
as well as closed and closed ($\mathrm{cc}$) channels.
Solutions (columns) of $\mymat{u}$ belonging to closed channels contain
unphysical (exponentially growing) $\mymat{F}^{(\mathrm{closed})}$ contributions.
Columns corresponding to open channels, however, only have exponentially decaying closed components
(proportional to $\mymat{G}^{(\mathrm{closed})}$),
which vanish asymptotically and carry no flux. This restricts the physically relevant
portion of the $\mymat{S}$-matrix to the open-open block
\begin{align}
    \mymat{S}_{\mathrm{oo}} = \left(\mathbbm{1}+i\mymat{K}_{\mathrm{oo}}\right)\left(\mathbbm{1}-i\mymat{K}_{\mathrm{oo}}\right)^{-1}\,\,.
    \label{eq:S_from_Koo}
\end{align}
Having obtained $\mymat{S}_{\mathrm{oo}}$, we can specialise the abstract definition
of the integral cross section stated in \cref{eq:integral_cross_section} to the coupled-basis
case for atom-diatom collisions, with channels indexed by $\multiindex{n}=(v,j,\ell)$.
Since different values of $J$ do not couple, $\mymat{S}$ takes on a block diagonal form,
with blocks $\mymat{S}^{J}$.
The initial and final diatomic rotational projection quantum numbers $m_j$ and $m_j'$
can be averaged and summed over, respectively, to obtain an expression due to Arthurs and Dalgarno
\cite{arthurs1960theory}
\begin{align}
    \sigma_{v'j'\leftarrow vj} = \frac{\pi}{(2j+1)k_{vj}^2}\sum_{J=0}^{\infty}(2J+1)\sum_{\ell=\lvert J-j\rvert}^{J+j}\sum_{\ell'=\lvert J-j'\rvert}^{J+j'}
    \left\lvert\delta_{\multiindex{n'}\multiindex{n}} - S^{J}_{\multiindex{n'}\multiindex{n}}\right\rvert^2\,\,,
    \label{eq:arthurs_dalgarno}
\end{align}
with both parity blocks contributing to each $J$.
We obtain the corresponding differential cross section from
\cref{eq:scattering_amplitude_P} by transforming $\mymat{S}^{J}$ 
back to the uncoupled basis $(v,j,m_{j},\ell,m_{\ell})$ 
using the Clebsch--Gordan coefficients $C_{jm_{j}\ell m_{\ell}}^{JM}$ we introduced above.
Since the incoming partial waves have $m_{\ell}=0$ (and therefore $M=m_{j}$), this yields
\begin{align}
    \frac{\infin{\sigma}_{v'j'\leftarrow vj}}{\infin{\Omega}} &= \frac{1}{2j+1}\sum_{m_{j}=-j}^{j}\sum_{m_{j}'=-j'}^{j'} \left\lvert f_{v'j'm_{j}'\leftarrow vjm_{j}}(\theta,\varphi)\right\rvert^{2}\,\,,
    \label{eq:dcs_atom_diatom} \\
    f_{v'j'm_{j}'\leftarrow vjm_{j}}(\theta,\varphi) &= \sum_{J=0}^{\infty}\sum_{\ell=\lvert J-j\rvert}^{J+j}\sum_{\ell'=\lvert J-j'\rvert}^{J+j'} \frac{\sqrt{4\pi(2\ell+1)}}{i2k_{vj}}\, i^{\ell-\ell'}\, C_{jm_{j}\ell 0}^{Jm_{j}}\, C_{j'm_{j}'\ell'(m_{j}-m_{j}')}^{Jm_{j}} \nonumber \\
    &\quad\times \left(S^{J}_{\multiindex{n'}\multiindex{n}} - \delta_{\multiindex{n'}\multiindex{n}}\right) Y_{\ell',m_{j}-m_{j}'}(\theta,\varphi)\,\,.
    \label{eq:f_atom_diatom}
\end{align}
By integrating \cref{eq:dcs_atom_diatom} over the solid angle, 
it is possible to recover \cref{eq:arthurs_dalgarno}, as should 
be expected (see also \cite{lester1976dynamics}). 
Note that the amplitude in \cref{eq:f_atom_diatom} is defined with 
the flux-normalised $\mymat{S}$-matrix and therefore differs from 
the one in \cref{eq:scattering_amplitude_P} by a factor 
$(k_{v'j'}/k_{vj})^{1/2}$, which is why the velocity ratio of 
\cref{eq:differential_cross_section} does not appear in \cref{eq:dcs_atom_diatom}.
Further, the differential cross section typically requires more total angular momenta $J$ to converge than the
corresponding integral cross section, since different 
$S^{J}_{\multiindex{n'}\multiindex{n}}$ terms interfere in the former,
whereas the latter is an incoherent sum over $J$.

\subsection{`Propagating' input wavepackets}

\label{subsec:propagation_trick}

Interestingly, it is possible to `propagate'
the overlap of an input wavefunction along
with the numerical solution to our eigensystem
by adapting a trick detailed in \cite{karman2015collision}
with a similar method also appearing in \cite{gadea1997nonradiative}.
The requirement to obtain the overlap with an input
wavefunction arises, for example, when instead preparing the
collision for a single, sharp collision energy, the initial
state is a superposition of different energies, which combine
to form an input wavepacket, which would have a localised
spatial distribution.
In this case the `propagated' input wavepacket would enter
in the expression for the collision cross section.
Concretely, this technique was applied to simulate a collision 
which had been launched by a Penning ionisation process, where 
the ejected electron can carry away a variable amount of energy, thus 
preparing the collision complex in a broad superposition of collision 
energies \cite{margulis2023tomography}, amongst other applications \cite{karman2015collision,gadea1997nonradiative}.
One of the limitations of utilising the renormalised
version of  Numerov's method
makes the use of the following trick advantageous.
Roughly speaking, by using the renormalised version of Numerov's method
as described above, we only obtain the ratio of the eigenstates between
neighbouring grid points, making it is necessary to iterate
over the spatial grid a second time after performing the boundary condition
matching in order to explicitly construct the non-asymptotic eigenbasis.
In order to calculate the overlap with the input wavefunction,
we would therefore have to expend additional computational resources in order to
explicitly reconstruct this eigenbasis, even if we
do not otherwise need it.

To be more specific, by starting at a small inter-particle distance $x_{\min}$
(where we assume all eigenfunctions vanish, since the potential
energy surface is assumed to be strongly repulsive)
the ratio matrix of the eigenstates
between two neighbouring grid points is propagated
outwards to the end of the numerical grid, $x_{\max}$, where the asymptotic
boundary condition matching can be performed to obtain $\matnumvar(x_{\max})$ .
All the while, it is necessary to store the ratios of
the non-asymptotic eigenstates $\matnumratiosc(x_{i})=\matnumvar(x_{i-1})\matnumvar^{-1}(x_{i})$
at every pair of neighbouring grid points in memory and,
starting with the last element of $\left\{\qmat_{i}\right\}$,
obtain the various $\matnumvar({x_i})$ by backwards iteration
over the spatial grid, after which it is then possible
to numerically integrate the product of the eigenfunction $\matnumvar(x)$ with
the initial wavefunction.
In contrast, if we are merely interested in evaluating
either the cross section or $\mymat{K}$-matrix as discussed in
\cref{subsec:K_and_S_matrices} (in the case of
a single, sharp collision energy $\totenergy$) and had a way to
`carry' the overlap information of the input wavefunction
outwards during the actual Numerov `propagation', it would
suffice to know the final $\qmat$-matrix at the end of the grid
and all other $\qmat$-matrices could be discarded, thus
dramatically decreasing the amount of memory required
per propagation as well as the subsequent backwards iteration
over the grid.

Fortunately, there is a way to carry out the overlap propagation
mentioned above!
To this end, consider $\mymat{u}$, the matrix of reduced radial coefficients ($\mymat{u}=R\mymat{U}$)
of the wavefunction (columns=solutions, rows=channels) in the channel basis
introduced in \cref{eq:non_asymptotic_expansion}
and $\myvec{\phi}=R\cdot \myvec{\Phi}$,
an ($R$-dependent) vector describing the input wavepacket
(could either be sharp in a specific channel or spread across multiple channels).
On a quantised spatial grid $\{R_{i} | i=1,\dots,n\}$ with $R_{\min}=R_{1}$ and $R_{\max}=R_{n}$ the overlap integral can be expressed
using a mid-point quadrature
\begin{align}
	\myvec{c} = \int_{R_{\min}}^{R_{\max}} \infin{R} \mymat{u}^{T}(R) \myvec{\phi}(R) \approx \sum_{i=1}^{n} \Delta_{R} \mymat{u}_{i}^{T} \myvec{\phi}_{i} = \myvec{c}_{n}\,\,.
	\label{eq:overlapintegral}
\end{align}
Above, the index $n$ reminds us that the full quadrature approximation to the
overlap is obtained when the sum runs over all $n$ grid points.
We also make use of the fact that $\mymat{u}$ is real under $\mymat{K}$-matrix normalisation, such that $\mymat{u}^{\dagger}=\mymat{u}^{T}$.
In order to develop a recursion relation, however, it will be
convenient to define the quantity
\begin{align}
    \myvec{d}_{m} = \left(\mymat{u}_{m}^{T}\right)^{-1} \Delta_{R} \sum_{j=1}^{m} \mymat{u}_{j}^{T} \myvec{\phi}_{j}\,\,,
    \label{eq:overlap}
\end{align}
where $m \le n$ is an intermediate stopping point
on the grid.
Splitting off the last summand, we have
\begin{align}
	\myvec{d}_{m}
	& = \Delta_{R} \sum_{j=1}^{m-1} \left(\mymat{u}_{m}^{T}\right)^{-1} \mymat{u}_{j}^{T} \myvec{\phi}_{j} + \Delta_{R}\myvec{\phi}_{m} \\
		& = \Delta_{R} \sum_{j=1}^{m-1} \left(\mymat{u}_{m}^{T}\right)^{-1} \underbrace{\mymat{u}_{m-1}^{T} \left(\mymat{u}_{m-1}^{T}\right)^{-1}\dots\mymat{u}_{j+1}^{T} \left(\mymat{u}_{j+1}^{T}\right)^{-1}}_{=\mathbbm{1}} \mymat{u}_{j}^{T} \myvec{\phi}_{j} + \Delta_{R}\myvec{\phi}_{m} \\
		& = \Delta_{R} \sum_{j=1}^{m-1} \underbrace{\left(\mymat{u}_{m}^{T}\right)^{-1}\mymat{u}_{m-1}^{T}}_{=\qmat_{m}^{T}} \dots \underbrace{\left(\mymat{u}_{j+1}^{T}\right)^{-1} \mymat{u}_{j}^{T}}_{=\qmat_{j+1}^{T}} \myvec{\phi}_{j}+ \Delta_{R}\myvec{\phi}_{m} \\
		& = \Delta_{R} \sum_{j=1}^{m-1} \left(\prod_{i=j+1}^{m} \qmat_{i}\right)^{T} \myvec{\phi}_{j} + \Delta_{R}\myvec{\phi}_{m}\,\,.
\end{align}
Above, since matrix multiplication is not commutative, the product symbol is to be interpreted as the multiplication of matrices possessing ascending indices from left to right.
We can then derive the recursion relation in $\myvec{d}$ by considering the $m+1$st term
\begin{align}
	\myvec{d}_{m+1} & = \Delta_{R} \sum_{j=1}^{m} \left(\prod_{i=j+1}^{m+1} \qmat_{i}\right)^{T} \myvec{\phi}_{j} + \Delta_{R}\myvec{\phi}_{m+1} \\
		& = \Delta_{R} \sum_{j=1}^{m} \qmat_{m+1}^{T} \left(\prod_{i=j+1}^{m} \qmat_{i}\right)^{T} \myvec{\phi}_{j} + \Delta_{R}\myvec{\phi}_{m+1} \\
		& = \qmat_{m+1}^{T} \underbrace{\left( \Delta_{R} \sum_{j=1}^{m-1} \left(\prod_{i=j+1}^{m} \qmat_{i}\right)^{T} \myvec{\phi}_{j} + \Delta_{R} \myvec{\phi}_{m} \right)}_{=\myvec{d}_{m}} + \Delta_{R}\myvec{\phi}_{m+1} \\
		& = \qmat_{m+1}^{T} \myvec{d}_{m} + \Delta_{R} \myvec{\phi}_{m+1}\,\,.
\end{align}
Since $\mymat{u}_{1}=0$ (we assume we are starting the propagation in the strongly repulsive area, region (I), where the wavefunction vanishes),
it should be noted that $\qmat_{2}=\mymat{u}_1\mymat{u}_{2}^{-1}=0$ and the recursion actually starts off with $\myvec{d}_{2}=\Delta_{R} \myvec{\phi}_{2}$.
To finally obtain the initial wavepacket overlaps $\myvec{c}$
after the `wavepacket propagation' procedure,
we must remember to multiply $\myvec{c}=\mymat{u}_{n}^{T}\myvec{d}_{n}$ (where $\mymat{u}_{n}$ is given by the
asymptotic boundary conditions, as it is the final grid point).
This procedure is mathematically equivalent to explicitly reconstructing $\mymat{u}(R)$
first, by starting with $\mymat{u}_{n}$ and applying each $\qmat_{j}$ from the
end of the grid backwards and then calculating the integral \cref{eq:overlapintegral}
but in terms of computation, we can save both memory and the effort of
reconstructing $\mymat{u}(R)$ explicitly in a second iteration over the
spatial grid.

\subsection{Convergence and consistency checks}
\label{subsec:convergence}
There are limitations inherent to the numerical solution of any problem,
predominantly due to the speed at which computers can perform calculations
and the available memory.
Specifically, we mean the central processing unit (CPU) clock speed and the
size of the random access memory (RAM) for conventional computer simulations,
since these determine how fast sequential operations can be performed.
For highly parallelisable problems (which collision simulations over a wide range
of energies, total angular momenta and parities usually are), other hardware
such as graphics processing units (GPUs) can be used, but this is a more
specialised topic that will not be discussed here.
In order to `fit' a collision simulation onto a computer (or a cluster of computers),
concessions have to be made with respect to the size of the basis, the number
of partial waves $J$ and the number of grid points.
Pitching these numbers too high will cause the calculation to run for
longer than we care to wait, or to crash if the required memory is not available.
In contrast, pitching too low will cause numerical errors that can reduce
an otherwise realistic description to numerical noise.
Finding the sweet spot in between is the goal of convergence checks.
Unfortunately, many of the numerical parameters can only be determined by
trial and error; nevertheless, we present some guidelines to streamline
this tedious process.

\paragraph{Step size}
We have seen from the derivation of Numerov's method that the
local error scales as $\mathcal{O}(h^6)$.
It can be shown, however, that due to errors accumulating across the grid,
the global error is $\mathcal{O}(h^4)$.
A good starting point is a step size of the order of ten grid points per local
de~Broglie wavelength $2\pi/k_{\mathrm{loc,max}}$, evaluated where the
local kinetic energy is largest.
This is the case at the bottom of the potential well, where
$k_{\mathrm{loc,max}}=\sqrt{2\mu(E_{\mathrm{well}}+\totenergy-\intenergy_{\min})}/\hbar$,
with $E_{\mathrm{well}}$ the well depth and $\totenergy-\intenergy_{\min}$ the kinetic
energy in the lowest channel, which is the largest.
Then, by halving $h$, the error should roughly be reduced by a factor of 16.

\paragraph{Grid boundaries}
$R_{\min}$ has to lie far enough inside the repulsive wall for the wavefunction 
of all included channels to be negligible, and $R_{\max}$ far enough out for
the interaction potential to be negligible compared to the centrifugal term (region (III)).
Since we match to Riccati--Bessel functions, which already account for the centrifugal
term, $R_{\max}$ does not need to lie in the truly asymptotic region (IV).
Both boundaries should be varied until the results no longer change.

\paragraph{Basis truncation}
The rovibrational basis has to be enlarged until the cross sections are stable.
This requires including energetically closed channels, whose potential curves
can have wells that dip below the collision energy.
Through virtual excitation, the results can be significantly influenced by their
presence or absence.
In particular, quasi-bound states supported by closed channels give rise to
Feshbach resonances, which cannot be described unless these channels are
explicitly included.
The necessary $J_{\max}$ depends on the maximum collision energy considered.
Starting from a given $J_{\max}$, it should be increased until further
contributions to the differential cross section become negligible, as these
typically converge more slowly than those to the integral cross section.

\paragraph{Built-in checks}
The open-open block $\mymat{K}_{\mathrm{oo}}$ obtained from \cref{eq:K_from_Q} 
should be real and symmetric to within numerical accuracy, provided $R_{\max}$ 
is large enough for the closed-channel components to have decayed.
Since $\mymat{S}_{\mathrm{oo}}$ from \cref{eq:S_from_Koo} is then automatically unitary and symmetric, the asymmetry of the computed $\mymat{K}_{\mathrm{oo}}$ is a sensitive diagnostic for numerical problems such as an insufficient step size or too small an $R_{\max}$.

\subsection{Discussion}

This section arguably counts amongst the most challenging of the tutorial but hopefully feels rewarding to the diligent reader in turn.
We have seen the total framework discussed in the previous sections come together to build a full picture of the pieces needed to
simulate a state-of-the-art quantum collision system.
In particular, we have generalised the (renormalised) Numerov's method to handle multiple channels and seen how we can calculate
the bound states and eigenenergies of the constituents of our scattering system using the Discrete variable representation.
We have made all of this concrete for the particular case of an atom colliding with a diatom, introducing the appropriate
Jacobi coordinate system and discussing the Hamiltonian and its properties and most crucially how to evaluate the matrix
elements of the interaction potential for our given bases.

Quantum channels of diatomic molecules are each asymptotically labelled by a rotational and vibrational quantum number.
The Hilbert space dimension (and therefore the size of the matrices the algorithm works on) directly relates to the (necessarily finite) number of
states we account for.
Even channels which are not energetically accessible asymptotically can have a tangible influence on the outcome of a collision and
convergence has to be sought carefully when truncating bases \cite{hutson1994coupled}.
On the other hand, memory and processing constraints limit the sizes of systems it is feasible to treat.
As the collision energy increases, more and more channels participate in the collision, further exacerbating the convergence problem.
Paradoxically, the increase in participating channels causes the effect of quantum coherences to average out, opening the door for
alternative methods, such as the quasi-classical trajectory (QCT) method \cite{karplus1965exchange} to accurately describe the problem whilst remaining tractable.
Predicting resonances and other effects quantitatively crucially relies on having a very accurate potential energy surface, which, as we saw in previous sections,
relies upon solving the electronic part of the problem separately.
This compartmentalised treatment of the electronic structure hinges on the method used with several different levels of theory to choose from,
as well as the size of the electron orbital basis set flowing into the calculation and can limit the quality of the cross sections obtained
in a scattering simulation \cite{jensen2017introduction}.

With some dedication, the union of the methods and techniques described in this section, which build upon the conceptual understanding of the
preceding sections, can be translated into a simulation programme, enabling the user to perform cutting-edge scientific research.
Without any further modifications, for example, we can fully simulate inelastic processes, allowing for the population transfer between
different channels.
This leads to rich and interesting effects which theory is able to predict and study, for example Feshbach resonances
(where the particles can form a quasi-bound complex at certain collision energies), which explicitly rely on a fully
quantum treatment.
The contents of this tutorial serve as a foundation upon which the authors themselves have built in order to study various atom-diatom collision systems
and can further be adapted to treat even more complex problems.
This is far from the end of the story, though: obtaining a result (i.e. data from a collision simulation) can be viewed as just the
first step on a journey of analysis and interpretation.
It is at this point that imagination and insight can take over to yield new levels of understanding of quantum collisions
and the promise of potential discoveries yet to be made.

\subsection{Section summary}

\begin{keyresults}

\kr{\label{kr:ch4-dvr}%
  Discrete variable representation can be used to reliably calculate the bound states $\chi_{vj}(r)$ and energies of
  diatomic molecules, in the case of `sinc-DVR' the off-diagonal matrix elements of the diatomic
   molecule's potential vanish whilst the kinetic energy matrix elements assume the simple form
  \begin{equation*}
    \bra{n}\partial_x^{2}\ket{m} =
    \begin{cases}
        -\frac{1}{3}\frac{\pi^2}{\Delta^2}, & n=m \\
        -\frac{2}{\Delta^2}\frac{(-1)^{n-m}}{(n-m)^{2}}, & n \neq m
    \end{cases} .
  \end{equation*}
  \emph{Cf.}~\cref{eq:dvr_second_derivative}.}

\kr{\label{kr:ch4-vmat}%
The expansion of the full atom-diatom interaction potential matrix elements in the channel basis is
given in terms of the diatomic basis functions, Legendre polynomials and the Percival-Seaton coefficient
(\cref{eq:percival_seaton})
\begin{align*}
    \bra{\multiindex{n'}}\op{V}\ket{\multiindex{n}}(R)
  =& \frac{r_{\max}-r_{\min}}{2} \sum_{\lambda\eta} V_{\lambda\eta}(R) \int \infin{\tilde{r}} \chi_{v'j'}^{*}(r(\tilde{r}))P_{\eta}(\tilde{r})\chi_{vj}(r(\tilde{r})) \nonumber \\
    &\times \bra{j'\ell'J}P_{\lambda}(\cos\theta)\ket{j\ell J}\,\,.
\end{align*}
\emph{Cf.}~\cref{eq:final_V_matrix_element}.}

\kr{\label{kr:ch4-recursion}%
  The single-term recursion relation for the multi-channel version of the renormalised Numerov's method for the coupled
  system of differential equations $\partial_{x}^{2} \matnumvar(x) = -\matnumlinsc(x) \matnumvar(x)$ is given by
  \begin{align*}
    \matnumratiosc(x_{i+1})^{-1} \approx & \left(\mathbbm{1}+\frac{h^2}{12}\matnumlinsc(x_{i+1})\right)^{-1} \nonumber \\
    &  \times  \Big(2(\mathbbm{1}- \frac{5h^2}{12}\matnumlinsc(x_{i})) -  (\mathbbm{1}+\frac{h^2}{12}\matnumlinsc(x_{i-1})) \matnumratiosc(x_{i}) \Big)  .
  \end{align*}
  \emph{Cf.}~\cref{eq:multichan_ren_recursion}.}

\kr{\label{kr:ch4-arthursdalgarno}%
  The integral cross section in block form (for conserved $J$) in the case of atom-diatom collisions as obtained in \cite{arthurs1960theory}
  \begin{align*}
    \sigma_{v'j'\leftarrow vj} = \frac{\pi}{(2j+1)k_{vj}^2}\sum_{J=0}^{\infty}(2J+1)\sum_{\ell=\lvert J-j\rvert}^{J+j}\sum_{\ell'=\lvert J-j'\rvert}^{J+j'}
    \left\lvert\delta_{\multiindex{n'}\multiindex{n}} - S^{J}_{\multiindex{n'}\multiindex{n}}\right\rvert^2\,\,.
  \end{align*}
  \emph{Cf.}~\cref{eq:arthurs_dalgarno}.
}

\end{keyresults}

\section*{Conclusions}
\addcontentsline{toc}{section}{Conclusions}

In this tutorial we have gone from the most general form of a Hamiltonian for two interacting particles to a full description
of setting up a quantum simulation of the collision.
To this end, we introduced the Born-Oppenheimer approximation, effectively separating the electronic problem from the problem of colliding nuclei.
We then exploited the translational invariance of the system to separate out the centre-of-mass motion.
By expanding the wavefunction in terms of asymptotic states and inserting it into the Schr{\"o}dinger equation,
we arrived at the coupled channel differential equations.

Next, we introduced the asymptotic boundary conditions in terms of Riccati--Bessel and Riccati--Hankel functions,
which solve the differential equation in the absence of the interaction potential but still accounting for the
centrifugal term.
Critically, it was seen that the relative (complex) amplitudes of the incoming and outgoing Riccati--Hankel functions encode
the information required to extract physically observable information such as the $\mymat{S}$-matrix as
well as differential and integral cross sections.

Taking a first step in the direction of a numerical treatment of the quantum scattering problem,
we then regarded the simple case of two structureless particles colliding, by only accounting for
a single quantum channel.
For this special case we saw how the scattering amplitude relates directly to the cross section
and how the scattering phase can be used to express several scalar quantities.
Through Taylor-expansion of the solution, we derived Numerov's method for solving second order differential
equations and saw how the stability could be improved by propagating the ratio of the solution in neighbouring points.

Finally, the tutorial culminates in a full quantum treatment of an atom-diatom system.
To this end we introduced the Jacobi coordinate system and, together with an expansion in the
diatomic and angular bases, reduced the problem to a set of coupled one-dimensional equations.
We saw how another numerical technique, the discrete variable representation could be used to
obtain eigenvalues and eigenstates of the diatomic molecule which were needed to evaluate the
interaction potential matrix elements as a function of the interparticle distance.
We then generalised Numerov's method to work with multichannel systems, which can be used to  obtain
the reactance matrix from the propagation.
We saw how to obtain the $\mymat{S}$-matrix from the open-open block of $\mymat{K}$, leading to the Arthurs--Dalgarno
expression for the integral cross section. Since total angular momentum and parity are conserved, the scattering problem
segments into numerically manageable blocks.

Hopefully, this tutorial will have covered many of the pitfalls and fuzzy areas that beginners
in scattering theory often have to deal with and which many textbooks tend not to cover in detail.
Throughout, we have tried to build a consistent bridge between quantum mechanical theory
and numerics.
Whilst it would be impossible to give every single detail, the philosophy of this tutorial is
to be as explicit as possible, chaining together concepts to build a coherent picture.
To name just a few, we have paid special attention to the basis states and the coordinates they
are associated with.
We have also seen what consequences the choice of unit amplitude versus unit flux normalisation
has for the formulation of observables, and how to go from one form to the other.
Further, paying attention to the ordering of matrix indices (rows=channels, columns=solutions) turned
out to be critically important for the physical interpretation of simulations.
Equipped with this framework, we expect the reader to be able to not just implement the relevant
equations in code but also be able to apply physical intuition in order to interpret results.

In order to keep the scope of this tutorial finite, certain topics were only briefly addressed
or even omitted entirely.
Reactive collisions, for example, require a way of modelling the breaking of a bond and the
creation of another, for instance by performing coupled channel simulations with a hyperspherical
coordinate system\cite{pack1987quantum}, or can be treated with fundamentally different methods such as time-dependent
wavepackets\cite{neuhauser1991application}.
We have not considered the effects that external fields have on our collisions, which makes
the derivations slightly more complex but is possible to do in principle as shown in Krems'
textbook \cite{krems2018molecules}.
Another huge topic, that was taken entirely for granted is that of potential energy surfaces.
In practice obtaining these is a numerically very demanding process, whilst also requiring a
high degree of expertise to obtain good results.
Even once individual points have been obtained, there is an art to fitting them to make a
continuous surface which can be sampled at arbitrary points, such as using the 
reproducing kernel Hilbert spaces (RKHS) method \cite{ho1996general,unke2017toolkit}.

Readers who have taken a foray into the world of collision simulations using this tutorial
thus have a springboard towards new topics, with plenty of directions to go in.
As computational capabilities grow, so too will the sizes and variety of systems to explore.
For production calculations, readers may also turn to established, well-tested packages such as MOLSCAT \cite{hutson2019molscat} or Hibridon \cite{alexander2023hibridon}, whose documentation should now be considerably easier to follow.

\section*{Acknowledgements}
We would like to thank Gerard Meijer for helpful comments on KPH's PhD thesis
as well as inspiring discussions.
In addition we would also like to thank both Ad van der Avoird and
Gerrit Groenenboom, who laid a lot of the foundations this work is based on. %

\paragraph{Funding information}
No specific funding.

\begin{appendix}

\section{Bessel functions}
\label{apdx:bessel}

\Cref{eq:riccati_bessel} is a special case of the Bessel differential equation
\begin{align}
    \left(x^{2} \partial_{x}^{2} + cx \partial_{x} + (x^2-\alpha^2)\right) y(x) = 0
    \label{eq:standard_bessel}
\end{align}
where $c$ can be either 0 (Riccati case), 1 (cylindrical case)
or 2 (spherical case) and
$\alpha^2 = \ell(\ell+1)$ is expressed in terms
of the order of the Bessel function $\ell$.
By substitution it can be shown that a solution $y$ to the $c=2$
equation can be multiplied by $x$ to obtain a solution to its $c=0$  counterpart.

The Bessel differential equation is a second order, homogeneous differential equation which has huge significance in vastly diverse areas of physics.
Before delving too far into the properties of Bessel
functions, which are for example derived and discussed
in great detail elsewhere \cite{arfken2013mathematical},
we will first turn our attention to solving
the spherical Bessel differential equation using a power
series expansion with a treatment which is loosely based on
\cite{ryan2008techniques}.
At this stage, it should be pointed out, that the usual
way of obtaining the spherical Bessel functions is to
first solve the cylindrical Bessel equation and then transform
it into its spherical counterpart using the substitution
$y_{\mathrm{spherical}}(x)=\frac{y_{\mathrm{cylindrical}}(x)}{x^{1/2}}$.
Here we shall attempt to construct a solution to the
spherical Bessel differential equation directly.
First, however, we shall consider a theorem concerning
more general second-order differential equations.

\subsection{Fuchs' Theorem}

\Cref{eq:standard_bessel} is a second-order, ordinary differential
equation of the standard form
\begin{align}
    \left(\partial_{x}^{2} + p(x)\partial_{x} + q(x)\right) y(x) = g(x)\,\,,
    \label{apeq:bessel_diff_eq}
\end{align}
where $x=0$ is a regular singular point (i.e. $p(x)$ or $q(x)$ are singular,
whilst $x p(x)$ and $x^2 q(x)$ are analytic), preventing a straight-forward power series
expansion centered around the origin.
In our case $p(x)=2/x$ and $q(x)=1-\alpha^2/x^2$.

Stated succinctly, Fuchs' theorem \cite{asmar2016partial} tells us
that around a regular singular point $x_{0}$ (here $x_{0}=0$) there is always
at least one solution of the form
\begin{align}
    y_1(x) = \sum_{k=0}^{\infty} a_{k} (x-x_{0})^{k+r},~r \in~\mathbb{R}
    \label{apeq:fuchs_primary_series}
\end{align}
and a second, linearly independent solution of the form
\begin{align}
    y_2(x) = C\, y_{1}(x) \ln{(x-x_{0})} + \sum_{k=0}^{\infty} b_{k} (x-x_{0})^{k+s},~s\in~\mathbb{R}
    \label{apeq:fuchs_secondary_series}
\end{align}
(where $C$ may vanish), each with a radius of
convergence
\begin{align}
    \mathrm{rad~conv}(y(x)) \ge \min \mathrm{rad~conv}(p(x),q(x),g(x))\,\,.
\end{align}
Additionally, the two solutions above are linearly independent,
which is an important consideration to make with regards to the
solution of differential equations using power series methods.

It should be pointed out that regular power series expansion
\textit{ansatz} could be unsuitable if the differential equation
possesses singularities and we seek a solution with an infinite radius
of convergence.

\subsection{Frobenius' method}

The idea behind Frobenius' method is to patch
singularities in power series expansions by shifting
the power series exponent by some $r \in \mathbb{R}$
\cite{ryan2008techniques}.

We shall proceed by substituting the \textit{ansatz}
$y(x)=x^r \sum_{k=0}^{\infty} A_{k}x^{k}$ into the
homogeneous differential equation
\begin{align}
  \left(\partial_{x}^{2} + p(x)\partial_{x} + q(x)\right) y(x) = 0\,\,,
\end{align}
where, conforming to \cref{apeq:bessel_diff_eq} we have $b(x)=xp(x)=2$ and
$c(x)=x^2q(x)=x^2-\ell(\ell+1)$ in the special case of the spherical Bessel
differential equation, leading to
\begin{align}
  & A_{0} x^{r-2} \left(r(r-1) +2r -\ell(\ell+1)\right)
  +A_{1} x^{r-1} \left((r+1)r + 2(r+1) - \ell(\ell+1)\right) \nonumber \\
  & +\sum_{k=2}^{\infty} \left[A_{k}\left((k+r-1)(k+r) + 2(k+r) -\ell(\ell+1)\right) + A_{k-2} \right] x^{k+r-2} = 0\,\,.
\end{align}
Comparing terms in powers of $x$ (which all vanish on the
right hand side of the equation), we see that
\begin{align}
  &A_{0} \left(r(r-1) + 2r - \ell(\ell+1)\right)  \nonumber \\
  &= A_{0} \left(r^2 -r +2r - \ell(\ell+1)\right) \nonumber \\
  &= A_{0} \left(r(r+1) - \ell(\ell+1)\right) \nonumber \\
  &= 0 \,\,,
\end{align}
which, assuming $A_{0} \neq 0$, is fulfilled for $r_1=\ell$ and $r_2=-\ell-1$.
In terms of the more general $b(x)$ and $c(x)$
is
\begin{align}
  A_{0} \left(r(r-1)+b(x_0)r+c(x_0)\right) = 0
\end{align}
This so-called `indicial equation' has two roots in $r$
\begin{align}
  r_{1/2} = -\frac{b(x_0)-1}{2} \pm \sqrt{\frac{(b(x_0)-1)^2}{4}-c(x_0)}\,\,.
\end{align}

Remembering that $r_1$ and $r_2$ are the shifts in the
power series \textit{ansatz} used to treat the singularity,
we note that the difference in roots can either an integer
number (possibly zero, for a repeated root) or a non-integer
real number.
When the difference in roots is an integer number, the two
corresponding power series are no longer linearly independent
solutions, requiring us to seek the second linearly independent
solution in terms of the solution constructed using Fuchs' theorem
\cref{apeq:fuchs_secondary_series}.

Similarly, comparing the $x^{r-1}$ terms, we have
\begin{align}
  A_{1} \left((r+1)r + 2(r+1) - \ell(\ell+1)\right) = 0\,\,.
\end{align}
Since, for instance, substituting in $r_1=\ell$, as determined above,
leads to
\begin{align}
  A_{1}\left((\ell+1)\ell + 2(\ell+1) - \ell(\ell+1)\right) = 2A_{1} (\ell + 1) = 0\,\,,
\end{align}
we require that $A_{1}=0$.

Finally, we have the recursion relation for the coefficients $A_k$ with $k \ge 2$
\begin{align}
  A_{k} = - \frac{A_{k-2}}{(k+r)(k+r-1) + 2(k+r) - \ell(\ell+1)}\,\,,
\end{align}
which, combined with $A_{1}=0$ immediately implies that $A_{2n+1}=0~\forall n\in \mathbb{N}_{+}$.

Further, under substitution of $r_{1}=\ell$, the recursion relation can
be simplified to
\begin{align}
  A_{k} = - \frac{A_{k-2}}{k(k+2\ell+1)}
\end{align}
or, looking at even terms specifically
\begin{align}
  A_{2\kappa} = - \frac{A_{2\kappa-2}}{2\kappa(2\kappa+2\ell+1)} = -\frac{A_{2\kappa-2}}{2^2\kappa(\kappa+\ell+1/2)}\,\,,
\end{align}
where $\kappa \in \mathbb{N}_{+}$.
Consequently we can express $A_{2\kappa}$ in terms of $A_{0}$ and a product of reciprocals,
which in turn can be expressed in terms of the Gamma function
\begin{align}
  \Gamma(z)=\int_{\mathbb{R}_{+}} \infin{t}\, t^{z-1}e^{-t}\,\,,
\end{align}
which converges absolutely when $\mathcal{R}(z)>0$ and fulfils the property
$\Gamma(z+1)=z\Gamma(z)$, making it a useful extension to the factorial function.
\begin{align}
  A_{2\kappa}
  &= \frac{A_{0}}{2^{2\kappa}} \prod_{m=0}^{\kappa-1} \frac{-1}{(\kappa-m)(\kappa-m+\ell+1/2)} \nonumber \\
  &= \frac{A_{0}(-1)^{\kappa}}{2^{2\kappa}\kappa!} \prod_{m=0}^{\kappa-1} \frac{1}{\kappa-m+\ell+1/2} \nonumber \\
  &= \frac{A_{0}(-1)^{\kappa}}{2^{2\kappa}\kappa!}
    \frac{1}{(\kappa+\ell+1/2)} \frac{1}{(\kappa-1+\ell+1/2)} \frac{1}{(\kappa-2+\ell+1/2)} \ldots \frac{1}{(\kappa-(\kappa-1)+\ell+1/2)} \nonumber \\
  &= \frac{A_{0}(-1)^{\kappa}}{2^{2\kappa}\kappa!} \frac{\Gamma(\ell+3/2)}{\Gamma(\kappa+\ell+3/2)}
  \,\,.
\end{align}

Substituting the above coefficients into the \textit{ansatz}
$y(x)$ (which we shall denote by $y(x)=j_{\ell}(x)$ from now on)
$j_{\ell}(x)=x^{\ell}\sum_{k=0}^{\infty} A_{k} x^{k}=x^{\ell}\sum_{\kappa=0}^{\infty} A_{2\kappa}x^{2\kappa}$
\begin{align}
  j_{\ell}(x)
  &= x^{\ell} \sum_{\kappa=0}^{\infty} \frac{A_{0}(-1)^{\kappa}}{2^{2\kappa}\kappa!} \frac{\Gamma(\ell+3/2)}{\Gamma(\kappa+\ell+3/2)} x^{2\kappa} \nonumber \\
  &= \sum_{\kappa=0}^{\infty} A_{0}2^{\ell} \frac{\Gamma(\ell+3/2)}{\Gamma(\kappa+\ell+3/2)} \frac{(-1)^{\kappa}}{\kappa!} \left(\frac{x}{2}\right)^{2\kappa+\ell} \nonumber \\
  &= \sqrt{\frac{\pi}{2x}} \sum_{\kappa=0}^{\infty} A_{0}\frac{2^{\ell+1}}{\sqrt{\pi}} \frac{\Gamma(\ell+3/2)}{\Gamma(\kappa+\ell+3/2)} \frac{(-1)^{\kappa}}{\kappa!} \left(\frac{x}{2}\right)^{2\kappa+\ell+1/2}
\end{align}
and finally choosing $A_0=\frac{\sqrt{\pi}}{2^{\ell+1}\Gamma(\ell+3/2)}$ leads to
\begin{align}
  j_{\ell}(x)
  &= \sqrt{\frac{\pi}{2x}} \sum_{\kappa=0}^{\infty} \frac{1}{\Gamma(\kappa+\ell+3/2)} \frac{(-1)^{\kappa}}{\kappa!} \left(\frac{x}{2}\right)^{2\kappa+\ell+1/2}\,\,.
\end{align}

In practice, this expansion proves difficult to evaluate
for specific values of $\ell$.
With some additional arithmetic, however, it is possible to show
that
\begin{align}
  j_{0}(x) = \frac{\sin{x}}{x}
\end{align}
and that
\begin{align}
  j_{1}(x) = \frac{\sin{x}}{x^2} - \frac{\cos{x}}{x}\,\,.
\end{align}

\subsection{Generating function}

\begin{figure}
  \centering
  \begin{tikzpicture}
    \node [minimum width=1cm,minimum height=1cm] (zerozero) at (0,0) {$0,\,0$};
    \node [anchor=south] (eqeq) at ([yshift=0.2cm]zerozero.center) {\begin{turn}{90}$=$\end{turn} $\,\,$ \begin{turn}{90}$=$\end{turn}};
    \node [anchor=south] (rs) at (eqeq.center) {$r,\,\,s$};

    \node [minimum width=1cm,minimum height=1cm,anchor=west] (zeroone) at (zerozero.east) {$0,\,1$};
    \node [minimum width=1cm,minimum height=1cm,anchor=west] (zerotwo) at (zeroone.east) {$0,\,2$};

    \node [minimum width=1cm,minimum height=1cm,anchor=north] (onezero) at (zerozero.south) {$1,\,0$};
    \node [minimum width=1cm,minimum height=1cm,anchor=north] (twozero) at (onezero.south) {$2,\,0$};

    \node [minimum width=1cm,minimum height=1cm,anchor=west] (oneone) at (onezero.east) {$1,\,1$};
    \node [minimum width=1cm,minimum height=1cm,anchor=west] (onetwo) at (oneone.east) {$1,\,2$};

    \node [minimum width=1cm,minimum height=1cm,anchor=west] (twoone) at (twozero.east) {$2,\,1$};
    \node [minimum width=1cm,minimum height=1cm,anchor=west] (twotwo) at (twoone.east) {$2,\,2$};

    \node [minimum width=1cm,minimum height=1cm,anchor=west] (dotszerothree) at (zerotwo.east) {$\ldots$};
    \node [minimum width=1cm,minimum height=1cm,anchor=west] (dotsonethree) at (onetwo.east) {$\ldots$};
    \node [minimum width=1cm,minimum height=1cm,anchor=west] (dotstwothree) at (twotwo.east) {$\ldots$};

    \node [minimum width=1cm,minimum height=1cm,anchor=north] (dotsthreezero) at (twozero.south) {$\vdots$};
    \node [minimum width=1cm,minimum height=1cm,anchor=north] (dotsthreeone) at (twoone.south) {$\vdots$};
    \node [minimum width=1cm,minimum height=1cm,anchor=north] (dotsthreetwo) at (twotwo.south) {$\vdots$};
    \node [minimum width=1cm,minimum height=1cm,anchor=north west] (dotsthreethree) at (twotwo.south east) {$\ddots$};

    \draw [color=gray] (zerozero.center) -- (dotsthreethree.center) node [pos=1.0,above,rotate=-45,color=red] {$n=0$};
    \draw [color=gray] (onezero.center) -- (dotsthreetwo.center)    node [pos=1.0,above,rotate=-45,color=red] {$n=1$};
    \draw [color=gray] (twozero.center) -- (dotsthreeone.center)    node [pos=1.0,above,rotate=-45,color=red] {$n=2$};

    \draw [color=gray] (zeroone.center) -- (dotstwothree.center)    node [pos=1.0,above,rotate=-45,color=red] {$n=-1$};
    \draw [color=gray] (zerotwo.center) -- (dotsonethree.center)    node [pos=1.0,above,rotate=-45,color=red] {$n=-2$};
  \end{tikzpicture}
  \caption{Summation scheme over $n=r-s$.}
  \label{fig:summation_scheme}
\end{figure}
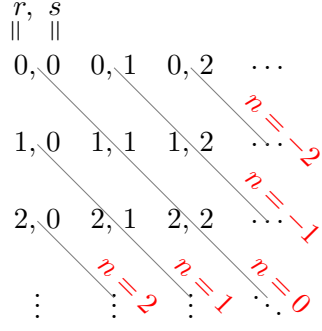

An alternate approach is to attempt to find a recurrence relation
which might allow higher-order Bessel function to be obtained from lower order ones.
This can be accomplished, for example, with the help of a so-called
`generating function' of the Bessel functions \cite{arfken2013mathematical}
\begin{align}
  g(x,t)
  &=  e^{\frac{x}{2}\left(t-\frac{1}{t}\right)} \nonumber \\
  &=  \sum_{r=0}^{\infty} \left(\frac{x}{2}\right)^r \frac{t^r}{r!} \sum_{s=0}^{\infty} (-1)^s \left(\frac{x}{2}\right)^s \frac{t^{-s}}{s!} \nonumber \\
  &=  \sum_{r,s=0}^{\infty} (-1)^{s} \left(\frac{x}{2}\right)^{r+s} \frac{t^{r-s}}{r!s!}\,\,.
\end{align}
The above pair of nested sums can be re-arranged in a clever way
by substituting the summation variable $r$ with $n=r-s$, leading the
corresponding value of $s$ to be constrained between $\max{(0,-n)}$ and
$\infty$ as indicated in \cref{fig:summation_scheme}.
The resulting expansion for the generating function is therefore
\begin{align}
  g(x,t)
  &= \sum_{n=-\infty}^{\infty} \sum_{s=\max{(0,-n)}}^{\infty} \frac{(-1)^{s}}{(n+s)!s!} \left(\frac{x}{2}\right)^{n+2s} t^{n} \nonumber \\
  &= \sum_{n=-\infty}^{\infty} J_{n}(x) t^{n}\,\,,
\end{align}
where
\begin{align}
  J_{n}(x) = \sum_{s=\max{(0,-n)}}^{\infty} \frac{(-1)^{s}}{(n+s)!s!} \left(\frac{x}{2}\right)^{n+2s}\,\,
\end{align}
are the (cylindrical) Bessel functions of first kind
(and of integer order, which can be
obtained with Frobenius' method in a way analogous to
how the spherical Bessel functions were obtained above).
By differentiating the generating function with respect to $t$ and
comparing powers of $t$, we can obtain a recurrence relation between
Bessel functions.
\begin{align}
  \sum_{n=-\infty}^{\infty} J_{n+1}(x) (n+1) t^{n} = \sum_{n=-\infty}^{\infty} J_{n}(x) n t^{n-1} = \partial_{t} g(x,t)
  &= \frac{x}{2}\left(1+\frac{1}{t^2}\right)e^{\frac{x}{2}\left(t-\frac{1}{t}\right)} \nonumber \\
  &= \frac{x}{2}\left(1+\frac{1}{t^2}\right) \sum_{n} J_{n}(x) t^{n} \nonumber \\
  &= \frac{x}{2} \sum_{n=-\infty}^{\infty} J_{n}(x) t^{n} + \frac{x}{2} \sum_{n=-\infty}^{\infty} J_{n}(x) t^{n-2} \nonumber \\
  &= \frac{x}{2} \sum_{n=-\infty}^{\infty} J_{n}(x) t^{n} + \frac{x}{2} \sum_{n=-\infty}^{\infty} J_{n+2}(x) t^{n}\,\,.
\end{align}
Comparing the first and last rows of the above set of equations for an arbitrary
power of $t$ it follows that
\begin{align}
  (n+1) J_{n+1}(x) = \frac{x}{2} J_{n} + \frac{x}{2} J_{n+2}(x)\,\,.
  \label{eq:base_rec_rel}
\end{align}
At face value, this recurrence relation is unhelpful, in the
context of spherical Bessel functions, since we have
assumed integer values of $n$ in its derivation, and the spherical
Bessel functions are given by
\begin{align}
  j_{n}(x) = \sqrt{\frac{\pi}{2x}}J_{n+\frac{1}{2}}(x)\,\,.
  \label{eq:spherical_to_cylindrical}
\end{align}

We can show, however, that any set of functions (i.e. even those
for which $n$ is not integer) fulfilling the
recurrence relation in fact also fulfils the Bessel differential
equation, thus leading us to a recurrence relation which is valid
regardless of whether the order is integer or not.
This can be accomplished by first obtaining a second recursion
relation, this time by differentiating the
generating function $g(x,t)$ with respect to $x$, giving us
\begin{align}
  \sum_{n=-\infty}^{\infty} J_{n}'(x) t^{n}
  = \partial_{x} g(x,t)
  &= \frac{1}{2}\left(t-\frac{1}{t}\right)e^{\frac{x}{2}\left(t-\frac{1}{t}\right)} \nonumber \\
  &= \frac{t}{2} \sum_{n=-\infty}^{\infty} J_{n}(x) t^{n} - \frac{1}{2t} \sum_{n=-\infty}^{\infty} J_{n}(x) t^{n} \nonumber \\
  &= \frac{1}{2} \sum_{n=-\infty}^{\infty} J_{n-1}(x) t^{n} -
  \frac{1}{2} \sum_{n=-\infty}^{\infty} J_{n+1}(x) t^{n}\,\,,
\end{align}
from which it follows that
\begin{align}
  J_{n}'(x) = \frac{1}{2} \left(J_{n-1}(x) - J_{n+1}(x)\right)\,\,.
  \label{eq:diff_rec_rel}
\end{align}
By adding and subtracting the two recursion relations
and performing the appropriate index shifts, we obtain
\begin{align}
  J_{n}(x) = \frac{n\pm 1}{x} J_{n\pm 1}(x) \pm J_{n\pm 1}'(x)\,\,.
\end{align}
Provided we had a some set of functions $Z_{\nu}(x)$, which fulfil
the above recursion relations then by combining
$x^2 \partial_x Z_{\nu}'(x)$, $x Z_{\nu}'(x)$ (left-hand-side
of \cref{eq:diff_rec_rel}) and $\nu^2 Z_{\nu}$ (left-hand-side of
\cref{eq:base_rec_rel})
gives
\begin{align}
  & x^2 Z_{\nu}''(x) + x Z_{\nu}'(x) - \nu^2 Z_{\nu} \nonumber \\
  &= \frac{x^{2}}{2} \left( Z_{\nu-1}'(x) - Z_{\nu+1}'(x) + \frac{1}{x}\left(Z_{\nu-1}(x) - Z_{\nu+1}(x)\right)
  -\frac{\nu}{x} \left(Z_{\nu-1}(x) + Z_{\nu+1}(x)\right) \right) \nonumber \\
  &= \frac{x^2}{2} \big( \underbrace{Z_{\nu-1}'(x) -\frac{\nu-1}{x} Z_{\nu-1}(x)}_{=-Z_{\nu}(x)}
  - \underbrace{\left(Z_{\nu+1}'(x) + \frac{\nu+1}{x} Z_{\nu+1}(x)\right)}_{=Z_{\nu}(x)}\big)\,\,,
\end{align}
i.e.,
\begin{align}
  x^2 Z_{\nu}''(x) + x Z_{\nu}'(x)+ (x^2  - \nu^2) Z_{\nu} = 0\,\,,
\end{align}
which is our original Bessel's differential equation!
We have thus shown that any functions satisfying the recursion
relations \cref{eq:base_rec_rel,eq:diff_rec_rel} are valid
solutions to the differential equation,
even when $\nu$ is a non-integer number.
As can be seen from their series representation, which is valid
for arbitrary $\nu$, the half-integer Bessel functions $J_{\ell+1/2}$
accordingly fulfil the recursion relations above \cite{arfken2013methods}.
Returning to \cref{eq:spherical_to_cylindrical},
where we had established that spherical Bessel functions can
be expressed as half-integer cylindrical Bessel functions,
one could in principle make use of the (non-integer) recursion relations
to start generating spherical Bessel functions of higher order.
The problem with starting at low $\ell$ and evaluating 
\begin{align*}
    j_{\ell+1}(x)&=\frac{2\ell+1}{x}j_{\ell}(x)-j_{\ell-1}(x) \\
    y_{\ell+1}(x)&=\frac{2\ell+1}{x}y_{\ell}(x)-y_{\ell-1}(x)\,\,,
\end{align*}
numerically is that it becomes unstable for $j_{\ell}(x)$ when $\ell >x$, on 
account of its rapid decay (rounding errors from the growing $y_{\ell}(x)$ solutions
quickly swamp it). 
For scattering calculations, which rely on Bessel functions that can fall into exactly
this regime, we therefore recommend the use of library routines, which implement fast 
and accurate algorithms to evaluate the Bessel functions in a reliable way.
This caveat also applies to the modified Bessel functions (as used for closed channels
in scattering calculations), which may also be found in mathematical software libraries.

\end{appendix}

\bibliography{refs}

\begin{thebibliography}{10}
\providecommand{\url}[1]{\texttt{#1}}
\providecommand{\urlprefix}{URL }
\expandafter\ifx\csname urlstyle\endcsname\relax
  \providecommand{\doi}[1]{doi:\discretionary{}{}{}#1}\else
  \providecommand{\doi}{doi:\discretionary{}{}{}\begingroup
  \urlstyle{rm}\Url}\fi
\providecommand{\eprint}[2][]{\url{#2}}

\bibitem{taylor2006scattering}
J.~R. Taylor,
\newblock \emph{Scattering Theory: The Quantum Theory of Nonrelativistic
  Collisions},
\newblock Dover Publications, Mineola, NY,
\newblock Reprint of the 1972 Wiley edition (2006).

\bibitem{child1996molecular}
M.~S. Child,
\newblock \emph{Molecular Collision Theory},
\newblock Dover Publications, New York,
\newblock Republication of the 1984 corrected reprint of the 1974 Academic
  Press edition (1996).

\bibitem{krems2018molecules}
R.~V. Krems,
\newblock \emph{Molecules in electromagnetic fields: from ultracold physics to
  controlled chemistry},
\newblock John Wiley \& Sons (2018).

\bibitem{hutson1994coupled}
J.~M. Hutson,
\newblock \emph{Coupled channel methods for solving the bound-state
  schrödinger equation},
\newblock Computer Physics Communications \textbf{84}(1), 1 (1994),
\newblock \doi{https://doi.org/10.1016/0010-4655(94)90200-3}.

\bibitem{hutson2019molscat}
J.~M. Hutson and C.~R. Le~Sueur,
\newblock \emph{{MOLSCAT}: A program for non-reactive quantum scattering
  calculations on atomic and molecular collisions},
\newblock Computer Physics Communications \textbf{241}, 9 (2019),
\newblock \doi{10.1016/j.cpc.2019.02.014}.

\bibitem{alexander2023hibridon}
M.~H. Alexander, P.~J. Dagdigian, H.-J. Werner, J.~K{\l}os, B.~Desrousseaux,
  G.~Raffy and F.~Lique,
\newblock \emph{{Hibridon}: A program suite for time-independent non-reactive
  quantum scattering calculations},
\newblock Computer Physics Communications \textbf{289}, 108761 (2023),
\newblock \doi{10.1016/j.cpc.2023.108761}.

\bibitem{sakurai2011modern}
J.~J. Sakurai and J.~Napolitano,
\newblock \emph{Modern Quantum Mechanics},
\newblock Addison-Wesley, 2nd edn. (2011).

\bibitem{cohen1986quantumv2}
C.~Cohen-Tannoudji, B.~Diu and F.~Laloe,
\newblock \emph{Quantum Mechanics, Volume 2}, vol.~2 (1986).

\bibitem{arfken2013methods}
G.~B. Arfken, H.~J. Weber and F.~E. Harris,
\newblock \emph{Mathematical Methods for Physicists},
\newblock Academic Press, Boston, 7th edn.,
\newblock ISBN 978-0-12-384654-9 (2013).

\bibitem{strang2016introduction}
G.~Strang,
\newblock \emph{Introduction to Linear Algebra},
\newblock Wellesley-Cambridge Press, 5th edn. (2016).

\bibitem{press2007numerical}
W.~H. Press, S.~A. Teukolsky, W.~T. Vetterling and B.~P. Flannery,
\newblock \emph{Numerical Recipes: The Art of Scientific Computing},
\newblock Cambridge University Press, Cambridge, 3rd edn. (2007).

\bibitem{pack1974space}
R.~T. Pack,
\newblock \emph{Space-fixed vs body-fixed axes in atom-diatomic molecule
  scattering. sudden approximations},
\newblock The Journal of Chemical Physics \textbf{60}(2), 633 (1974).

\bibitem{born1927quantentheorie}
M.~Born and R.~Oppenheimer,
\newblock \emph{Zur quantentheorie der molekeln},
\newblock Annalen der Physik \textbf{389}(20), 457 (1927),
\newblock \doi{https://doi.org/10.1002/andp.19273892002},
\newblock
  \eprint{https://onlinelibrary.wiley.com/doi/pdf/10.1002/andp.19273892002}.

\bibitem{hartree1928wavei}
D.~R. Hartree,
\newblock \emph{The wave mechanics of an atom with a non-coulomb central field.
  part i. theory and methods},
\newblock In \emph{Mathematical Proceedings of the Cambridge Philosophical
  Society}, vol.~24, pp. 89--110. Cambridge University Press (1928).

\bibitem{bartlett2007coupled}
R.~J. Bartlett and M.~Musia{\l},
\newblock \emph{Coupled-cluster theory in quantum chemistry},
\newblock Reviews of Modern Physics \textbf{79}(1), 291 (2007),
\newblock \doi{10.1103/RevModPhys.79.291}.

\bibitem{szalay2012multiconfiguration}
P.~G. Szalay, T.~Muller, G.~Gidofalvi, H.~Lischka and R.~Shepard,
\newblock \emph{Multiconfiguration self-consistent field and multireference
  configuration interaction methods and applications},
\newblock Chemical reviews \textbf{112}(1), 108 (2012).

\bibitem{ho1996general}
T.-S. Ho and H.~Rabitz,
\newblock \emph{A general method for constructing multidimensional molecular
  potential energy surfaces from ab initio calculations},
\newblock The Journal of chemical physics \textbf{104}(7), 2584 (1996),
\newblock \doi{10.1063/1.470946}.

\bibitem{unke2017toolkit}
O.~T. Unke and M.~Meuwly,
\newblock \emph{Toolkit for the construction of reproducing kernel-based
  representations of data: Application to multidimensional potential energy
  surfaces},
\newblock Journal of Chemical Information and Modeling \textbf{57}(8), 1923
  (2017),
\newblock \doi{10.1021/acs.jcim.7b00090}.

\bibitem{meuwly_rkhs_software}
O.~T. Unke and M.~Meuwly,
\newblock \emph{{RKHS}: A {Fortran} toolkit for reproducing kernel {Hilbert}
  space interpolation},
\newblock \url{https://github.com/MMunibas/RKHS},
\newblock Accessed on 16. September 2026 (2017).

\bibitem{messiah1974quantum}
A.~Messiah,
\newblock \emph{Quantum mechanics, Volume 1},
\newblock North Holland Publishing Company (1974).

\bibitem{nolting2012grundkurs}
W.~Nolting,
\newblock \emph{Grundkurs Theoretische Physik 5/2: Quantenmechanik-Methoden und
  Anwendungen},
\newblock Springer (2012).

\bibitem{lester1976dynamics}
W.~A. Lester~Jr,
\newblock \emph{The n coupled-channel problem},
\newblock In \emph{Dynamics of Molecular Collisions: Part A}, pp. 1--32.
  Springer (1976).

\bibitem{tiemann2020beyond}
E.~Tiemann, P.~Gersema, K.~K. Voges, T.~Hartmann, A.~Zenesini and S.~Ospelkaus,
\newblock \emph{Beyond born-oppenheimer approximation in ultracold atomic
  collisions},
\newblock Phys. Rev. Res. \textbf{2}, 013366 (2020),
\newblock \doi{10.1103/PhysRevResearch.2.013366}.

\bibitem{johnson1977new}
B.~R. Johnson,
\newblock \emph{New numerical methods applied to solving the one-dimensional
  eigenvalue problem},
\newblock The Journal of Chemical Physics \textbf{67}(9), 4086 (1977).

\bibitem{haber2018physics}
H.~Haber,
\newblock \emph{Expansion of plane waves in spherical harmonics, physics 215
  handout},
\newblock
  \url{https://web.archive.org/web/20230327171441/http://scipp.ucsc.edu/~haber/ph215/PlaneWaveExpansion.pdf},
\newblock Archived on 27. March 2023 (2018).

\bibitem{karman2014renormalized}
T.~Karman, L.~Janssen, R.~Sprenkels and G.~C. Groenenboom,
\newblock \emph{A renormalized potential-following propagation algorithm for
  solving the coupled-channels equations},
\newblock The Journal of Chemical Physics \textbf{141}(6), 064102 (2014).

\bibitem{littlejohn2021time}
R.~G. Littlejohn,
\newblock \emph{Time reversal, physics 221a, notes 22},
\newblock
  \url{https://web.archive.org/web/20230627210640/https://bohr.physics.berkeley.edu/classes/221/notes/timerev.pdf},
\newblock Archived on 27. June 2023 (2021).

\bibitem{althorpe2003quantum}
S.~C. Althorpe and D.~C. Clary,
\newblock \emph{Quantum scattering calculations on chemical reactions},
\newblock Annual review of physical chemistry \textbf{54}(1), 493 (2003).

\bibitem{friedrich2013scattering}
H.~Friedrich,
\newblock \emph{Scattering theory}, vol. 872,
\newblock Springer (2013).

\bibitem{vigo2005variable}
J.~Vigo-Aguiar and H.~Ramos,
\newblock \emph{A variable-step numerov method for the numerical solution of
  the schr{\"o}dinger equation},
\newblock Journal of mathematical chemistry \textbf{37}, 255 (2005).

\bibitem{johnson1973multichannel}
B.~R. Johnson,
\newblock \emph{The multichannel log-derivative method for scattering
  calculations},
\newblock Journal of Computational Physics \textbf{13}(3), 445 (1973),
\newblock \doi{https://doi.org/10.1016/0021-9991(73)90049-1}.

\bibitem{manolopoulos1986improved}
D.~E. Manolopoulos,
\newblock \emph{An improved log derivative method for inelastic scattering},
\newblock The Journal of Chemical Physics \textbf{85}(11), 6425 (1986),
\newblock \doi{10.1063/1.451472}.

\bibitem{liu1978effect}
W.-K. Liu, J.~Grabenstetter, R.~Le~Roy and F.~McCourt,
\newblock \emph{Effect of asymmetric isotopic substitution on atom--diatom
  potentials},
\newblock The Journal of Chemical Physics \textbf{68}(11), 5028 (1978).

\bibitem{landau1991quantum}
L.~D. Landau and E.~M. Lifshitz,
\newblock \emph{Quantum mechanics: Non-relativistic theory}, vol.~3,
\newblock Pergamon Press, third edn. (1991).

\bibitem{groenenboom1993combining}
G.~C. Groenenboom and D.~T. Colbert,
\newblock \emph{Combining the discrete variable representation with the
  s-matrix kohn method for quantum reactive scattering},
\newblock The Journal of Chemical Physics \textbf{99}(12), 9681 (1993).

\bibitem{groenenboom2001discrete}
G.~C. Groenenboom,
\newblock \emph{The discrete variable representation},
\newblock
  \url{https://web.archive.org/web/20220124011015/https://www.theochem.ru.nl/files/dbase/gcg2001.pdf},
\newblock Archived on 12. December 2022 (2001).

\bibitem{lill1982discrete}
J.~Lill, G.~Parker and J.~Light,
\newblock \emph{Discrete variable representations and sudden models in quantum
  scattering theory},
\newblock Chemical Physics Letters \textbf{89}(6), 483 (1982).

\bibitem{colbert1992novel}
D.~T. Colbert and W.~H. Miller,
\newblock \emph{A novel discrete variable representation for quantum mechanical
  reactive scattering via the s-matrix kohn method},
\newblock The Journal of Chemical Physics \textbf{96}(3), 1982 (1992).

\bibitem{meyer2017numerical}
H.-D. Meyer,
\newblock \emph{Numerical methods of quantum dynamics discrete variable
  representation (dvr) integrators},
\newblock
  \url{https://web.archive.org/web/20231216103255/https://www.pci.uni-heidelberg.de/tc/usr/mctdh/lit/NumericalMethods.pdf},
\newblock Archived on 16. December 2023 (2017).

\bibitem{abramowitz1988handbook}
M.~Abramowitz, I.~A. Stegun and R.~H. Romer,
\newblock \emph{Handbook of mathematical functions with formulas, graphs, and
  mathematical tables},
\newblock American Association of Physics Teachers (1988).

\bibitem{percival1957partial}
I.~Percival and M.~Seaton,
\newblock \emph{The partial wave theory of electron-hydrogen atom collisions},
\newblock In \emph{Mathematical Proceedings of the Cambridge Philosophical
  Society}, vol.~53, pp. 654--662. Cambridge University Press (1957).

\bibitem{wormer2006angular}
P.~Wormer,
\newblock \emph{Angular momentum theory},
\newblock
  \url{https://web.archive.org/web/20230110064826/https://www.theochem.ru.nl/~pwormer/teachmat/angmom.pdf},
\newblock \doi{10.1007/978-0-387-26308-3_2},
\newblock Archived on 10. January 2023 (2006).

\bibitem{arthurs1960theory}
A.~Arthurs and A.~Dalgarno,
\newblock \emph{The theory of scattering by a rigid rotator},
\newblock Proceedings of the Royal Society of London. Series A. Mathematical
  and Physical Sciences \textbf{256}(1287), 540 (1960).

\bibitem{karman2015collision}
T.~Karman, A.~van~der Avoird and G.~C. Groenenboom,
\newblock \emph{Collision-induced absorption with exchange effects and
  anisotropic interactions: Theory and application to $\mathrm{H}_2-
  \mathrm{H}_2$},
\newblock The Journal of Chemical Physics \textbf{142}(8) (2015).

\bibitem{gadea1997nonradiative}
F.~Gad{\'e}a, H.~Berriche, O.~Roncero, P.~Villarreal and G.~Delgado~Barrio,
\newblock \emph{Nonradiative lifetimes for $\mathrm{LiH}$ in the $a$ state
  using adiabatic and diabatic schemes},
\newblock The Journal of Chemical Physics \textbf{107}(24), 10515 (1997).

\bibitem{margulis2023tomography}
B.~Margulis, K.~P. Horn, D.~M. Reich, M.~Upadhyay, N.~Kahn, A.~Christianen,
  A.~van~der Avoird, G.~C. Groenenboom, M.~Meuwly, C.~P. Koch and
  E.~Narevicius,
\newblock \emph{Tomography of feshbach resonance states},
\newblock Science \textbf{380}(6640), 77 (2023),
\newblock \doi{10.1126/science.adf9888},
\newblock \eprint{https://www.science.org/doi/pdf/10.1126/science.adf9888}.

\bibitem{karplus1965exchange}
M.~Karplus, R.~N. Porter and R.~D. Sharma,
\newblock \emph{Exchange reactions with activation energy. i. simple barrier
  potential for (h, h2)},
\newblock The Journal of Chemical Physics \textbf{43}(9), 3259 (1965),
\newblock \doi{10.1063/1.1697301},
\newblock
  \eprint{https://pubs.aip.org/aip/jcp/article-pdf/43/9/3259/18840257/3259_1_online.pdf}.

\bibitem{jensen2017introduction}
F.~Jensen,
\newblock \emph{Introduction to computational chemistry},
\newblock John wiley \& sons (2017).

\bibitem{pack1987quantum}
R.~T. Pack and G.~A. Parker,
\newblock \emph{Quantum reactive scattering in three dimensions using
  hyperspherical ({APH}) coordinates. {T}heory},
\newblock The Journal of Chemical Physics \textbf{87}, 3888 (1987).

\bibitem{neuhauser1991application}
D.~Neuhauser, M.~Baer, R.~S. Judson and D.~J. Kouri,
\newblock \emph{The application of time-dependent wavepacket methods to
  reactive scattering},
\newblock Computer physics communications \textbf{63}(1-3), 460 (1991).

\bibitem{arfken2013mathematical}
G.~B. Arfken, H.~J. Weber and F.~E. Harris,
\newblock \emph{Chapter 14 - bessel functions},
\newblock In G.~B. Arfken, H.~J. Weber and F.~E. Harris, eds.,
  \emph{Mathematical Methods for Physicists (Seventh Edition)}, pp. 643--713.
  Academic Press, Boston, seventh edition edn.,
\newblock ISBN 978-0-12-384654-9,
\newblock \doi{https://doi.org/10.1016/B978-0-12-384654-9.00014-1} (2013).

\bibitem{ryan2008techniques}
S.~M. Ryan,
\newblock \emph{Techniques for theoretical physics, frobenius series
  solutions},
\newblock
  \url{https://web.archive.org/web/20240116143251/https://www.maths.tcd.ie/~ryan/teaching/11404/ODEs5.pdf},
\newblock Archived on 16. January 2024.

\bibitem{asmar2016partial}
N.~H. Asmar,
\newblock \emph{Partial differential equations with Fourier series and boundary
  value problems},
\newblock Courier Dover Publications (2016).

\end{thebibliography}

\nolinenumbers

\end{document}